\documentclass[useamsfonts]{pasj01}

\usepackage{graphicx}
\usepackage{todonotes}
\usepackage{xcolor}
\usepackage{verbatim}
\usepackage{natbib}
\usepackage{bm}
\usepackage{multirow}
\usepackage{array}

\begin{document}
\SetRunningHead{Bosch et al}{The HSC Pipeline}

\title{The Hyper Suprime-Cam Software Pipeline}

\author{
  James Bosch\altaffilmark{1*},
  Robert Armstrong\altaffilmark{1},
  Steven Bickerton\altaffilmark{2,3},
  Hisanori Furusawa\altaffilmark{4},
  Hiroyuki Ikeda\altaffilmark{4},
  Michitaro Koike\altaffilmark{4},
  Robert Lupton\altaffilmark{1},
  Sogo Mineo\altaffilmark{4},
  Paul Price\altaffilmark{1},
  Tadafumi Takata\altaffilmark{4,5},
  Masayuki Tanaka\altaffilmark{4},
  Naoki Yasuda\altaffilmark{3},
  Yusra AlSayyad\altaffilmark{1},
  Andrew C. Becker\altaffilmark{6},
  William Coulton\altaffilmark{7},
  Jean Coupon\altaffilmark{8},
  Jose Garmilla\altaffilmark{1,9},
  Song Huang\altaffilmark{3,10},
  K. Simon Krughoff\altaffilmark{6},
  Dustin Lang\altaffilmark{11},
  Alexie Leauthaud\altaffilmark{3,9},
  Kian-Tat Lim\altaffilmark{12},
  Nate B. Lust\altaffilmark{1},
  Lauren A. MacArthur\altaffilmark{1},
  Rachel Mandelbaum\altaffilmark{13},
  Hironao Miyatake\altaffilmark{3,14},
  Satoshi Miyazaki\altaffilmark{4,5},
  Ryoma Murata\altaffilmark{3,15},
  Surhud More\altaffilmark{3},
  Yuki Okura\altaffilmark{16,17},
  Russell Owen\altaffilmark{6},
  John D. Swinbank\altaffilmark{1},
  Michael A. Strauss\altaffilmark{1},
  Yoshihiko Yamada\altaffilmark{4},
  Hitomi Yamanoi\altaffilmark{4}
}
\altaffiltext{1}{Department of Astrophysical Sciences, Princeton University, 4 Ivy Lane, Princeton, NJ 08544}
\altaffiltext{2}{Orbital Insight, 100 W. Evelyn Ave. Mountain View, CA 94041}
\altaffiltext{3}{Kavli Institute for the Physics and Mathematics of the Universe (Kavli IPMU, WPI), University of Tokyo, Chiba 277-8582, Japan}
\altaffiltext{4}{National Astronomical Observatory of Japan, 2-21-1 Osawa, Mitaka, Tokyo 181-8588, Japan}
\altaffiltext{5}{Department of Astronomy, School of Science, Graduate University for Advanced Studies (SOKENDAI), 2-21-1, Osawa, Mitaka, Tokyo 181-8588, Japan}
\altaffiltext{6}{Astronomy Department, University of Washington, Seattle, WA 98195}
\altaffiltext{7}{Department of Physics, Princeton University, Jadwin Hall, Princeton, NJ 08544}
\altaffiltext{8}{Department of Astronomy, University of Geneva, ch. d'\'Ecogia 16, 1290, Versoix, Switzerland}
\altaffiltext{9}{Massive Dynamics, One Palmer Square Suite 530, Princeton, NJ 08542}
\altaffiltext{10}{Department of Astronomy and Astrophysics, University of California, Santa Cruz, 1156 High Street, Santa Cruz, CA 95064 USA}
\altaffiltext{11}{Dunlap Institute and Department of Astronomy \& Astrophysics, University of Toronto, 50 St George St, Toronto, ON M5S 3H4, Canada}
\altaffiltext{12}{SLAC National Accelerator Laboratory, 2575 Sand Hill Road, Menlo Park, CA 94025}
\altaffiltext{13}{McWilliams Center for Cosmology, Department of Physics, Carnegie Mellon University, Pittsburgh, PA 15213, USA}
\altaffiltext{14}{Jet Propulsion Laboratory, California Institute of Technology, Pasadena, CA 91109, USA}
\altaffiltext{15}{Department of Physics, University of Tokyo, Tokyo 113-0033, Japan}
\altaffiltext{15}{RIKEN High Energy Astrophysics Laboratory,　2-1 Hirosawa, Wako, Saitama 351-0198, Japan}
\altaffiltext{16}{RIKEN BNL Research Center, Bldg. 510A, 20 Pennsylvania Street, Brookhaven National Laboratory, Upton, NY 11973}

\altaffiltext{*}{Corresponding Author}
\email{jbosch@astro.princeton.edu}

\newcommand{\eg}{{\textit{e.g.}}}
\newcommand{\ie}{{\textit{i.e.}}}
\newcommand{\secref}[1]{Section~\ref{sec:#1}}
\newcommand{\figref}[1]{Figure~\ref{fig:#1}}
\newcommand{\eqnref}[1]{Eqn.~(\ref{eqn:#1})}
\newcommand{\appendixref}[1]{Appendix~\ref{sec:#1}}

\newcommand\note[1]{\todo[color=yellow, inline, size=\small]{#1}}
\newcommand{\rachel}[1]{{\textcolor{blue}{(RM: #1)}}}



\maketitle

\begin{abstract}
In this paper, we describe the optical imaging data processing pipeline developed for the Subaru Telescope's Hyper Suprime-Cam (HSC) instrument.  The HSC Pipeline builds on the prototype pipeline being developed by the Large Synoptic Survey Telescope's Data Management system, adding customizations for HSC, large-scale processing capabilities, and novel algorithms that have since been reincorporated into the LSST codebase.  While designed primarily to reduce HSC Subaru Strategic Program (SSP) data, it is also the recommended pipeline for reducing general-observer HSC data.  The HSC pipeline includes high level processing steps that generate coadded images and science-ready catalogs as well as low-level detrending and image characterizations.
\end{abstract}

\section{Introduction}
\label{sec:introduction}

One of the most important developments in modern astronomy is the steadily increasing survey power of optical telescopes, which has been driven in large part by a generation of wide field-of-view CCD imaging cameras.  The science enabled by modern wide imaging surveys with such instruments is extremely broad, because large samples of objects are important for both finding rare objects and making precise measurements of noisy quantities.

Constructing a scientifically useful large catalog from a wide-field photometric survey is a major challenge in its own right, however.  Searches for rare objects such as high-redshift quasars or strong gravitational lenses are easily swamped by false positives due to processing problems such as spurious detections or unmasked instrumental artifacts.  While follow-up observations of these objects may be necessary for both confirmation and to extract information about \eg~reionization and dark matter substructure (respectively), even producing a candidate sample small enough to vet by-eye (let alone propose for follow-up observations) from a large survey requires careful image processing to avoid these false positives.  Similarly, studies of statistical signals such as cosmological weak lensing or galaxy clustering are very sensitive to even small systematic errors in galaxy shape measurements and photometry.  Because the tolerance for systematic errors is essentially proportional to the expected statistical error, the quality (and generally the sophistication) of processing must improve as survey area and depth increase to avoid wasting some of the survey's statistical power.

Most modern surveys have responded to this challenge by building a dedicated pipeline responsible for producing a catalog that can be used for most science without further pixel-level processing, developed by a team of scientists focused on developing new algorithms and understanding the features of the data.  The Hyper Suprime-Cam (HSC) team has taken this approach with both its Subaru Strategic Program (SSP) Survey \citep{2017arXiv170405858A} and general-use observers, which use the same pipeline.  We have accomplished this by building the HSC Pipeline as a customization of the Large Synoptic Survey Telescope (\citealt{2009arXiv0912.0201L}) Data Management (DM) software stack \citep{2015arXiv151207914J}, which in turn builds on an algorithmic and conceptual foundation inherited from the Sloan Digital Sky Survey (SDSS) \textit{Photo} Pipeline \citep{2001ASPC..238..269L}.  A large fraction of the HSC pipeline team is also part of LSST DM, and this relationship is highly symbiotic; HSC provides an invaluable early testbed for algorithms and software being developed for LSST, while LSST provides a much broader team of developers and support scientists than is available purely within the smaller HSC collaboration.

Both the HSC Pipeline and the LSST DM stack are open source software, released under the GNU General Public License (Version 3).  This paper does not attempt to provide a tutorial for downloading, installing, and using the HSC Pipeline or the LSST DM stack.  Readers should instead consult the LSST DM\footnote{\texttt{http://dm.lsst.org}} or HSC\footnote{\texttt{http://hsc.mtk.nao.ac.jp/ssp/pipedoc}} websites for that information.

\subsection{HSC Overview}
\label{sec:hsc-overview}

Hyper Suprime-Cam is a 1.77 $\mathrm{deg}^2$ imaging camera at the prime focus of the Subaru telescope.  It includes 116 2k$\times$4k CCDs (104 science sensors, 4 guide sensors, and 8 focus sensors) with 0.168\arcsec\ pixels.  Some of the most important science it enables comes from the primary associated survey, the HSC-SSP.  The HSC-SSP is a multi-layer survey comprised of a 1400 $\mathrm{deg}^2$ Wide layer, a 27 $\mathrm{deg}^2$ Deep layer, and a 3.5 $\mathrm{deg}^2$ UltraDeep layer, with depths of approximately  (5$\sigma$ point source, AB magnitude) $r\sim 26$, $r\sim 27$, and $r\sim 28$, respectively.  All three layers utilize $grizy$ broad-band filters, with four additional narrow-band filters also used in the Deep and UltraDeep layers.  The HSC instrument is more fully described in \citet{hsc-instrument} and \citet{hsc-dewar}, while the SSP is described in \citet{2017arXiv170405858A}.  The HSC Pipeline is also a major component of the on-site quality analysis system used by observers \citep{hsc-onsite}.

Approximately 100 $\mathrm{deg}^2$ of the Wide survey were observed to full depth in all bands as of November 2015, and are included in the first public data release (PDR1) produced by the software version (4.0) described in this paper.  This data release is described in detail in \citet{2017arXiv170208449A}, which also includes additional validation of the HSC Pipeline specific to the SSP dataset.  HSC-SSP data releases are served via a web portal and database described in \citet{hsc-db}.

Because the SSP wide observing strategy is depth-first, the survey already goes significantly deeper than other ongoing ground-based optical surveys such as the Dark Energy Survey (DES; \citealt{2016MNRAS.460.1270D}) and the Kilo Degree Survey (KiDS; \citealt{2013ExA....35...25D}).  This puts HSC in a qualitatively different regime for image processing algorithms, simply because a large fraction of objects are blended (\ie~they have significantly overlapping isophotes at the surface brightness limit of the survey).  While astronomical image processing pipelines have always had to deal with blending at some level, at shallow depths blends are rare enough that they can simply be omitted from most analyses.  At HSC (and LSST) depths, that sort of cut is simply unacceptable: 58\% of objects in the HSC Wide survey are blended, in the sense that they were detected as part of an above-threshold region of sky containing multiple significant peaks in surface brightness.  Because most of the blended objects are galaxies, algorithms developed for crowded stellar fields (\eg~\citealt{1987PASP...99..191S}) are not applicable.  One advantage HSC has in this regard is its spectacular seeing (0.6\arcsec\ median FWHM in the $i$-band), but this can also make this \textit{deblending} challenge more severe; neighbors that might have been hopelessly unresolved in worse seeing may now be possible to deblend, but these are in general the hardest blends to handle because we are limited more by the unknown morphologies of the objects than the (known) point-spread function.

Like DES, KiDS, and LSST, many of HSC's science goals are cosmological; we seek to measure the distribution and evolution of structure in the universe and use these to infer the properties of dark matter and dark energy.  One of the most important cosmological probes in photometric surveys is weak gravitational lensing \citep[\eg][]{2008ARNPS..58...99H}, which uses the shapes of distant galaxies to measure the foreground matter distribution via the coherent distortions of those shapes.  These distortions are extremely subtle compared with many observational effects, such as convolution with the point-spread function (PSF) and distortion due to the optics, and this puts extremely stringent requirements on the image processing pipeline.  At present, all image processing for weak lensing is performed by the main HSC Pipeline; there is no independent or follow-up pipeline for weak lensing.  The HSC weak lensing team has put considerable additional effort into validating the HSC Pipeline's data products, however, as well as an extensive simulation effort to calibrate its results.  We refer the reader to \citet{hsc-shear} for more information on the performance of the HSC Pipeline for weak lensing.

\subsection{Terminology}
\label{sec:terminology}

We adopt the term \emph{visit} from LSST to denote a single exposure (for LSST, a visit is actually a pair of back-to-back exposures of the same pointing, but these are combined in the earliest stages of processing).  Raw and processed CCD-level datasets are thus identified by the combination of a visit ID and CCD ID.  Most HSC observations (including all SSP observations) involve multiple dithered exposures at each pointing.  We process all visits that overlap each area of sky together, primarily by building coadded images.

The HSC/LSST pipeline divides the sky (or some region of interest on the sky) into \emph{tracts}.  Each tract is a rectangular region with a common map projection, which is further subdivided into \emph{patches}, which share the tract coordinate system but define smaller regions that are a convenient size for processing and storing images.  Tracts overlap on the sky, and patches overlap within a tract; we process these overlap regions multiple times, and then resolve duplicates at the catalog level.

While the pipeline itself allows essentially arbitrary tract definitions, SSP productions use a variant of the \emph{rings} sky tessellation initially developed by the Pan-STARRS team \citep{2016arXiv161205245W}, which splits the sky into a number of constant-declination rings (120, in our case) and then divides these in right ascension into approximately square overlapping tracts of constant size.  These tracts are approximately 1.68 $\mathrm{deg}$ on a side (slightly larger than the HSC field of view) and they overlap by at least 1\arcmin\ on each side.  Each tract is divided into 9$\times$9 patches that are 4k$\times$4k pixels (0.03434 $\mathrm{deg}^2$), with 100-pixel overlaps; these coadd pixel are the same size on the sky (0.168\arcsec) as the raw data pixels.

\subsection{Organization of this Paper}
\label{sec:organization}

\secref{architecture} describes the HSC Pipeline's software architecture (which is largely inherited directly from LSST DM).  Readers interested only in the algorithms from a scientific perspective could consider skipping this section.  \secref{pipelines} describes the HSC Pipeline at the highest level, and is organized according to the actual sequencing of pipeline stages.  Lower-level algorithms that are novel, unusual, or particularly complex are described more fully in \secref{algorithms}, which also includes demonstrations of their performance.  Overall assessment of the pipeline's performance and future plans are discussed in \secref{future-work-and-conclusions}.

\section{Software Architecture}
\label{sec:architecture}

The HSC Pipeline described here is a fork of the original LSST DM software stack that began to slowly diverge from the main LSST codebase in early 2014.  Over the course of 2016, the features introduced on the HSC side were fully reintegrated into the LSST stack, including the ability to fully process raw HSC data, and all future HSC-SSP data releases will be produced with a minimally-modified version of the main LSST stack.  This paper thus describes the current algorithmic capabilities of the LSST stack in a general sense, as these have been driven largely by HSC development recently, but it explicitly focuses on the HSC Pipeline implementation and the features (and shortcomings) present in the version used to process the recent SSP PDR1 \citep{2017arXiv170208449A}.  All SSP internal data releases thus far have also been produced with the HSC Pipeline fork.

Both the LSST Data Management codebase and the HSC Pipeline are written in a combination of C++ and Python.  Flexibility and extensibility are core development goals at all levels of the codebase: we have attempted to make virtually every algorithm not just configurable but replaceable (even by external code).  This flexibility is important for LSST in part because many algorithmic choices in LSST processing still require significant research and are under active development.  As the HSC Pipeline explicitly demonstrates, it also makes it possible to begin to commission LSST's pipelines on similar precursor datasets well before LSST commissioning begins.

\subsection{Tasks: Configurable Python Algorithms}

High-level algorithms are written in Python, primarily as subclasses of our \texttt{Task} class.  Each \texttt{Task} subclass defines a set of configuration options as an associated Python \texttt{Config} class, but otherwise \texttt{Task}s are essentially arbitrary Python callables.  \texttt{Task}s may also be nested, with a parent \texttt{Task} delegating work to one or more ``subtasks'', and this nesting is reflected in the associated \texttt{Config} classes; a top-level \texttt{Config} instance holds the configuration for all subtask \texttt{Config}s, forming a complete tree for that unit of processing.  Our configuration files are simply Python files that set the attributes of the configuration tree.

Our configuration system also allows \emph{any} subtask in the hierarchy to be \textit{retargeted} to a different \texttt{Task}  with the same signature without changing any code, simply by adding a Python \texttt{import} statement to a configuration file to load the replacement \texttt{Task} and calling a \texttt{retarget} method on the configuration node representing a subtask.  Retargeting is our simplest extensibility mechanism, and it is used by the HSC Pipeline to customize the generic detrending algorithms provided by LSST's Instrument Signature Removal (ISR) \texttt{Task} with a variant specialized for HSC.  All other HSC Pipeline tasks are the same \texttt{Task}s used by default in the LSST DM stack (though in many cases these \texttt{Tasks} were first developed on the HSC Pipeline fork and then ported back to the LSST codebase).

The most frequently used points of customization are actually in lower-level algorithms that are not written as \texttt{Task}s.  These specialized customization points include algorithms for PSF estimation (see \secref{psf-modeling}) and source measurement (\secref{source-measurement}).  As with \texttt{Task}s, a user-supplied algorithm can be added to the pipeline simply by \texttt{importing} it in a Python-parsed configuration file; this adds the algorithm to the list (called a \texttt{Registry}) of available algorithms for a pipeline step, allowing it to be enabled in the same way internal algorithms are selected.

\subsection{Defining the Python/C++ Boundary}

Low-level algorithms that perform direct pixel-level processing or other computationally intensive operations are written in C++.  We also define most of our low-level primitive classes and data structures (\eg~images, PSF models, geometric regions) in C++, making them available to Python code using SWIG\footnote{\raggedright We have nearly completed a switch from SWIG (\texttt{http://www.swig.org}) to pybind11 (\texttt{http://pybind11.readthedocs.io}); the latest version of the LSST DM codebase uses pybind11 instead.}.  This is an important philosophical difference from other astronomical software projects that mix Python and C/C++ but define all class interfaces in Python (and pass only simple primitive types such as numbers, strings, and pointers to C/C++).  This difference has limited our ability to utilize third-party Python libraries (because we frequently need access to the functionality they provide in C++ as well), and our broader Python/C++ boundary layer represents a sometimes complex category of code that the more Python-centric philosophy largely avoids.  But in return, our approach allows us to use the full object-oriented power of C++ when writing some of our most complex and challenging algorithms, making that code more readable and more maintainable without sacrificing performance.  It also makes our code more reusable: by providing both Python and C++ APIs, we allow both Python-centric projects and those that heavily utilize C++ to build on our software.

\subsection{I/O and Provenance}
\label{sec:io-and-provenance}

The HSC Pipeline uses LSST DM's \emph{Data Butler} concept to organize and execute all input and output.  The butler manages the file format and path of every concrete data product consumed or created by the pipeline (including intermediates), allowing algorithmic code to read or write these data products with just a dataset name (\eg~\texttt{src}, for source catalog) and a \emph{data ID} that describes the unit of data as a set of key-value pairs (\eg~\texttt{visit=1228, ccd=40}).  The butler allows different cameras to define their own key-value pairs, customize file paths, and associate camera-specific metadata with data IDs.  These customizations are invisible to the algorithmic code in \texttt{Tasks}, allowing nearly all code to be completely camera-independent.  The names of butler datasets provide a convenient and precise way to refer to pipeline data products, and we will use these names in \secref{pipelines} when declaring the inputs and outputs of each high-level processing stage.  Note that \citet{2017arXiv170208449A} refers to these data products via their file names, and the mapping between these and the butler dataset names is in some cases confusing (for regrettable historical reasons).  Table~\ref{tbl:data-products} shows the dataset and file names for the HSC Pipeline's main data products.

The butler organizes the data products created or consumed by individual runs of the pipeline into \emph{data repositories}.  All data products in a repository are produced with the same versions and configuration of the pipeline.  Repositories can be chained together -- the outputs from one stage of the pipeline can be saved in one repository, and the next stage of the pipeline can use this repository for its inputs while writing its outputs to a new repository.  Multiple runs of one pipeline stage with \eg~different configurations but common inputs thus map naturally to a tree of data repositories.  All repositories containing pipeline outputs are ultimately chained to a root input repository containing the raw data, calibrations, and metadata information.  HSC data repositories are currently implemented as directories with symbolic links to the repository directories they are chained to, but this too is customizable and hidden from algorithmic code; the butler automatically traverses the tree of chained repositories when retrieving data products.

\section{Pipelines}
\label{sec:pipelines}

\begin{figure}
    \includegraphics[width=0.48\textwidth]{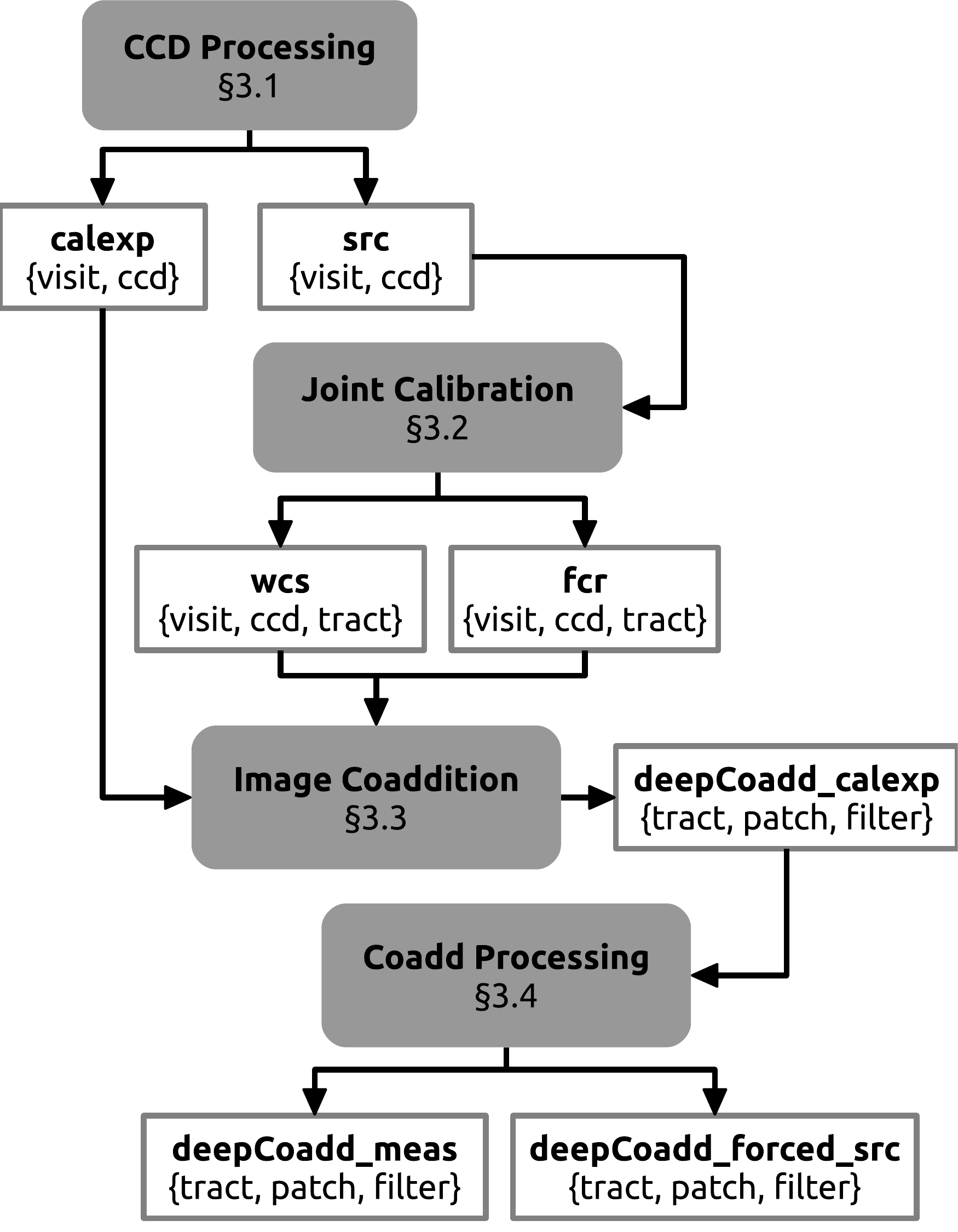}
    \caption{
        Conceptual flow of processing in the HSC Pipeline.  Filled rectangles are the four high-level stages of the pipeline discussed in \secref{pipelines}.  Unfilled rectangles show the most important data products and their granularity.  Dataset names are those used by the data butler (see \secref{io-and-provenance} and Table~\ref{tbl:data-products}).
    }
    \label{fig:pipeline-diagram}
\end{figure}

The HSC Pipeline is currently comprised of four high-level stages:
\begin{itemize}
\item in \textit{CCD Processing} (\secref{ccd-processing}) we detect sources on each CCD image and use them to calibrate and characterize the image;
\item in \textit{Joint Calibration} (\secref{joint-calibration}) we use repeat observations of sources in different visits to further constrain the astrometric and photometric calibrations;
\item in \textit{Image Coaddition} (\secref{image-coaddition}) we combine visit images into coadds;
\item and in \textit{Coadd Processing} (\secref{coadd-processing}) we detect and measure objects on the coadd images.
\end{itemize}
These stages are focused on making measurements of the deep, static sky; while the LSST codebase includes some support for image subtraction (to detect transient objects) and visit-level forced photometry (to measure light-curves of variable objects), these are not yet run regularly in HSC-SSP data release processing and are not described further here.  \figref{pipeline-diagram} shows the relationships between these steps and the datasets they produce and consume.

\begin{table*}
    \tbl{Data Products}{%
    \footnotesize
    \begin{tabular}{ l l }
        \hline
        Name & File Pattern \\
        \hline
        \texttt{src} &
        \texttt{\$(pointing)/\$(filter)/output/SRC-\$(visit)-\$(ccd).fits} \\
        \texttt{calexp} &
        \texttt{\$(pointing)/\$(filter)/corr/CORR-\$(visit)-\$(ccd).fits} \\
        \texttt{wcs} &
        \texttt{\$(pointing)/\$(filter)/corr/\$(tract)/wcs-\$(visit)-\$(ccd).fits} \\
        \texttt{fcr} &
        \texttt{\$(pointing)/\$(filter)/corr/\$(tract)/fcr-\$(visit)-\$(ccd).fits} \\
        \texttt{deepCoadd\_calexp} &
        \texttt{deepCoadd/\$(filter)/\$(tract)/\$(patch)/calexp-\$(filter)-\$(tract)-\$(patch).fits} \\
        \texttt{deepCoadd\_meas} &
        \texttt{deepCoadd-results/\$(filter)/\$(tract)/\$(patch)/meas-\$(filter)-\$(tract)-\$(patch).fits} \\
        \texttt{deepCoadd\_forced\_src} &
        \texttt{deepCoadd-results/\$(filter)/\$(tract)/\$(patch)/forced-src-\$(filter)-\$(tract)-\$(patch).fits} \\
        \hline
    \end{tabular}}\label{tbl:data-products}
    \begin{tabnote}
        \normalsize
        Primary data products produced by the pipeline, with their dataset name and file location relative to the root of a data repository.
    \end{tabnote}
\end{table*}

Throughout all pipeline steps, we approach the problem of missing or bad data from the perspective that we should attempt to process everything and use catalog flags and image masks to indicate results that may be unreliable.  Our mask images use each bit of the integer pixel values to represent a different ``mask plane'', and for most mask planes we set a corresponding flag field in our catalog for any object that contains a pixel with that bit set.  The most common mask planes and flags are listed in Table~\ref{tbl:masks-and-flags}.

\begin{table*}
    \tbl{Mask Planes and Catalog Flags}{%
    \footnotesize
    \begin{tabular}{ l l m{9cm} }
        \hline
        Catalog Flag & Mask Plane & Description \\
        \hline
        \noalign{\smallskip}
            \texttt{flags\_pixel\_bad} &
            \texttt{BAD} &
            Object overlaps a sensor defect. \\
        \noalign{\smallskip}
            \texttt{flags\_pixel\_bright\_object\_any} &
            \multirow{2}{*}{\texttt{BRIGHT\_OBJECT}} &
            \multirow{2}{*}{\parbox{9cm}{
                A very bright object nearby may have negatively affected background subtraction or detection \citep[see][]{2017arXiv170500622C}.
            }} \\
            \texttt{flags\_pixel\_bright\_object\_center} & & \\
        \noalign{\smallskip}
            \texttt{flags\_pixel\_clipped\_any} &
            \texttt{CLIPPED} &
            Object is in a region where one or more input images had pixels rejected (see \secref{safe-clipping}). \\
        \noalign{\smallskip}
            \texttt{flags\_pixel\_cr\_any} &
            \multirow{2}{*}{\texttt{CR}} &
            \multirow{2}{*}{\parbox{9cm}{
                Object overlaps a cosmic ray (see \secref{cr-detection}).
            }} \\
            \texttt{flags\_pixel\_cr\_center} & & \\
        \noalign{\smallskip}
            \texttt{flags\_pixel\_edge} &
            \texttt{EDGE} &
            Object was near the edge of a CCD or coadd patch and may be truncated. \\
        \noalign{\smallskip}
            \texttt{flags\_pixel\_interpolated\_any} &
            \multirow{2}{*}{\texttt{INTERP}} &
            \multirow{2}{*}{\parbox{9cm}{
                Object overlaps a pixel that was set by interpolating its neighbors (see \secref{bad-pixel-interpolation}).
            }} \\
            \texttt{flags\_pixel\_interpolated\_center} & & \\
        \noalign{\smallskip}
            \texttt{flags\_pixel\_saturated\_any} &
            \multirow{2}{*}{\texttt{SAT}} &
            \multirow{2}{*}{\parbox{9cm}{
                Object overlaps a saturated pixel.
            }} \\
            \texttt{flags\_pixel\_saturated\_center} & & \\
        \noalign{\smallskip}
            \texttt{flags\_pixel\_suspect\_any} &
            \multirow{2}{*}{\texttt{SUSPECT}} &
            \multirow{2}{*}{\parbox{9cm}{
                Object overlaps a pixel whose value was above the level where our linearity correction is reliable.
            }} \\
            \texttt{flags\_pixel\_suspect\_center} & & \\
        \noalign{\smallskip}
        \hline
    \end{tabular}}\label{tbl:masks-and-flags}
    \begin{tabnote}
        \normalsize
        Image mask planes and their corresponding flags.  Flags with the \texttt{\_center} suffix indicate that the mask plane bit was set in the central 3$\times$3 pixels of an object.  All other flags indicate mask bits set anywhere in the object's deblended \texttt{Footprint} (see Sections~\ref{sec:detection} and~\ref{sec:deblending}).  Each source measurement algorithm (\secref{source-measurement}) also adds its own set of diagnostic flags to the catalog.  These are fully documented in the headers of the pipeline's output files.
    \end{tabnote}
\end{table*}

\subsection{CCD Processing}
\label{sec:ccd-processing}

In CCD Processing, we process each CCD in a visit to produce a calibrated image and catalog.  Each CCD is currently processed completely independently; while some algorithms (\eg~background and PSF modeling) can be improved by considering all CCDs in a visit together, this was not implemented in the version used for HSC-SSP releases.

This pipeline consists of a long sequence of semi-iterative steps: many operations have (weak) circular dependencies, so some steps are repeated to take into account improved inputs generated by previous steps.  Most of these steps are explained in greater detail in \secref{algorithms}, so we focus here on the order of operations and the relationships between them.

The steps are as follows.

\begin{enumerate}

  \item \label{itm:processCcd-isr} We run Instrument Signature Removal (ISR; \secref{isr}), to assemble raw amplifier-level images into CCD images, perform basic detrending (\ie~flat-field, bias, and dark correction), and apply nonlinearity, brighter-fatter (\secref{brighter-fatter-correction}), and crosstalk corrections.

  \item \label{itm:processCcd-cr-1} We detect pixels affected by cosmic rays (\secref{cr-detection}) and mask them.  This algorithm uses a circular Gaussian with 1\arcsec\ FWHM width as a stand-in for the PSF model we have not yet estimated, and hence these detections are preliminary.

  \item \label{itm:processCcd-repair-1} We repair bad pixels and those affected by saturation by using a linear predictive code to interpolate the values of nearby good pixels (\secref{bad-pixel-interpolation}).  We defer repairing pixels affected by cosmic rays to step~\ref{itm:processCcd-cr-2} in order to make use of the PSF model.

  \item \label{itm:processCcd-background-1} We estimate and subtract an initial model of the sky background (\secref{background-subtraction}).

  \item \label{itm:processCcd-detection-1} We run a maximum-likelihood detection algorithm (\secref{detection}) with a $50\sigma$ threshold and the same $1^{\prime\prime}$ FWHM Gaussian as above as the smoothing filter.  Because we are interested only in obtaining enough bright stars to use for PSF modeling and astrometric and photometric calibration, the fact that this filter may be significantly different from the true PSF is not a serious concern.

  \item \label{itm:processCcd-background-2} We re-estimate and subtract the background, ignoring pixels marked as belonging to detections in step~\ref{itm:processCcd-detection-1}.

  \item \label{itm:processCcd-measurement-1} We run centroiding, adaptive moment shape algorithms, and aperture photometry algorithms (\secref{source-measurement}) on the detected sources from step~\ref{itm:processCcd-detection-1}.  We do not deblend prior to measurement at this stage, since our PSF modeling algorithm requires unblended sources.

  \item \label{itm:processCcd-calibration-1} We use a custom implementation of the ``Optimistic B'' algorithm of \citet{2007PASA...24..189T} to match these to an external reference catalog.  We explicitly correct source positions to account for a static approximation to the optical distortion before matching.  For HSC-SSP data release processing, we use the Pan-STARRS PV1 catalog \citep{2016arXiv161205560C}.  We use these matches to fit both a per-CCD magnitude zero-point and a gnomonic world coordinate system transform (TAN WCS) with 3rd-order polynomial distortions.  We undo the approximate static distortion correction before fitting so the full distortion is included in the fitted model.

  \item \label{itm:processCcd-star-selection} We use the source measurements from step~\ref{itm:processCcd-measurement-1} (particularly adaptive moments) to filter out galaxies and most blends, yielding a catalog of secure stars for use in PSF estimation.  We reserve 20\% of these stars as a validation sample.

  \item \label{itm:processCcd-psf-determination} We construct a PSF model from images of the other 80\% of our secure star catalog, using outlier rejection to remove any barely-resolved galaxies or blends that passed previous cuts.  Most of this work is delegated to a restructured version of the public \texttt{PSFEx} code \citep{2013ascl.soft01001B}.  The details of the star selection and PSF measurement algorithms are described more fully in \secref{psf-modeling}.

  \item \label{itm:processCcd-cr-2} We repeat cosmic ray detection (step~\ref{itm:processCcd-cr-1}), and re-run our pixel interpolation algorithm (step~\ref{itm:processCcd-repair-1}) to repair pixels affected by cosmic rays.

  \item \label{itm:processCcd-measurement-2} We re-run source measurement (step~\ref{itm:processCcd-measurement-1}) on all detected sources, this time utilizing our full suite of measurement algorithms (many of which require or perform better with an accurate PSF model).  We also repeat aperture flux and adaptive moments measurements because these depend on centroid measurements that in turn utilize the PSF model.

  \item \label{item:processCcd-measure-apcorr} We estimate aperture corrections (\secref{aperture-corrections}) for each photometry algorithm by modeling the spatial variation of the ratio of each algorithm's flux to the flux we use for photometric calibration, by default a 4\arcsec diameter circular aperture.  The aperture corrections are determined using the same sample of stars used to construct the PSF model, and then applied to the measurements for all sources.

  \item \label{itm:processCcd-calibration-2} We re-run our astrometric and photometric calibration (step~\ref{itm:processCcd-calibration-1}), refining the results from that step by using the improved measurements from step~\ref{itm:processCcd-measurement-2}.

  \item \label{itm:processCcd-detection-2} We construct a deeper catalog (repeating step~\ref{itm:processCcd-detection-1}), this time using a detection threshold of $5\sigma$ and a Gaussian smoothing filter with the same size as the PSF model.

  \item \label{itm:processCcd-background-3} We once again re-estimate and subtract the background (repeating step~\ref{itm:processCcd-background-1}), using the deeper catalog from step~\ref{itm:processCcd-detection-2} to mask out pixels containing detected objects.

  \item We deblend (\secref{deblending}) all above-threshold regions containing multiple peaks.

  \item \label{itm:processCcd-measurement-3} We run our full suite of source measurement algorithms a third time, now on sources in the deeper catalog, using the deblended pixels, and apply the aperture corrections measured in step~\ref{item:processCcd-measure-apcorr}.

\end{enumerate}

CCD Processing generates two primary data products, which we refer to as \texttt{calexp} (for ``calibrated exposure'') and \texttt{src} (for ``source catalog'').  The \texttt{calexp} dataset includes the detrended, background-subtracted image, an integer mask image, and an image containing a per-pixel estimate of the variance.  Individual bits in the integer mask pixels correspond to different kinds of features, such as saturation or cosmic rays (see Table~\ref{tbl:masks-and-flags}); the association with is dynamic and is documented in the image headers.  The \texttt{calexp} dataset also contains several smaller but generally more complex objects representing the PSF model, the astrometric calibration (WCS), the photometric calibration (magnitude zero-point), and the aperture corrections.  Background models are currently stored in a separate dataset (\texttt{calexpBackground}).

In addition to measurement outputs and diagnostic flags (from the final measurement stage on the deeper catalog in step~\ref{itm:processCcd-measurement-3}), the \texttt{src} catalog also holds objects called \texttt{Footprints} that record the exact above-threshold detection region.  These are similar to a per-source version of SExtractor's ``segmentation map'' \citep{1996A&AS..117..393B}.  For deblended sources the \texttt{Footprints} are extended to hold the corresponding array of deblended pixel values.


\subsection{Joint Calibration}
\label{sec:joint-calibration}

After individually processing all CCDs, we can refine their astrometric and photometric calibrations by requiring consistent positions and fluxes for sources that appear on different parts of the focal plane in different visits.  This allows us to use many more stars than are available in the shallower reference catalog for internal calibration, providing constraints on smaller spatial scales as well as stronger constraints overall.

For both the astrometric and photometric fit, we begin with a spatial match between all of the CCD-level \texttt{src} catalogs that overlap a particular tract in a particular band (each tract$+$band combination is fit independently).  We also include external reference catalog sources in this match to tie to absolute astrometric and photometric systems, but we do not require that all matches have a reference catalog source.  We select only point sources (via their \textit{extendedness}; see \secref{star-galaxy-classification}) that were not blended.

Our astrometric model includes a 9th-order polynomial distortion that is continuous over the full focal plane composed with distinct translation and rotation transforms for every CCD.  We use the same translation and rotation parameters for all visits in the per-tract fit.  While these parameters should be a function of time or telescope configuration rather than sky position, their variation is currently not significant compared to other sources of uncertainty and we have sufficient information to constrain them well in a single tract, so we do not yet need to move away from this simpler approach.  We do fit a different polynomial distortion for each visit, as this term models both the optical distortion (which is mostly stable, but may depend subtly on the telescope configuration) and distortions due to the atmosphere (which vary from visit to visit).  We use standard polynomials instead of an orthogonal polynomial set (\eg~Chebyshev polynomials) to permit the results to be written to disk using the popular TAN-SIP \citep{2005ASPC..347..491S} FITS convention.

\begin{figure}
    \includegraphics[width=0.5\textwidth]{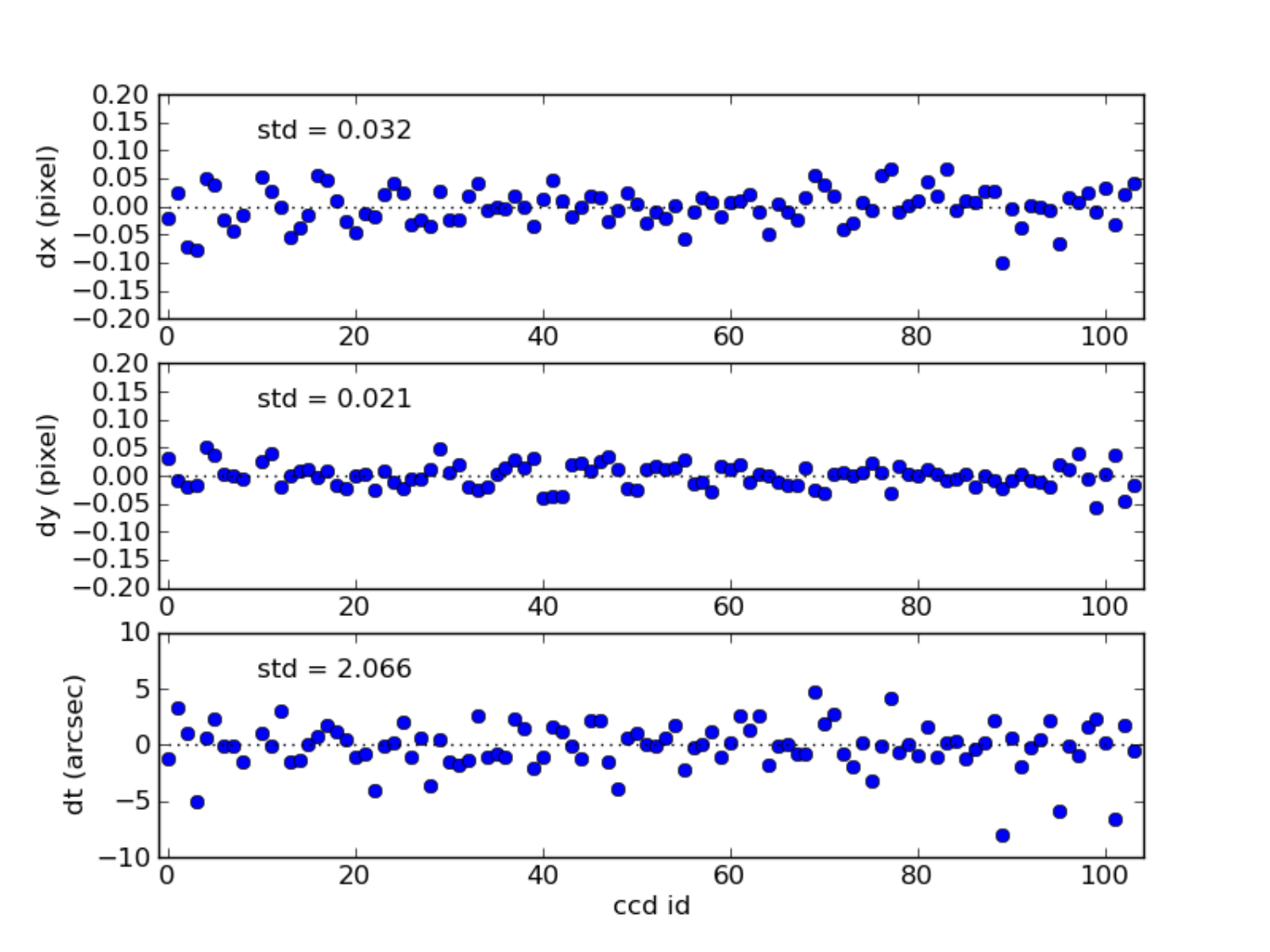}
    \caption{
        Differences in CCD position ($x$: upper, $y$: middle) and rotations (bottom) in the focal plane from joint astrometric fitting in different tracts in $i$.
        \label{fig:mosaic_ccdpos}
    }
\end{figure}

For the photometric fit, the model is composed of a per-visit zero-point, a per-visit 7th-order Chebyshev polynomial over the focal plane, and a constant per-CCD scaling.  This combines a correction derived from stars to the flat field and a correction for nonphotometric conditions, which are degenerate in this model.  We also include a correction for deviations from the nominal on-sky area of each pixel derived from the local Jacobian of the astrometric model.  We do not attempt to remove nonphotometric data; because we fit a different Chebyshev polynomial scaling for each visit, this term naturally accounts for both non-uniform illumination of the flat-field screen (which is approximately constant across visits) and gray atmospheric extinction (which is not).

We fit the astrometric and photometric models separately, starting with the astrometric fit.  We treat the true fluxes and positions of the stars as additional parameters to be fit, and use a dense multi-threaded LU decomposition with partial pivoting to solve the normal equations at each iteration.  This scales poorly when the number of visits in a tract is very large (as in \eg~SSP UltraDeep fields), and an alternate implementation using sparse matrices is currently being developed.

The details of these mathematical models and how we solve them are described more fully in Appendix~\ref{sec:jointcal-details}. 

\subsection{Image Coaddition}
\label{sec:image-coaddition}

The traditional approach to processing multi-epoch surveys for static-sky science is to build coadds: images from different observations are resampled onto a common grid and averaged together to construct a single deeper image.  Many common approaches to coaddition either degrade the data or introduce systematic errors, however, and recent surveys focused on weak lensing have avoided these problems by instead using models fit simultaneously to images from all epochs for their most sensitive measurements (CFHTLens, \citealt{2013MNRAS.429.2858M}; DES, \citealt{2016MNRAS.460.2245J}; and KiDS, \citealt{2015MNRAS.454.3500K}).  The HSC Pipeline has instead focused on building coadds without introducing systematic errors or discarding information.  While it remains to be seen whether this approach will work as the HSC-SSP survey area grows and our tolerance for systematic errors decreases, at the level required for first-year science we believe we have achieved this goal.

The coadds generated by the HSC Pipeline are computed as a direct weighted average of the resampled CCD images -- we do not match the PSFs of the input images before combining them.  This combination is not optimal (see \eg~\citealt{2015arXiv151206879Z}), but its signal-to-noise ratio (S/N) and effective PSF are equivalent to those of a single long exposure with the same total exposure time (and observing conditions) if the exposure time is used to set the weights of the input images.  We actually use the inverse of the mean variance of the input images to set the weights, so our coadd S/N is actually slightly better than the hypothetical long exposure.  This direct coaddition algorithm is significantly simpler to implement than the optimal algorithm, and can easily handle missing data.  It also approaches optimality in the limit that the PSFs of all input visits are the same.  In the HSC-SSP Wide layer, the mean loss in S/N due to suboptimal coaddition is only $1.7\%$ (with a standard deviation of $1.5\%$).

Prior to combining CCD images, we resample them to a common pixel grid using 3rd-order Lanczos interpolation.  We resample each bit plane of the integer mask images using a 2$\times$2 top-hat filter to avoid spreading out masked areas too much.  While this means some pixels originally affected by cosmic rays, saturation, or image defects (and repaired via interpolation; steps~\ref{itm:processCcd-repair-1} and~\ref{itm:processCcd-cr-2} in \secref{ccd-processing}) do contribute to the coadd without being masked on the coadd, this ensures that these contributions to a given coadd pixel are always small compared to contributions to the same pixel from both other epochs and non-interpolated pixels from the same epoch.

The final step of coaddition is to set the \texttt{BRIGHT\_OBJECT} mask plane in the coadded images from a catalog of geometric regions whose processing may be affected by bright stars.  The construction of these regions is described in \citet{2017arXiv170500622C}.

Coadds for each patch (see \secref{terminology}) and filter are built independently.  Because patches overlap, and the relative weights of visits are determined separately for each patch, the coadd pixel values in the overlap regions are not identical.  These differences are not scientifically significant, but they can slightly complicate the resolution of duplicate detections between patches described in \secref{coadd-processing}.

\subsubsection{PSF Coaddition}
\label{sec:psf-coaddition}

Even slightly different PSFs in different visits make it essentially impossible to correctly model the effective PSF of the coadd using the coadded images of stars: even small dithers to fill the gaps between CCDs will yield discontinuities in the effective PSF on the coadd.  As the number of dithers increases, the regions of the coadd with a continuous effective PSF become smaller, and it becomes increasingly unlikely that each region will contain even one star usable for PSF modeling.

Instead, the HSC pipeline includes an implementation of the approach of \citet{2011PASP..123..596J}, in which the existing PSF models for the input images are resampled and combined with the same coordinate transformations and weights that are applied to the image data.  \citet{2014ApJ...794..120A} followed a similar approach, but coadded the PSF at predefined positions on a grid and then fit a smoothly-varying PSF model to these (despite the fact that the PSF being modeled may contain discontinuities).  We instead evaluate the PSF only at the position of each object to be measured, and we perform this evaluation ``on demand''.  Along with the image data for each coadd patch, we store the PSF models, WCS transforms, combination weights, and boundaries of all of the CCD images that went into it.  When an image of the effective coadd PSF is requested at a point on the coadd (by, \eg, one of the source measurement algorithms described in \secref{source-measurement}), we transform that point to the coordinate system of each of the input CCD images that went into the coadd at that point, evaluate their PSF models at that point, resample them to the coadd coordinate system, and finally combine them (with appropriate weights) to generate an image of the effective coadd PSF.  We cache the results so repeated evaluations of the PSF model at a single point will not require additional computation.

We typically assume that the model PSF at the center of an object is appropriate for the entirety of the object.  Usually the spatial variation is sufficiently slow that this approximation is perfectly valid even for large objects, but it is invalid where the number of inputs to the image change discontinuously.  This can happen both at the edges of input images and in regions where one or more pixels were not included in the coadd due to masking or clipping (see \secref{safe-clipping}).  Because there is no well-defined PSF for the entire object in these cases, we simply attempt to flag such objects as having been measured with such PSFs by setting the \texttt{flags\_pixels\_clipped\_any} flag.  In most cases, these PSF inaccuracies are quite small, and science cases that do not require careful control over the PSF generally should not need to filter out these objects.  Unfortunately, a bug in the current version of the pipeline prevents the flag from being set on objects at CCD boundaries or containing inputs rejected due to cosmic rays or known bad pixels.  This affects approximately 18\% of the objects in the HSC-SSP Wide layer, though this fraction is quite sensitive to the subjective definition of the area covered by an object (we have used the detection \texttt{Footprint} region here; see \secref{detection}).  This will be fixed in a future release.

\subsubsection{Safe Clipping}
\label{sec:safe-clipping}

For PSF model coaddition to be valid, the operation used to combine all input pixels at each point on the coadd image must be strictly linear -- robust estimators such as the median or sigma-clipped means cannot be used.  Nonlinear estimators do not just prevent PSF coaddition from working, however; they prevent the coadd image from even having a well-defined effective PSF.  Any estimator that rejects individual pixel outliers will tend to reject pixels in the cores of stars on the best-seeing exposures, and brighter stars will experience more rejection, giving them a different profile than fainter stars.  It should be emphasized that this occurs even in the absence of noise, and even with extremely weak outlier rejection (\eg~clipping at $10\sigma$).  All robust estimators start from the ansatz that all input values are drawn from the same underlying distribution, and convolution with different PSFs means that they are not.

\begin{figure*}
    \includegraphics[width=0.33\textwidth]{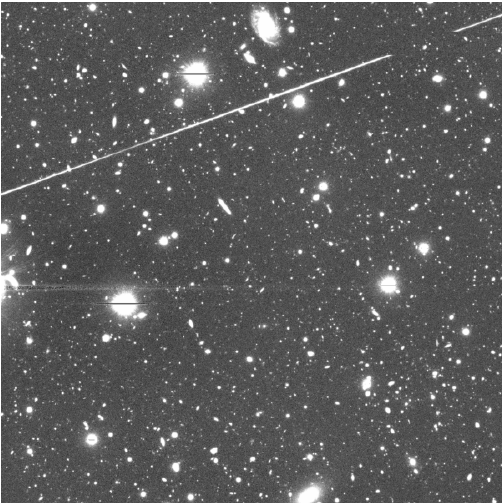}
    \includegraphics[width=0.33\textwidth]{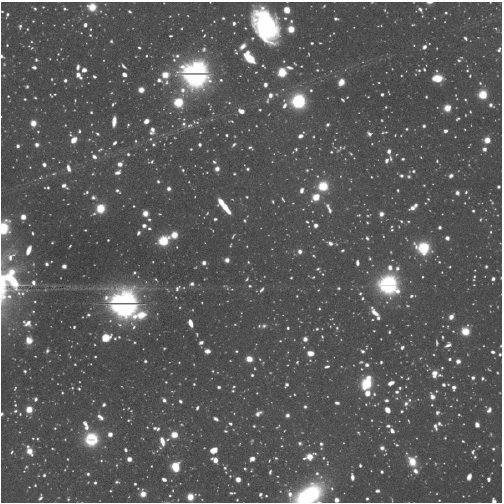}
    \includegraphics[width=0.33\textwidth]{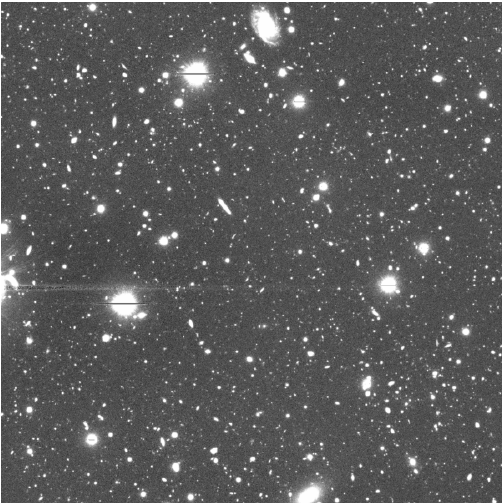}\\
    \includegraphics[width=0.33\textwidth]{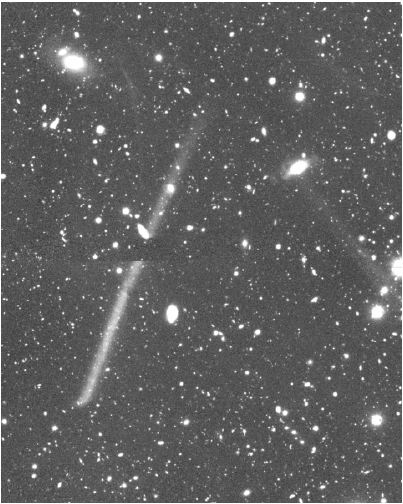}
    \includegraphics[width=0.33\textwidth]{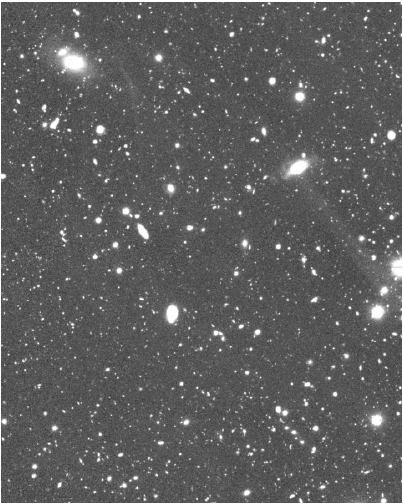}
    \includegraphics[width=0.33\textwidth]{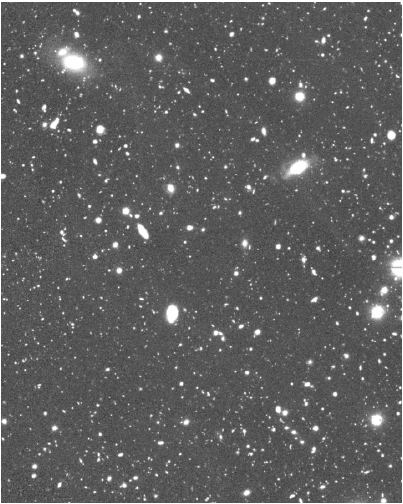}
    \caption{
        \label{fig:safe-clipping-demo}
        Examples of coaddition algorithms in the presence of image artifacts: a satellite trail (top) and a pair of optical ghosts (bottom).  The images in the left column were computed with a direct mean, in which the PSF is preserved but single-image artifacts clearly print through to the coadd.  The middle column images were created with an iterative 3$\sigma$ clip, which invalidates the PSF and still fails to remove the wings of the artifact.  The right column images were produced with our ``safe clipping'' algorithm, which does a better job of removing the artifacts and preserves the PSF in the rest of the image.  Note that none of the methods fully remove the rightmost ghost in the lower panel.
    }
\end{figure*}

To detect image artifacts that were not detected morphologically in individual visit CCD processing, then, we need a more sophisticated approach that works beyond the level of individual pixels.  Our starting point is to first make both a direct mean coadd with no outlier rejection and an iterative $1.5\sigma$-clipped mean coadd, and difference them.  We then run our usual maximum-likelihood detection algorithm (\secref{detection}) on the difference with an extremely low $2\sigma$ threshold.  The above-threshold regions (coadd difference regions; CDRs) will include both true artifacts and PSF inconsistencies due to clipping.  To avoid rejecting pixels due to PSF differences we compare the CDRs to the sources detected in single-visit CCD Processing (\secref{ccd-processing}, step~\ref{itm:processCcd-detection-2}).  For each CDR-visit pair, we compute the ``overlap fraction'': the number of pixels shared by the CDR and detected sources from that visit divided by the total number of pixels in the CDR.  We consider three cases:

\begin{itemize}
  \item If only one visit has an overlap fraction with a CDR greater than $0.45$, we set the \texttt{CLIPPED} mask bit in the CDR region on that visit; this means the CDR is probably a satellite trail, ghost, or other artifact that appears in only a single image.
  \item If no more than two visits have an overlap fraction with a CDR greater than $0.15$ and at least four visits contributed to the coadd at the position of the CDR, we set the \texttt{CLIPPED} mask bit in the CDR region in those visits.  The low overlap fraction in this case is intended to identify artifacts that intersect real astrophysical sources, but it will miss cases where the artifact occupies a relatively small fraction of the above-threshold region.
  \item In all other cases, we ignore the CDR.  Essentially all differences due to PSF inconsistencies will fall into this category, as they occur on stars or galaxies that are present in every image.
\end{itemize}
In all of the above cases, we also discard any CDRs that are more than $50\%$ occupied by pixels already masked as bad pixels, saturated, or cosmic rays in single-visit CCD Processing (\secref{ccd-processing}, steps~\ref{itm:processCcd-isr} and~\ref{itm:processCcd-cr-2}).

We then look for cases where the clipped area may not be large enough because a single large artifact (frequently satellite trails or optical ghosts) was detected as multiple CDRs because some pixels dipped below our threshold.  We look for single-visit sources that already have more than 100 pixels marked \texttt{CLIPPED}; in these cases we mark the full single-visit detection as \texttt{CLIPPED} as well.

Finally, we build a new coadd using a direct mean that ignores all pixels marked \texttt{CLIPPED} in the visit images.  We also set \texttt{CLIPPED} in the coadd mask image in these regions, where the complete set of input images was not used.  We also ignore pixels marked as \texttt{CR} or \texttt{BAD} (see Table~\ref{tbl:masks-and-flags}) in CCD Processing.

This ``safe clipping'' procedure is a complicated, highly heuristic algorithm that has been tuned to work with HSC-SSP data, especially in the Wide layer; the small dithers (typically $<10$\arcmin) and large number of exposures in the SSP Deep and UltraDeep layers \citep[see][]{2017arXiv170405858A} make it harder for to detect and remove artifacts in those fields.  The main failure mode in Wide processing is a failure to remove some satellite trails and optical ghosts that significantly overlap real bright objects; these have CDRs with large overlap fractions on multiple visits.  A more general replacement algorithm that more directly utilizes image differences is currently under development.

\subsection{Coadd Processing}
\label{sec:coadd-processing}

\begin{figure}
    \includegraphics[width=0.48\textwidth]{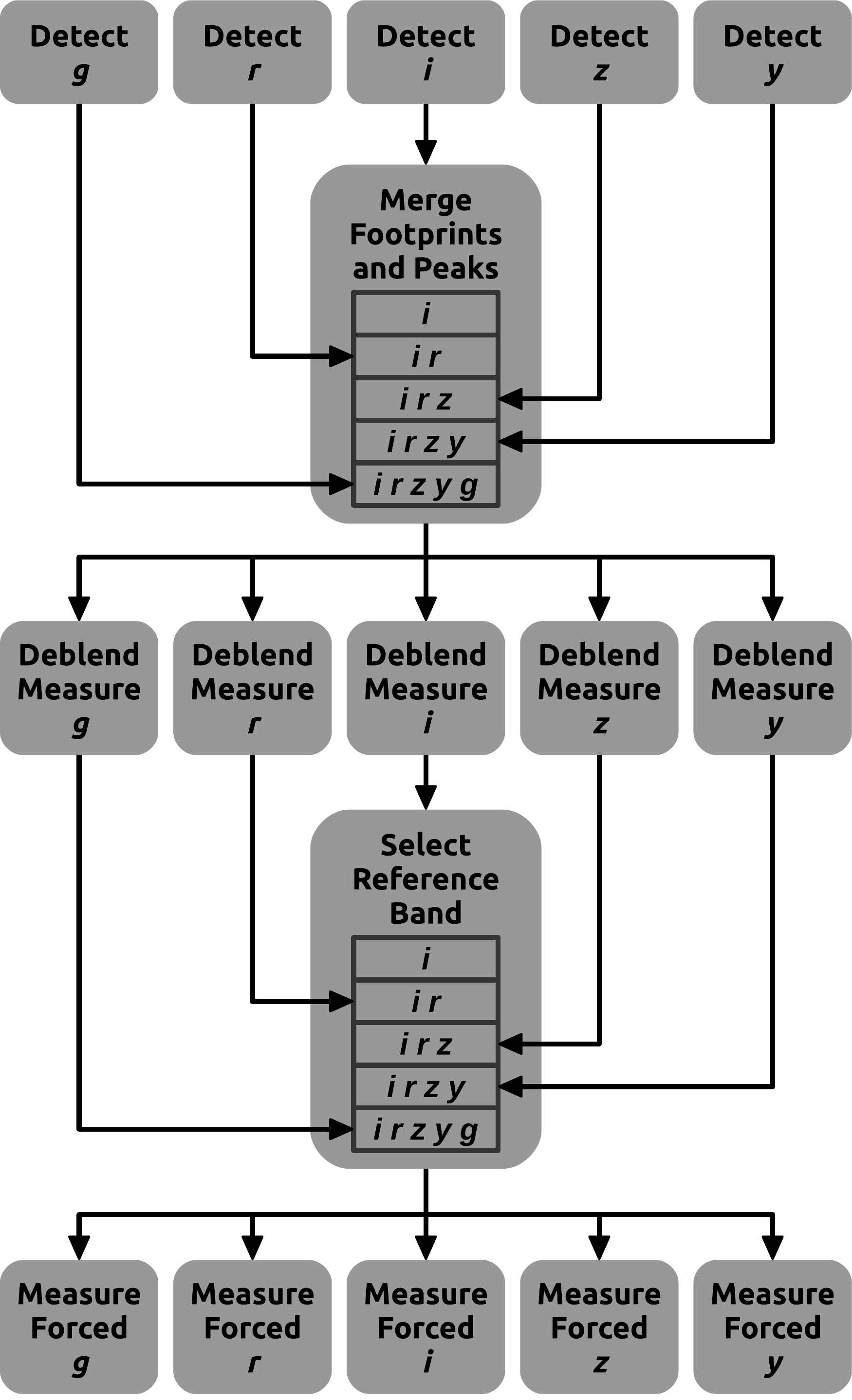}
    \caption{
        Diagram of the coadd processing flow described in \secref{coadd-processing}.  As we detail in the text, we start by detecting sources in each band independently, then merge these detections to build a consistent cross-band object list.  We then detect and measure in each band separately.  Finally, we consider all bands together to select a reference band for each object and then perform forced photometry in each band.  This algorithm is naturally extended to include narrow-band filters (in the order described in the text) in the Deep and UltraDeep layers.
    }
    \label{fig:coadd-processing-diagram}
\end{figure}

In Coadd Processing, we detect, deblend, and measure objects on the coadd images.  To maximize cross-band consistency in these measurements, we do this in five steps, alternating between processing images of each band independently and processing catalogs from all bands jointly: we detect in each band separately, merge the detections across bands, deblend and measure in each band, identify a reference band for each object, and finally perform forced measurements in each band using the reference-band positions and shapes.  Each patch is processed completely independently.  \figref{coadd-processing-diagram} provides a summary of the processing flow.

The first step is to simply run our maximum-likelihood detection algorithm (\secref{detection}) independently on the coadd image from each band with a 5$\sigma$ threshold.  As in CCD Processing, we use a circular Gaussian smoothing filter matched to the RMS size of the effective coadd PSF.  This generates for each band a set of above-threshold regions we call \texttt{Footprints}.  Each \texttt{Footprint} contains one or more positions corresponding to peaks in the smoothed image.  We also perform one more round of background subtraction (\secref{background-subtraction}) to account for the new, deeper detections moving out of the definition of the background.  Unlike previous background subtractions, we use a single bin for each 4k$\times$4k patch and simply subtract that constant.

In the next step, we merge the \texttt{Footprints} and the peak positions within them across bands.  The bands are processed sequentially in what we call their \textit{priority order}, which we define as \textit{irzyg} for the broad-band filters, followed by the 921, 816, 1010, 387, and 515 (nm) narrow-band filters.  This ordering is simply a heuristic, but it approximately goes from higher S/N bands (for typical objects) to lower S/N.  The algorithm starts with the \texttt{Footprints} and lists of peaks from the highest-priority band (\textit{i}), and proceeds as follows for the other bands.
\begin{enumerate}
  \item We merge the \texttt{Footprint}s from the previous step with any overlapping \texttt{Footprint}s from the current band, defining the merged \texttt{Footprint} to contain the union of all pixels contained by all \texttt{Footprints} that went into the merge.  Any \texttt{Footprints} in the current set that now overlap are similarly merged.  As a result, this step may decrease the number of distinct merged \texttt{Footprints}, but the total area covered by all \texttt{Footprints} can only increase.
  \item If a peak in the new band is at least $1^{\prime\prime}$ from all peaks from the previous step, we add a new peak.
  \item If a peak in the new band is less than $0.3^{\prime\prime}$ from the nearest peak from the previous step, we mark the peak as having been detected in the new band while maintaining the position from the previous step.
  \item If a peak in the new band is between $0.3^{\prime\prime}$ and $1^{\prime\prime}$ from the nearest peak from the previous step, it is ignored: we cannot conclusively identify the peak as the same source, and we let the higher-priority band's set of peaks take precedence.
\end{enumerate}
After the merging is complete, we attempt to cull peaks that are likely spurious detections on the outskirts of bright objects, brought above the detection threshold by flux from the bright object.  We cull peaks that were detected in only one band that belong to \texttt{Footprints} in which there are 30 brighter peaks (in that band) or in which more than 70\% of the peaks are brighter.  This culling is intended to be conservative; it preserves many spurious detections to avoid any significant chance of removing real objects.  The final output of this step is a catalog of merged \texttt{Footprints} and peaks that is consistent across bands.


We return to processing the images for each band independently in the third step.  We run our deblending algorithm (\secref{deblending}) on each \texttt{Footprint} from the merged catalog.  This creates an object catalog containing an undeblended \emph{parent} object for every \texttt{Footprint} and a deblended \emph{child} object for every peak (in all subsequent steps, the list of objects includes both).  Because the set of \texttt{Footprints} and peaks is the same for all bands, we guarantee that the catalog contains the same objects in every band, but the deblending is otherwise completely independent in each band.  After deblending, we run the full suite of source measurement algorithms (\secref{source-measurement}) on all objects, yielding independent measurements in every band for each object.  We then resolve duplicates in the overlap regions between tracts and patches by setting the \texttt{detect\_is-tract-inner} and \texttt{detect\_is-patch-inner} flags on the version of each duplicate that is closest to a tract and/or patch center (respectively).  This catalog of independent per-band object measurements (\texttt{deepCoadd\_meas}) is one of our most important output data products.  Note that many measurements in this catalog will be performed in a band in which the object was not even detected, with predictably poor results for algorithms with many degrees of freedom.

To address this limitation, the fourth step in Coadd Processing determines a \emph{reference band} for each object by considering both the bands in which it was detected and its PSF-weighted S/N (see \secref{psf-photometry}).  The reference band for an object must be one with sufficient S/N to yield reliable measurements (or the most reliable measurements possible) even for algorithms with many degrees of freedom, but it is not simply the highest S/N band; we also wish to maximize the number of objects with the same reference band.  To that end, we consider each band in the same priority order defined in the second step of Coadd Processing, and define that band to be the reference band for an object if all of the following are true:
\begin{itemize}
\item the object is marked as detected in this band;
\item the object's PSF S/N in this band is at least 10 \emph{or} no band has PSF S/N greater than 10 and this band's PSF S/N is at least 3 higher than the last band considered;
\item the PSF (\secref{psf-photometry}, Kron (\secref{kron-photometry}), and CModel photometry (\secref{cmodel-photometry}) algorithms succeeded in this band or did not succeed in any bands.
\end{itemize}
In the SSP Wide layer, 57\% of objects have $i$ as their reference band, followed by 21\% for $r$, 13\% for $g$, 6\% in $z$, and 3\% in $y$.  In the UltraDeep layer, approximately 1\% of objects use one of the narrow-band filters as their reference band, with the other bands following approximately the same proportions as the Wide layer.

The final step is another run of the source measurement suite, but this time in \emph{forced} mode: we hold all position and shape parameters fixed to the values from the previous measurement \emph{in the reference band}.  This ensures that the forced measurements are consistent across bands and use a well-constrained position and shape, which is particularly important for computing colors from differences between magnitudes in different bands.  Consistent colors also require accounting for the different effective coadd PSFs in different bands.  The PSF and CModel photometry algorithms convolve a model for the true object's morphology with the effective PSF model at the position of the object, which maximizes S/N but depends on the correctness of the assumed underlying morphology.  The seeing-matched variants of aperture and Kron photometry algorithms instead convolve the coadd images prior to measurement to match their PSFs, which reduces S/N but is insensitive to morphology.  We refer the reader to \secref{source-measurement} for the details of all of these algorithms.  This final step in Coadd Processing produces the \texttt{deepCoadd\_forced\_src} dataset.

\section{Algorithms}
\label{sec:algorithms}

\subsection{Instrument Signature Removal}
\label{sec:isr}

The Instrument Signature Removal (ISR) component of the pipeline seeks to correct the raw images for features introduced by the CCD camera.  For ISR we use the \texttt{ip\_isr} package, which contains multiple elements that can be used in a standard or custom order.  For HSC, we use the following elements, applied in order:
\begin{itemize}
    \item Saturation detection: pixels exceeding the maximum correctable level are flagged as \texttt{SUSPECT}, while pixels exceeding the saturation level (either the full well depth or the range of the analog-to-digital converter; \citealt{hsc-instrument}) are flagged as \texttt{SAT}.
    \item Overscan removal: we fit (with a single rejection iteration) the overclock pixels along the rows with a 30th order Akima spline and subtract the fit from the science pixels.  This leaves zero-length exposures (bias frames) generally featureless except for features manifesting as a function of column.
    \item CCD assembly: the amplifier images comprising the CCD are assembled (applying necessary flips) into a single image for each CCD.
    \item Bias correction: we subtract an average of multiple (typically around 10) overscan-corrected zero-length exposures.  These overscan-corrected bias frames are generally featureless except for glowing pixels on \texttt{ccd=33} and a ridge line running the length of the rows between the second and third amplifiers.
    \item Linearity correction: we apply a cubic polynomial (with coefficients determined per amplifier) to correct for non-linearity.  The polynomial coefficients are determined from a fit to a series of dome flats with different exposure times.
    \item Crosstalk correction: we correct for intra-CCD crosstalk using a single correction matrix for all CCDs, and flag pixels for which the correction was significant. The correction matrix was determined during instrument commissioning from the ratios of the responses of different amplifiers to bright sources on the sky.
    \item Brighter-fatter correction: see \secref{brighter-fatter-correction}.
    \item Dark correction: we subtract an average of multiple (typically 3--5) long exposures with the shutter closed, scaled by the ``darktime'' (time since CCD wipe). The dark is generally featureless except for glowing pixels on \texttt{ccd=33}.
    \item Flat correction: we divide by the dome flat frames.  The dome flat is the average of multiple (typically around 10) exposures, normalized to a mean of unity over the camera.  We use dome flats because they are easy to obtain, easy to repeat with the same conditions, and the incandescent lamp spectrum roughly corresponds to that of our science sources (at least, better than does the night sky). Dome flat sequences are typically acquired at least once for each available filter during a run.
    \item Defect masking: we flag as \texttt{BAD} pixels which we have identified as defects.  The HSC CCDs are of high cosmetic quality, and there are few defects, although there are a few amplifiers (out of 448) that have gone bad since commissioning.
    \item Fringe correction: for exposures acquired using the red $y$-band or NB0921 filters, we subtract a scaled fringe frame.  The fringe frame is the average (with objects masked, and rejection of outliers) of science exposures from a variety of fields observed throughout one night.  The scaling is determined by measuring the average value within many ($10^{4}$) small regions over the image on the science exposure and the fringe frame.
\end{itemize}

The ideal result is an image for which the counts are proportional to flux from the sky when conditions are photometric.  This ideal is not actually achieved by the ISR component because the flat-fields inevitably contain scattered light and so do not perfectly reflect the response of the camera.  Furthermore, the Subaru telescope currently uses multiple dome lamps mounted on the top-end near the flat-field screen which, combined with the vignetting in the camera, means that different CCDs respond to different combinations of the lamps and therefore the flat-field pattern can change with time as the lamps age differently.
These shortcomings are addressed by the Joint Calibration component (\secref{joint-calibration}) of the pipeline.

\subsection{Brighter-Fatter Correction}
\label{sec:brighter-fatter-correction}
It has been recently observed for thick-depleted CCDs that the size and shape of the PSF depend upon the intensity of the source \citep{2014JInst...9C3048A,2015JInst..10C5032G}.  The effect has been attributed to the fact that a lateral electric field is generated as charge accumulates in the potential well.  This electric field causes charges to be laterally displaced, altering the field lines and causing brighter stars to be larger and more elliptical than faint stars.  This causes problems for science cases like weak lensing that rely on using bright stars to estimate the PSF model for fainter galaxies.

\begin{figure*}
\begin{center}
\includegraphics[width=8cm]{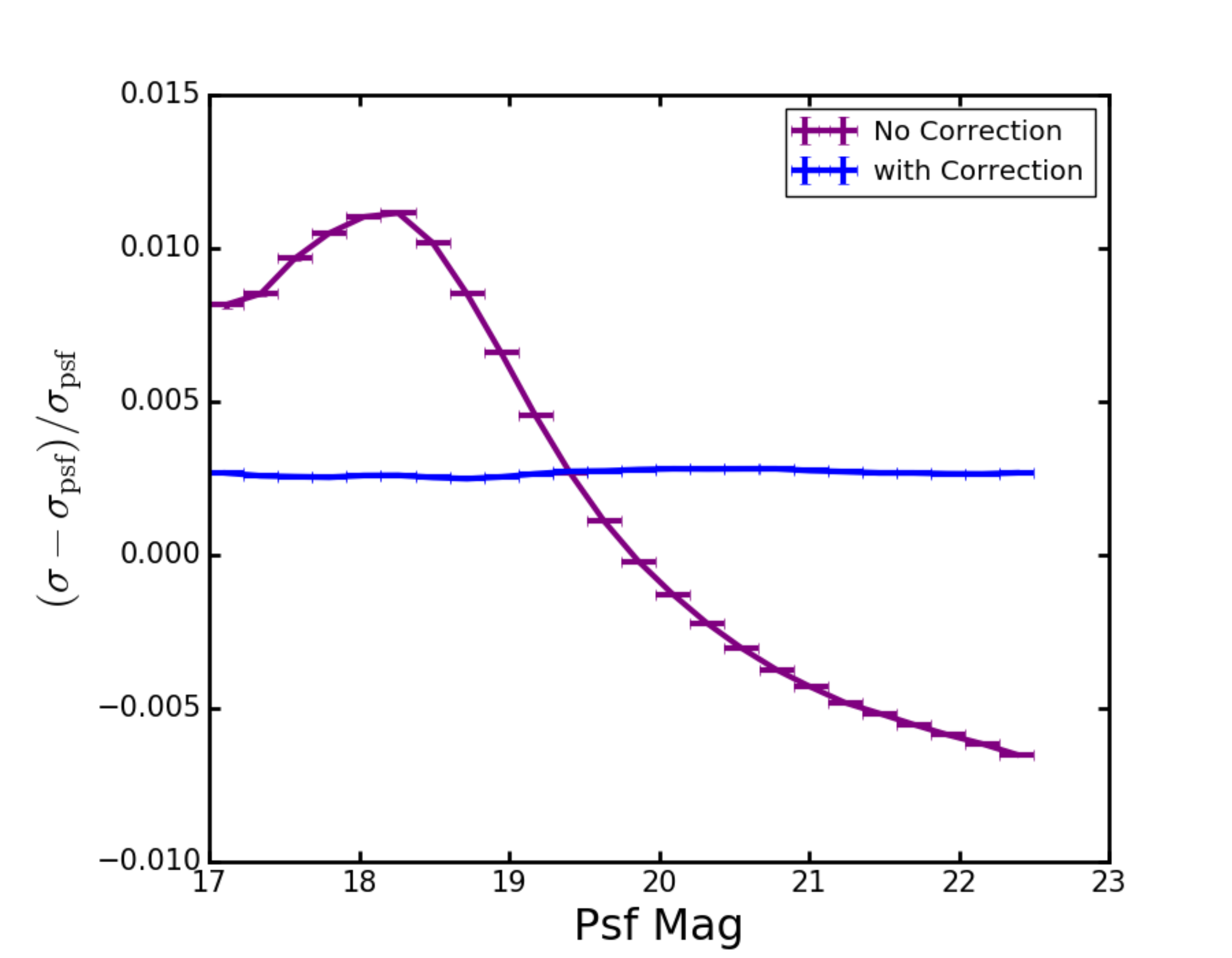}
\includegraphics[width=8cm]{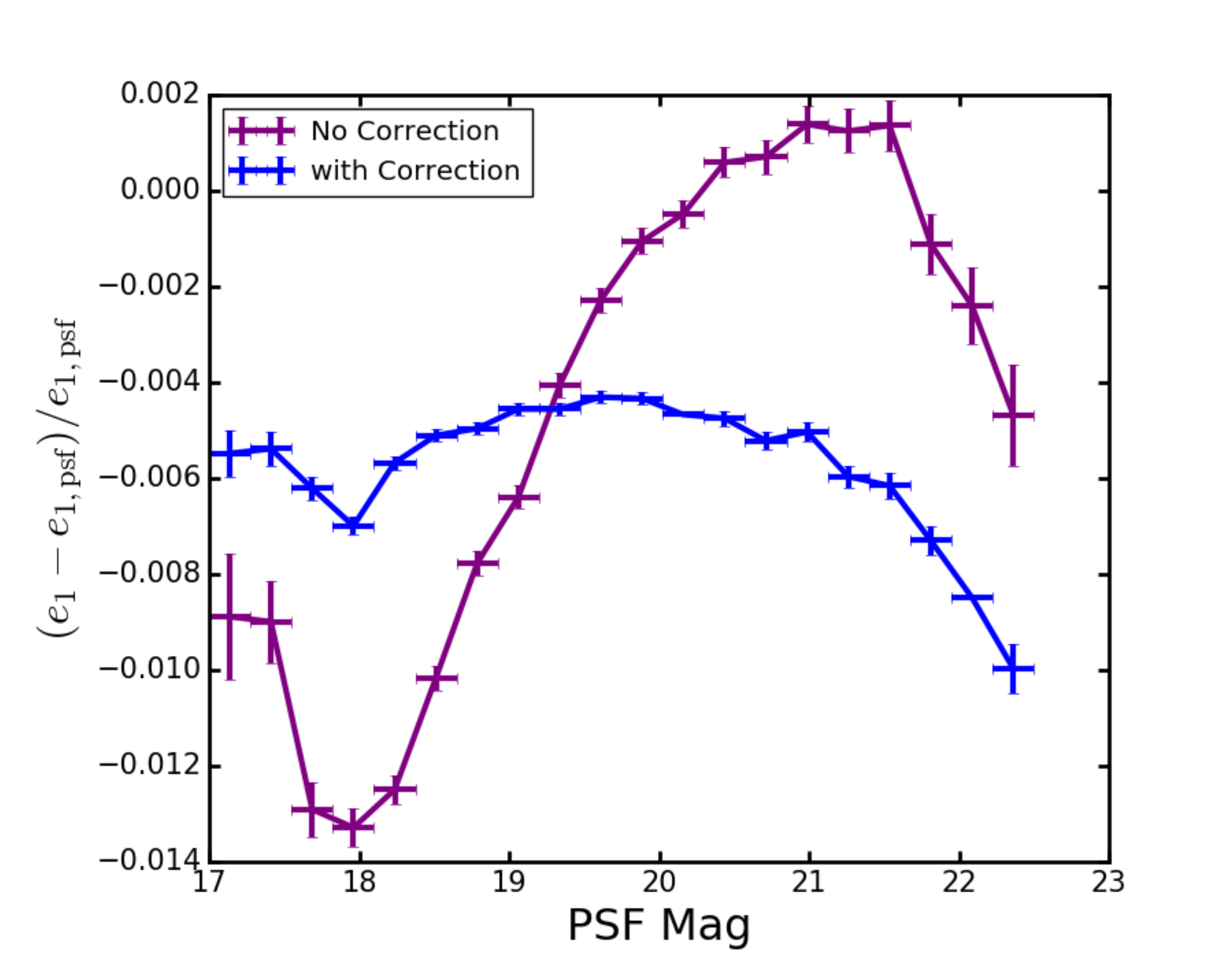}
\includegraphics[width=8cm]{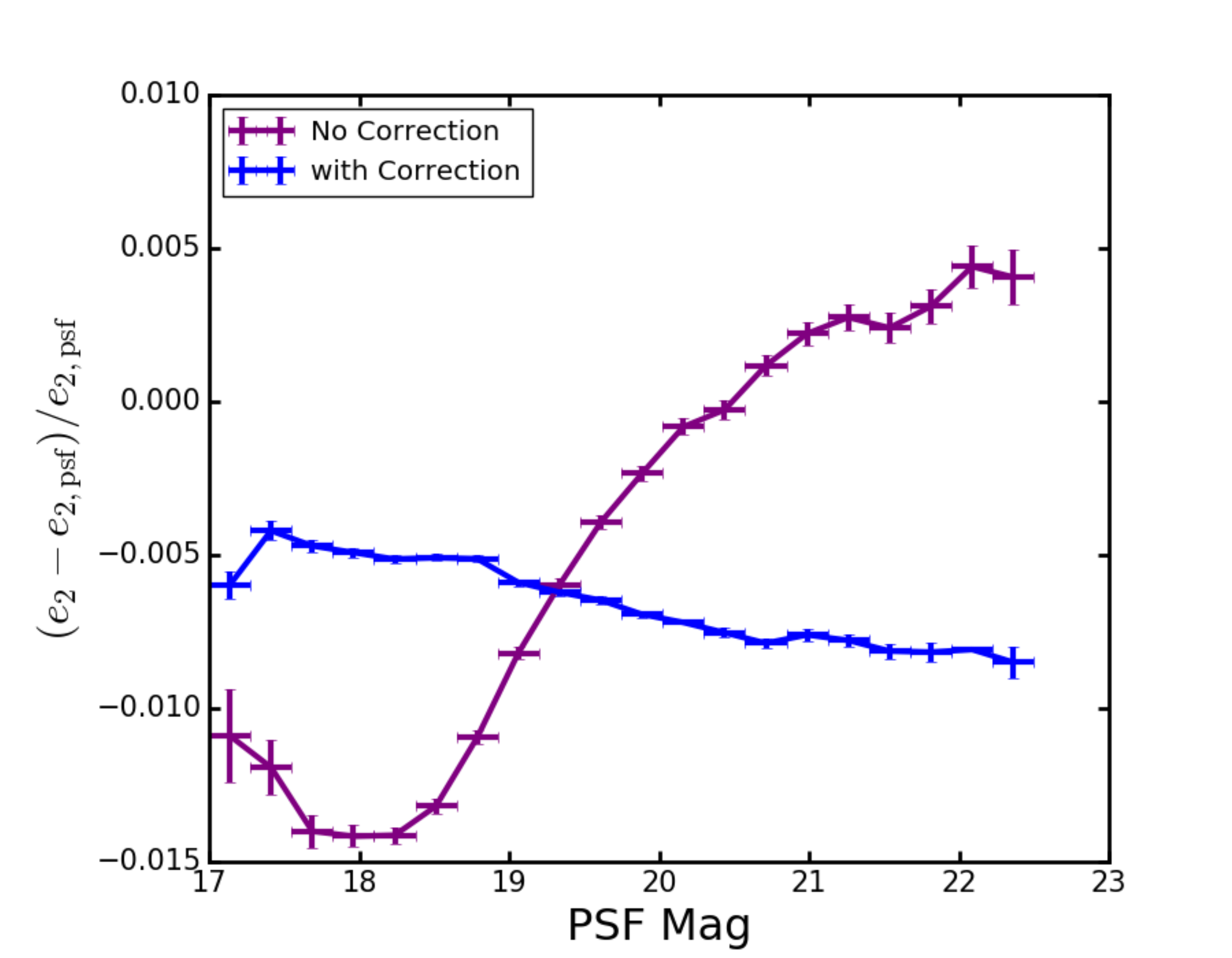}
\includegraphics[width=8cm]{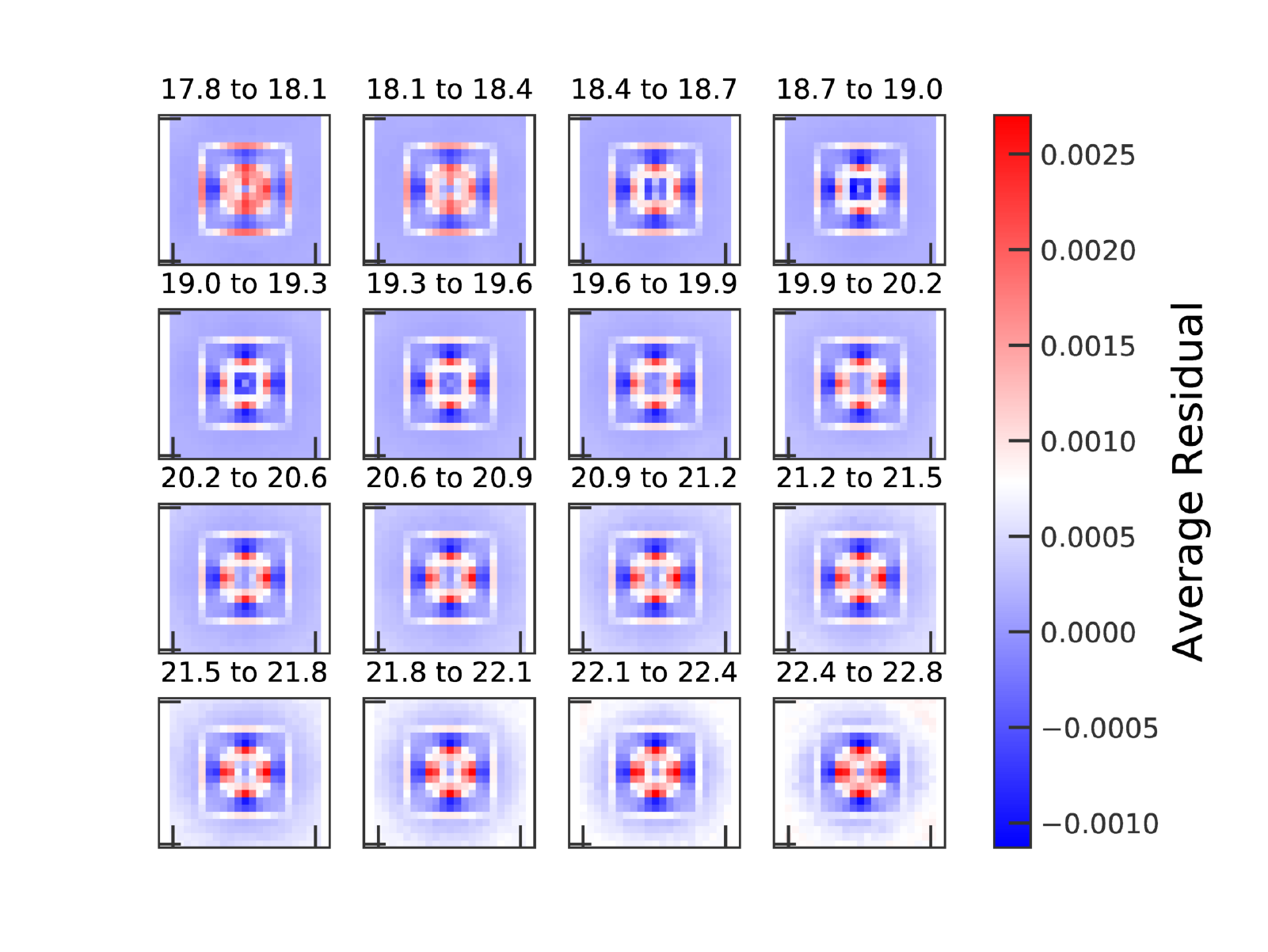}
\end{center}
  \caption{Residuals of the PSF models for size and ellipticity for the S16a $i$-band data as a function of
    PSF magnitude, both before (blue) and after (purple) brighter-fatter correction.  The lower right shows
    the stacked residuals of the corrected image and the PSF model as a function of PSF magnitude, where the
    brightness decreases from top left to bottom right. }
\label{fig:brighter-fatter}
\end{figure*}

We have devised a correction strategy in \citet{Coulton2017} to reapportion the flux.  In this strategy we
assume that the electric field has zero curl and can thus be written as the gradient of a potential.  This
potential is  determined from a fit to the noise correlations in flat field data.  In
\figref{brighter-fatter}, we show the performance of the correction algorithm on all of the $i$-band single
epoch data in the S16a internal data release by comparing the difference between the measured and model PSF
sizes and ellipticities with and without the correction.  We also include a plot of the difference between the
corrected stellar images and the PSF model where we have stacked the stars together in magnitude bins.  The correction reduces the large magnitude-dependent
bias for the size and ellipticity\footnote{Ellipticity is defined throughout this paper using the
  \textit{distortion} parameterization; see \appendixref{ellipse-parameterizations}}.  There is also an
overall magnitude-independent bias that is not corrected.  Further work in \citet{Coulton2017} was able to correct some of these biases by narrowing the magnitude range of selected stars and improving the recovery of the potential, but these changes were not available in the S16a internal release or PDR1.  Since we are comparing to the PSF model, there may be additional errors coming from PSF modeling itself (see \secref{psf-modeling}).  It should be noted that these corrections are nevertheless sufficient to meet the requirements for weak lensing science as demonstrated in \citet{hsc-shear}.

\subsection{PSF Modeling}
\label{sec:psf-modeling}

To perform PSF modeling we use a version of the PSFEx software \citep{2013ascl.soft01001B} modified for use as a library in the LSST DM codebase.  The modifications remove the necessity of running SExtractor beforehand, but we have not implemented the full set of features available with the nominal version; we limit its capability to performing a polynomial fit of the PSF as a function of position on a single CCD.  The modifications allow us to separate out the task of selecting stars from modeling the PSF.  The standard version performs both of these tasks, while we perform the star selection with our own algorithm and leave only the modeling to PSFEx.

\begin{figure}
\includegraphics[width=0.5\textwidth]{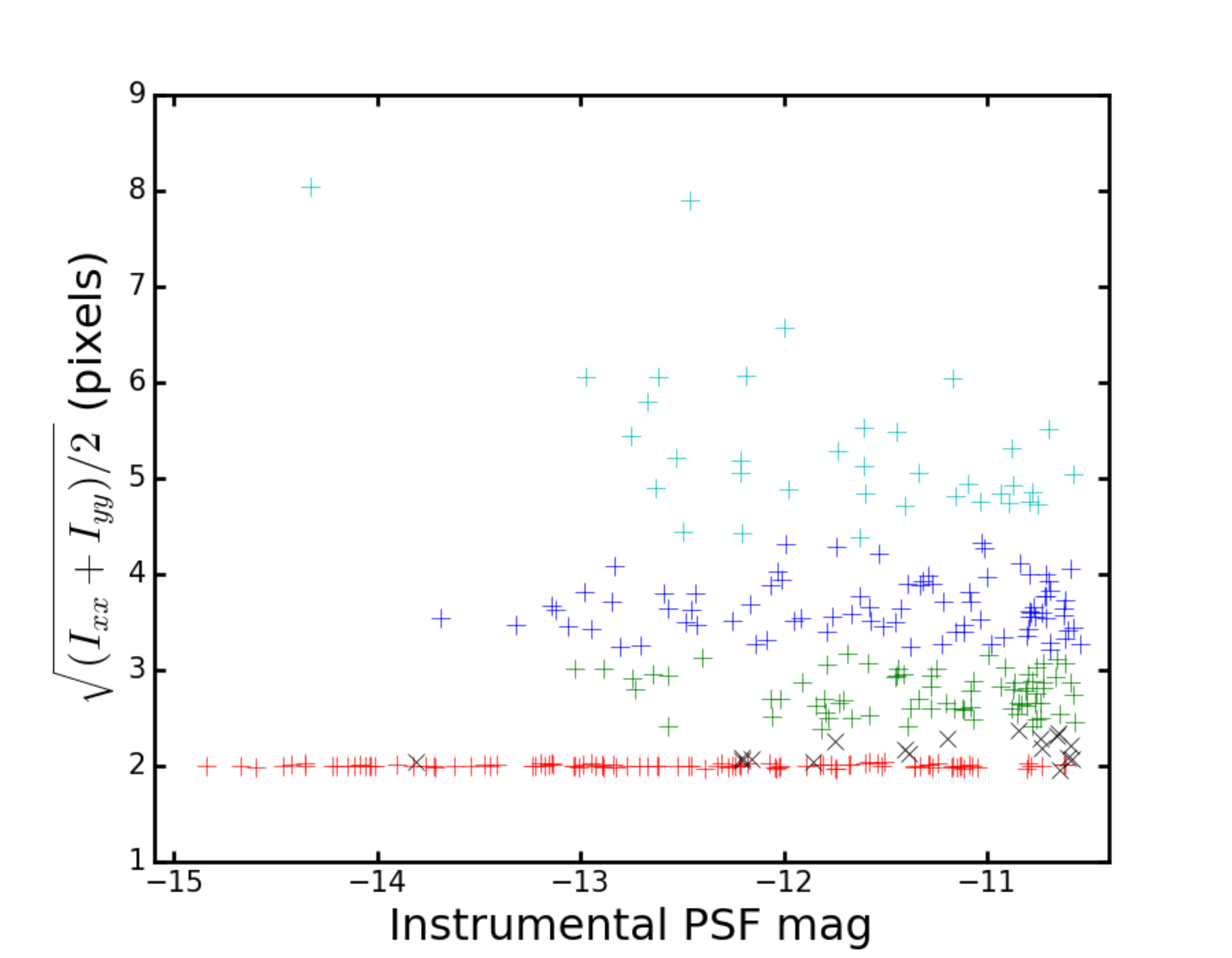}
\caption{Selection of stars based on the ``trace radius'' size (y axis) and PSF magnitude (x axis), in a
  single CCD.  Each of the different colors represents the classification from the k-means clustering
  algorithm.  The red objects are selected as PSF candidates.  The black objects are those that were rejected
  from the initial selection.  Worse seeing increases the scatter in the clusters but (at the levels we see in the SSP data) does not inhibit our ability to determine the clusters.}
\label{fig:star_selection}
\end{figure}

\begin{figure}
\begin{center}
\includegraphics[width=0.5\textwidth]{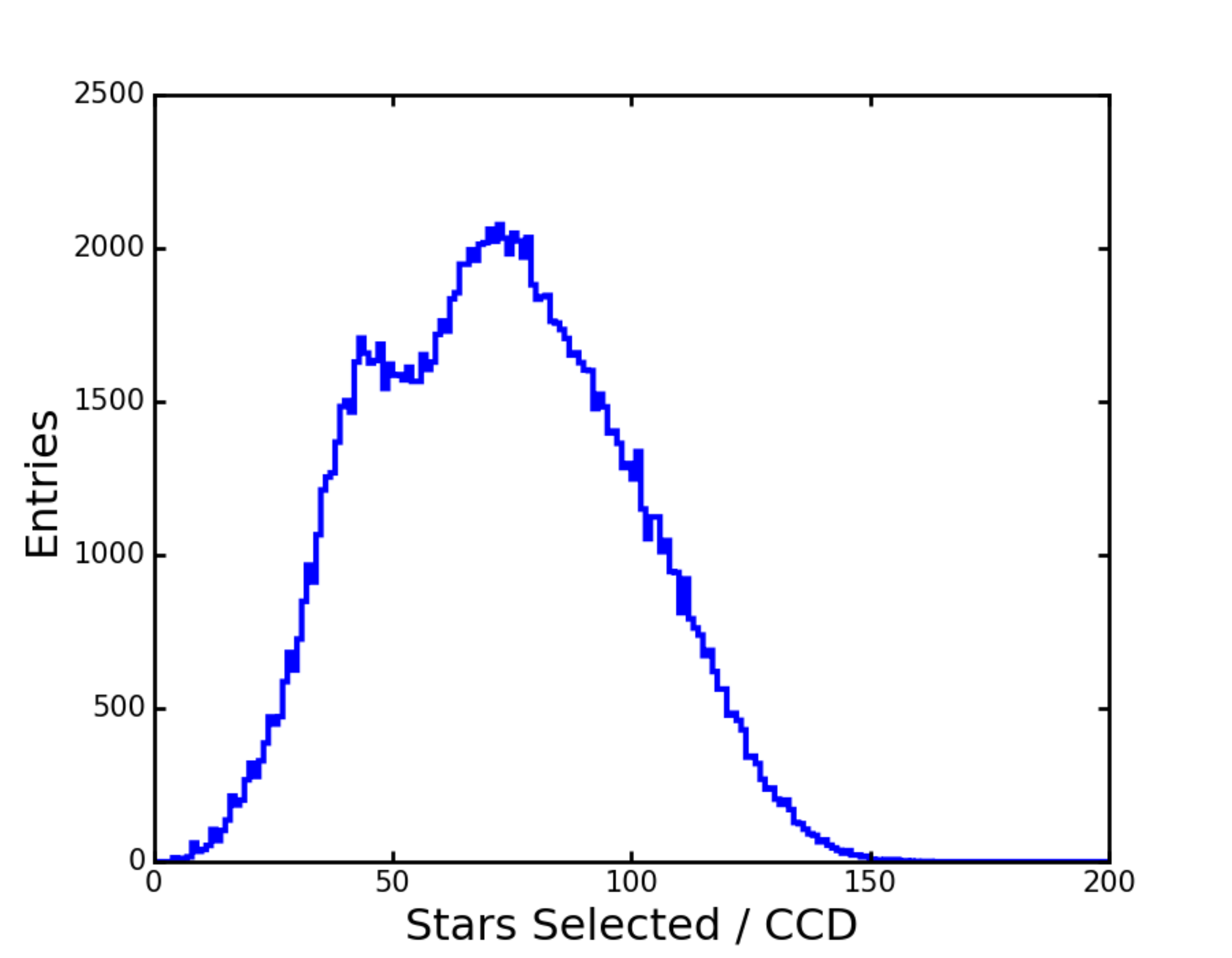}
\end{center}
  \caption{The number of stars selected per CCD in $i$-band images in the HSC-SSP Wide layer.
 }
\label{fig:nstar}
\end{figure}

We choose the set of stars to feed PSFEx based on a clustering analysis of the measured sizes of objects in each CCD (measured via
second image moments; see \secref{shapes}).  We use a k-means clustering algorithm on the sizes with a fixed set of
four clusters to isolate the stellar locus, after first limiting the source catalog to objects brighter than
12500 counts (approximately magnitude 22.3 in $i$).  The first cluster is initialized with the position of the
10th percentile from the sizes of all the measured objects.  The subsequent three clusters are then initially
spaced by intervals of 0.15 pixels. Each object in the catalog is assigned to the cluster whose center is
closest to its measured size, and as we iterate the position of each cluster is computed as the average size of its
members.  We iterate until objects do not move from one cluster to another.  The objects from
the cluster with the smallest mean size are used as the PSF candidate sample. A final rejection is done
on the sample by clipping objects outside the interquartile range centered on the median.  Figure~\ref{fig:star_selection} shows an example of the four different clusters as a function of size and magnitude for a single CCD.  The red objects are those selected as stars.  As a cross-validation sample we reserve 20$\%$ of the objects from the modeling.  \figref{nstar} shows the number of stars selected per CCD for the $i$-band visits.  On average, we select $\sim$72 stars to be used in the PSF modeling on each CCD.

Postage stamps (41$\times$41 pixels) are extracted from the images around each star selected for use in PSF estimation.  Since we have not run deblending yet, we mask out secondary peaks (see \secref{detection}) and regions suspected to be influenced by neighboring objects.  These stamps are then fed to PSFex as the initial set of stars.  PSFEx is configured to use \texttt{BASIS\_TYPE}=\texttt{PIXEL\_AUTO} which chooses the sampling of the PSF model as the native pixel size, unless the seeing goes below 3 pixels ($\sim$0.5\arcsec), in which case the model is oversampled by a factor of two.  We use \texttt{BASIS\_NUMBER}=20, indicating that the full modeling is done on the central 20$\times$20 pixels (or 10$\times$10 when oversampling).  For the flux normalization, we use an aperture flux with 12 pixel radius, which is large enough to encompass the majority of the flux and still be high S/N.  We fit the PSF model to each CCD completely independently, using a second order polynomial to interpolate between stars the values of each pixel in the kernel.

To assess the quality of the PSF modeling we fit Gaussian moments as described in \secref{shapes} to the stars and the PSF model
and compare their sizes and ellipticities.  Another useful quality metric is the correlation function of PSF model ellipticity residuals \citep{2010MNRAS.404..350R},
\begin{eqnarray}
\rho_1(r) &=& \sum_{\rm{pairs}\,i,j} \bigg[(e_1-e_{1,\mathrm{psf}})_i*(e_1-e_{1,\mathrm{psf}})_j \bigg. \nonumber\\
     & \quad & \quad\quad\quad + \, \bigg. (e_2-e_{2,\mathrm{psf}})_i*(e_2-e_{2,\mathrm{psf}})_j\bigg]
\label{eqn:rho1}
\end{eqnarray}
where the sum is over all pairs separated by a given distance $r$.  Using the coadded PSF models described in \secref{psf-coaddition}, we can run the same test on the stars and PSF models on the coadds.  To select a set of stars on the coadds we require that a given star be selected as a PSF star in $\ge$20$\%$ of the individual visits to allow for chip gaps, etc.

\begin{figure*}
\begin{center}
\includegraphics[width=8cm]{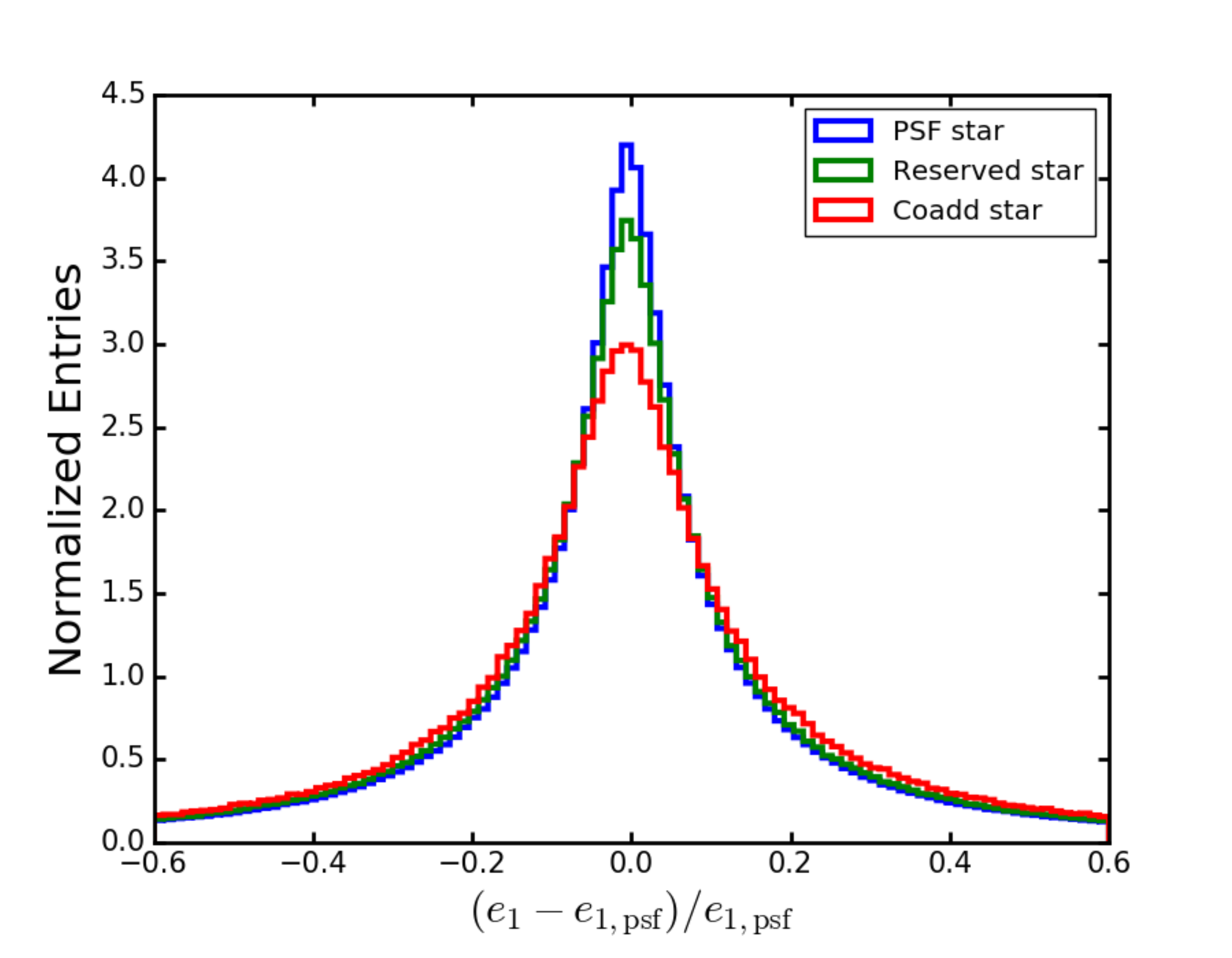}
\includegraphics[width=8cm]{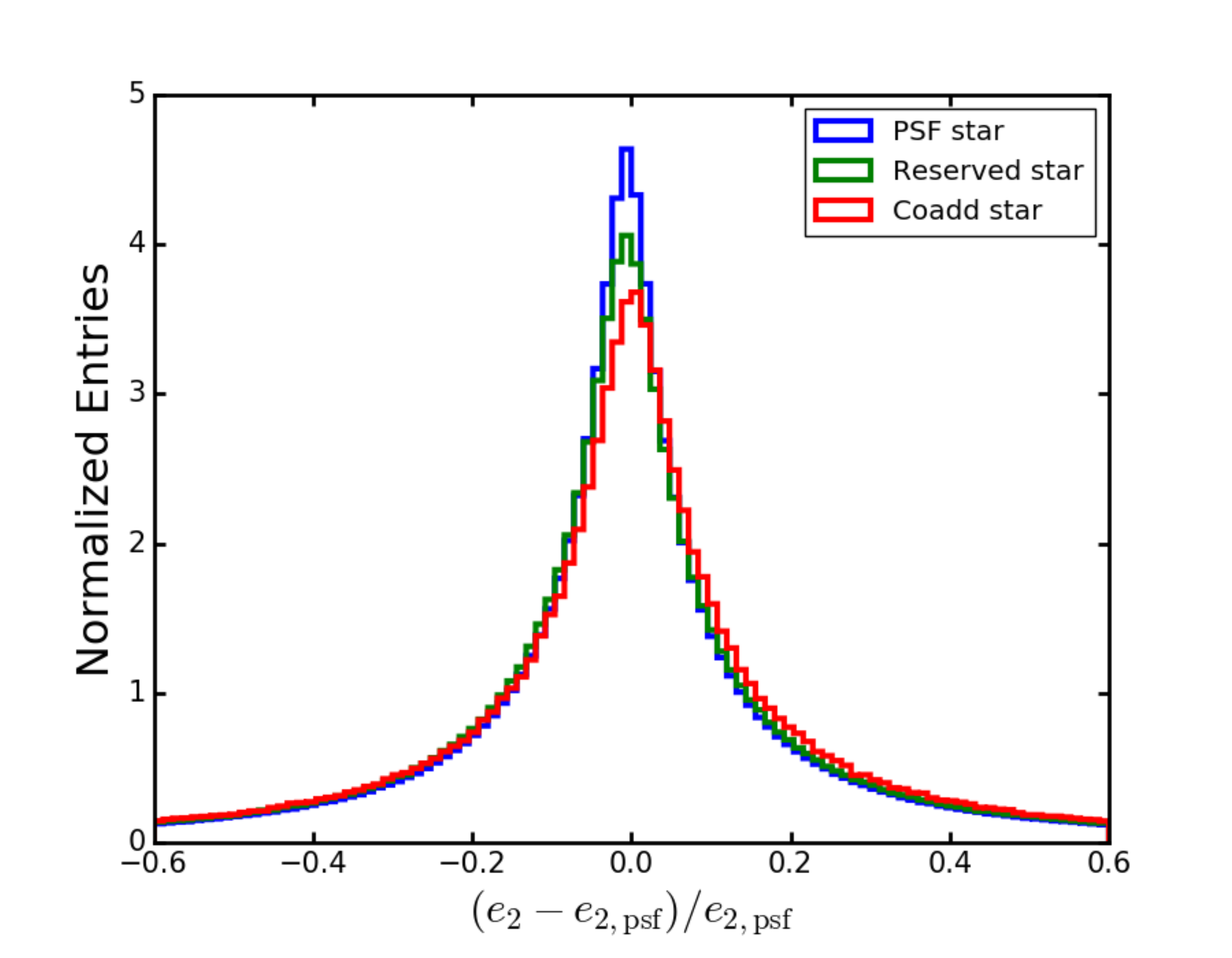}
\includegraphics[width=8cm]{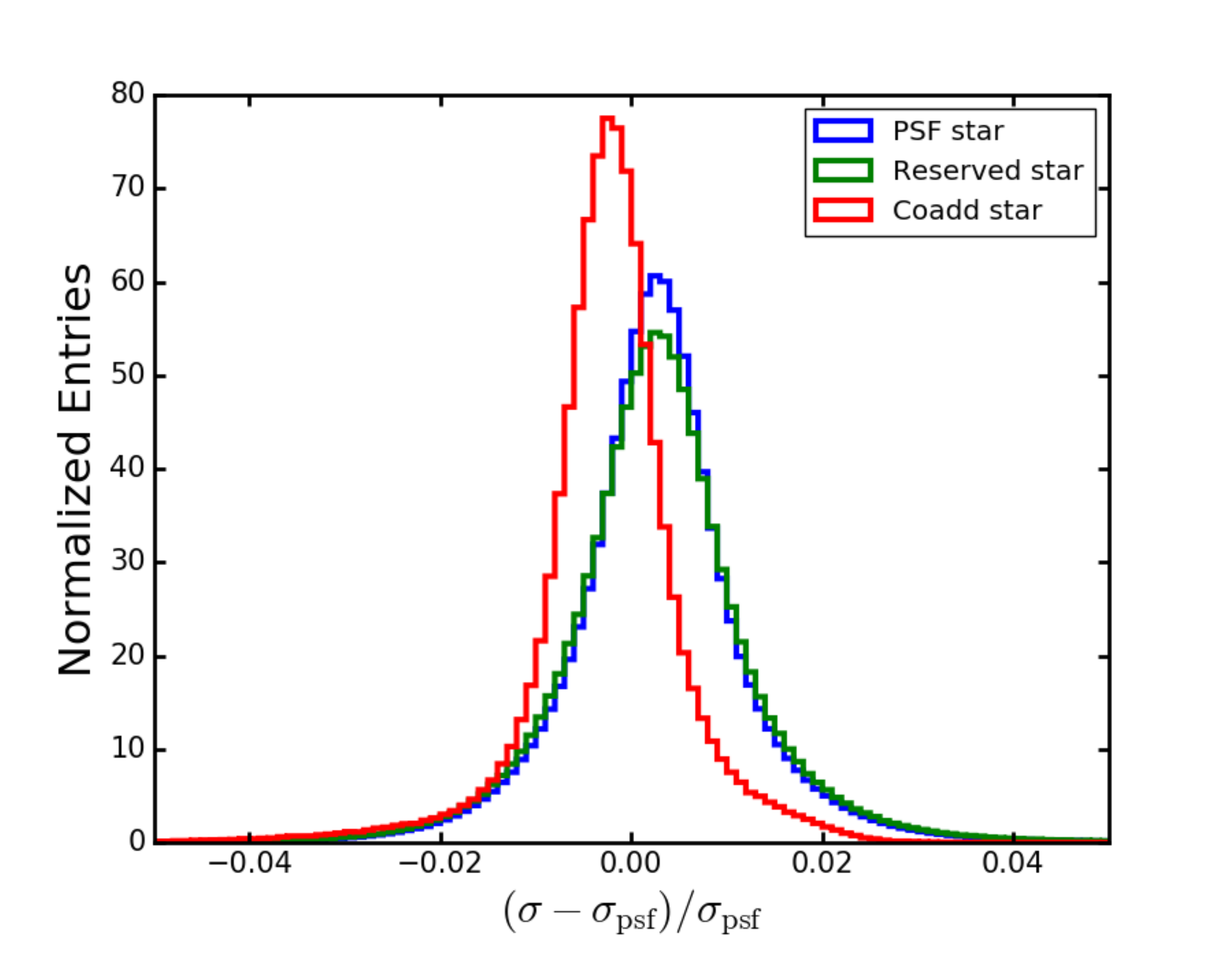}
\includegraphics[width=8cm]{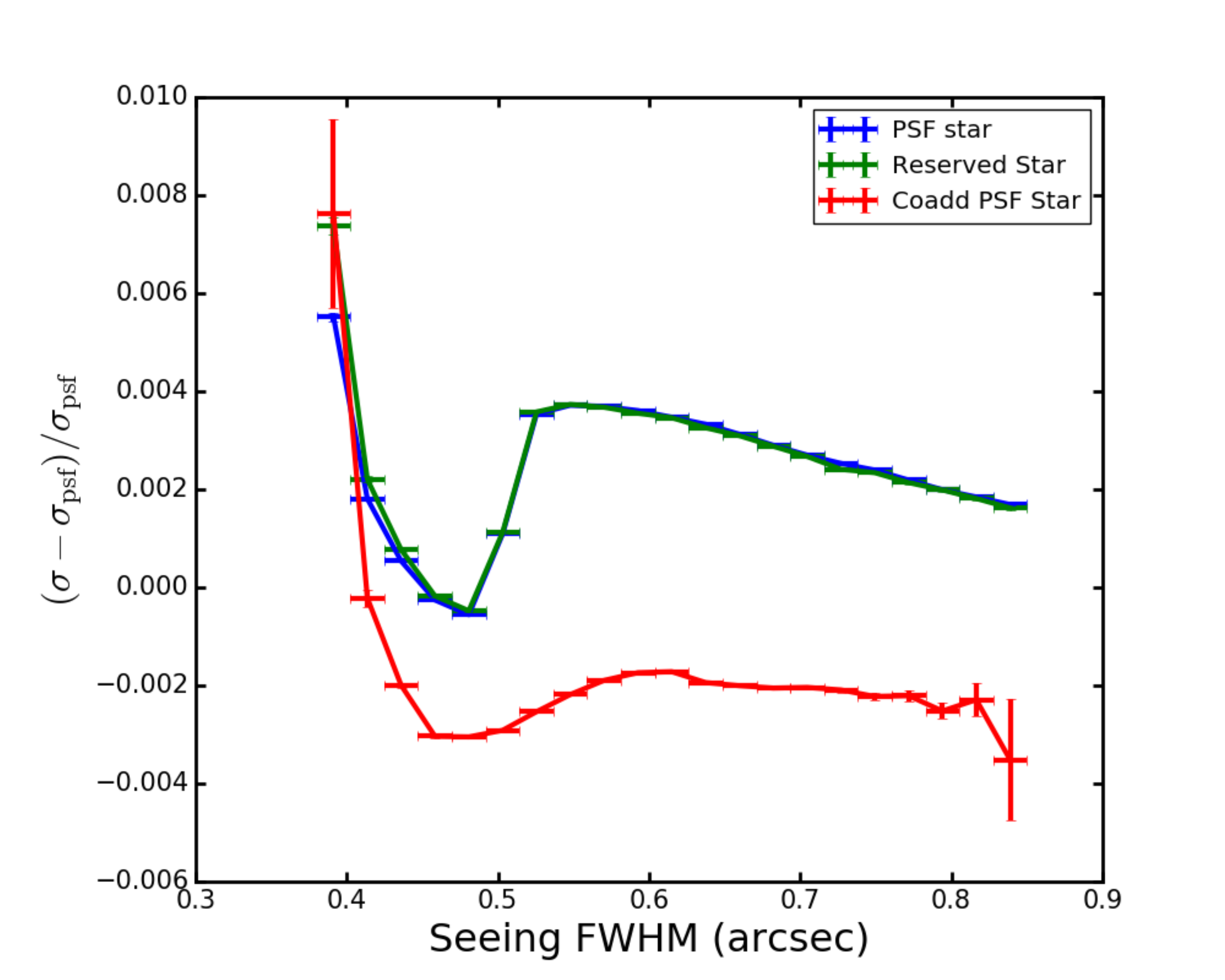}
\end{center}
  \caption{Size and ellipticity residuals for the $i$-band visits for three different samples: PSF stars,
    reserved stars and coadd stars.  The upper panels show histograms of the difference between the
    ellipticity of a star and the PSF model evaluated at the same position for the $e_1$ (pixel-aligned) and
    $e_2$ (diagonal) ellipticity components.  The lower left panel shows the histogram of PSF-star fractional
    residuals of the size, $\sigma$, for the same sample, and the lower right panel shows the fractional size residuals as a function of size.  Ellipticities are defined using the distortion convention (see \appendixref{ellipse-parameterizations}).
 }
\label{fig:psf_hist}
\end{figure*}

\begin{figure*}
\begin{center}
\includegraphics[width=8cm]{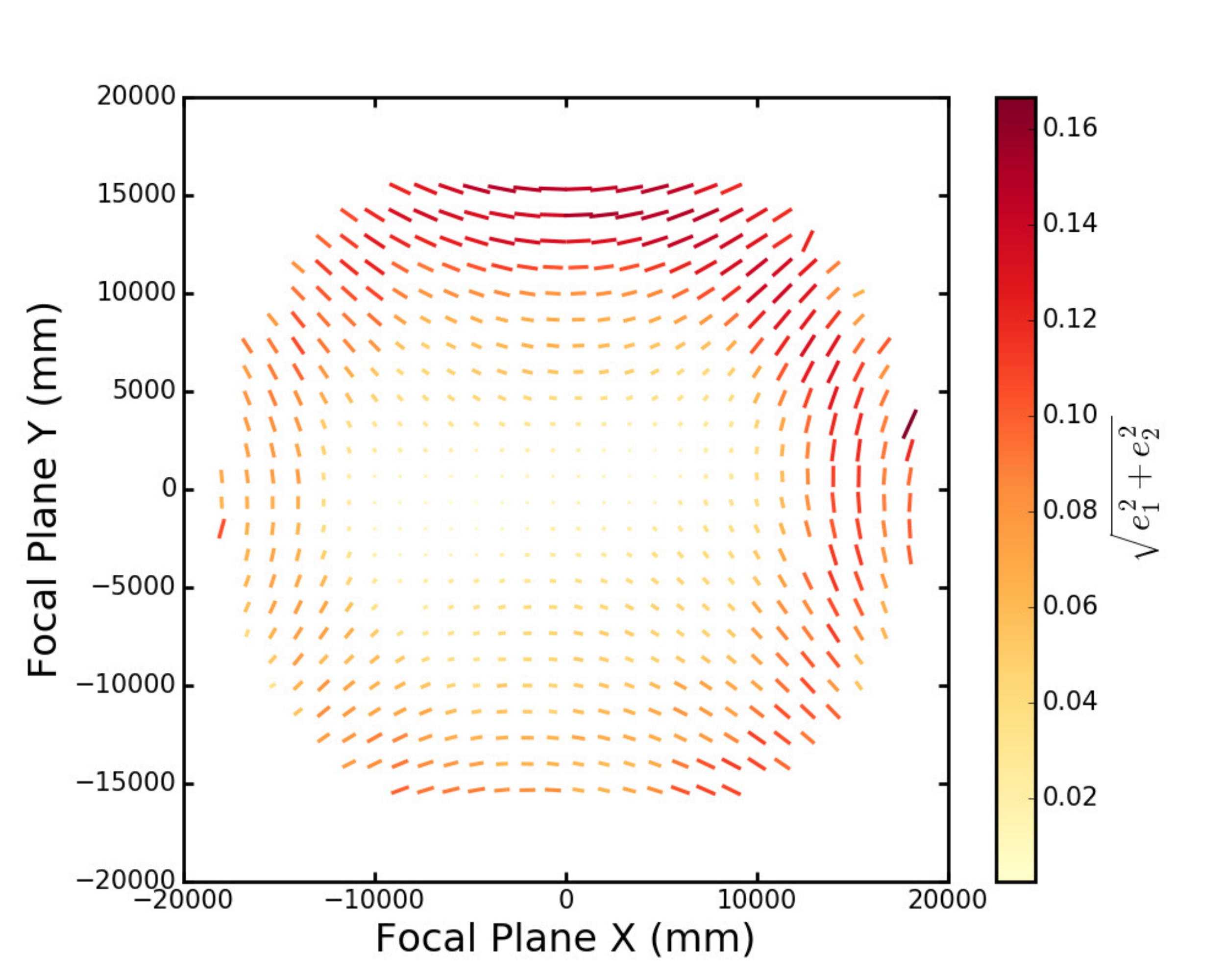}
\includegraphics[width=8cm]{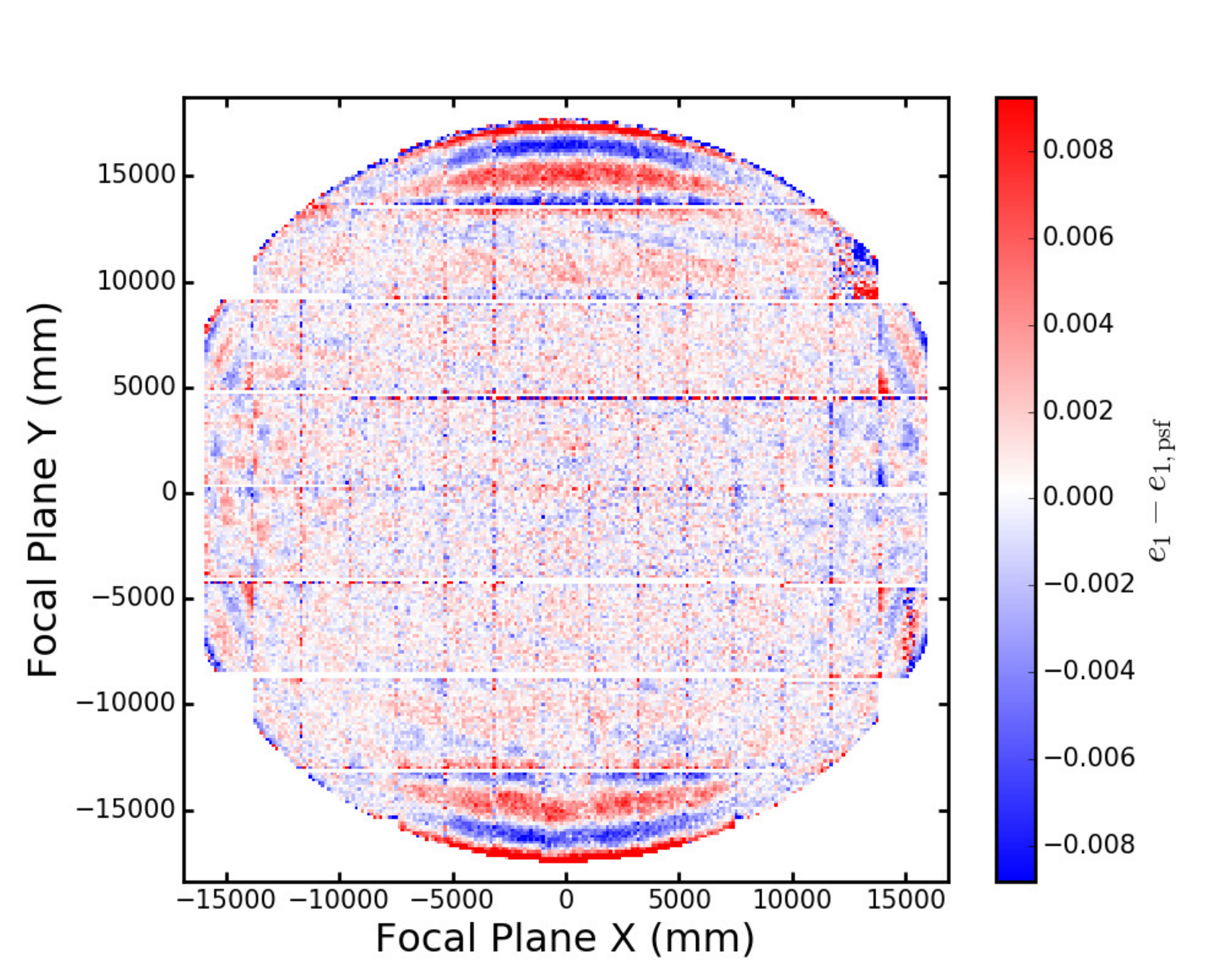}
\includegraphics[width=8cm]{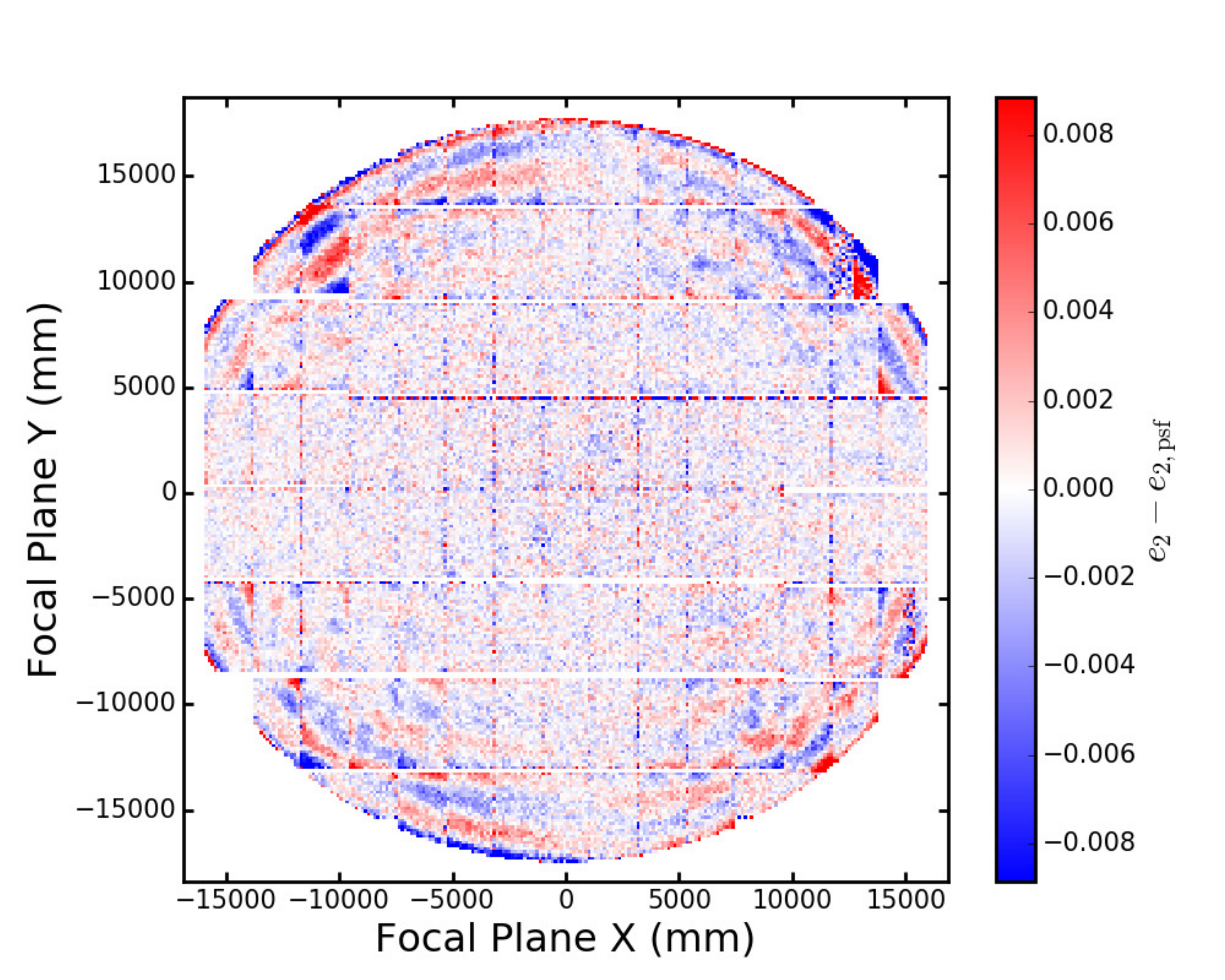}
\includegraphics[width=8cm]{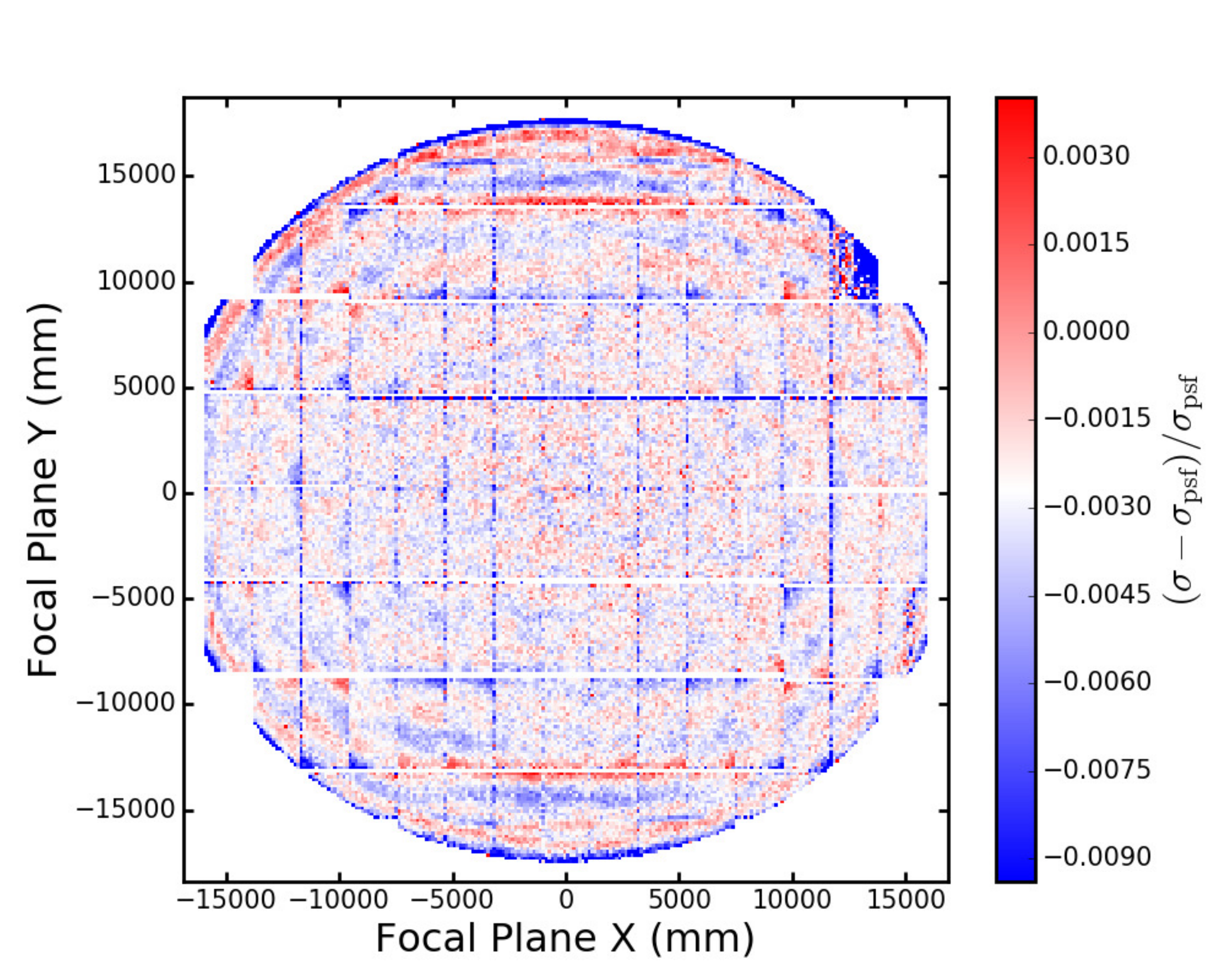}
\end{center}
  \caption{Ellipticity and size metrics as a function of position on the focal plane averaged over all of the
    $i$-band data.  The upper left panel
    shows the average ellipticity of the PSF a function of position on the focal plane.  The upper right and
    lower left panels (respectively) show the difference between the PSF model and stars in the $e_1$ and
    $e_2$ ellipticity components.  The relative difference in size, $\sigma$, is shown in the lower right
    panel.  Ellipticities are defined using the distortion convention (see \appendixref{ellipse-parameterizations}).
  }
\label{fig:psf_focal}
\end{figure*}

\begin{figure*}
\begin{center}
\includegraphics[width=8cm]{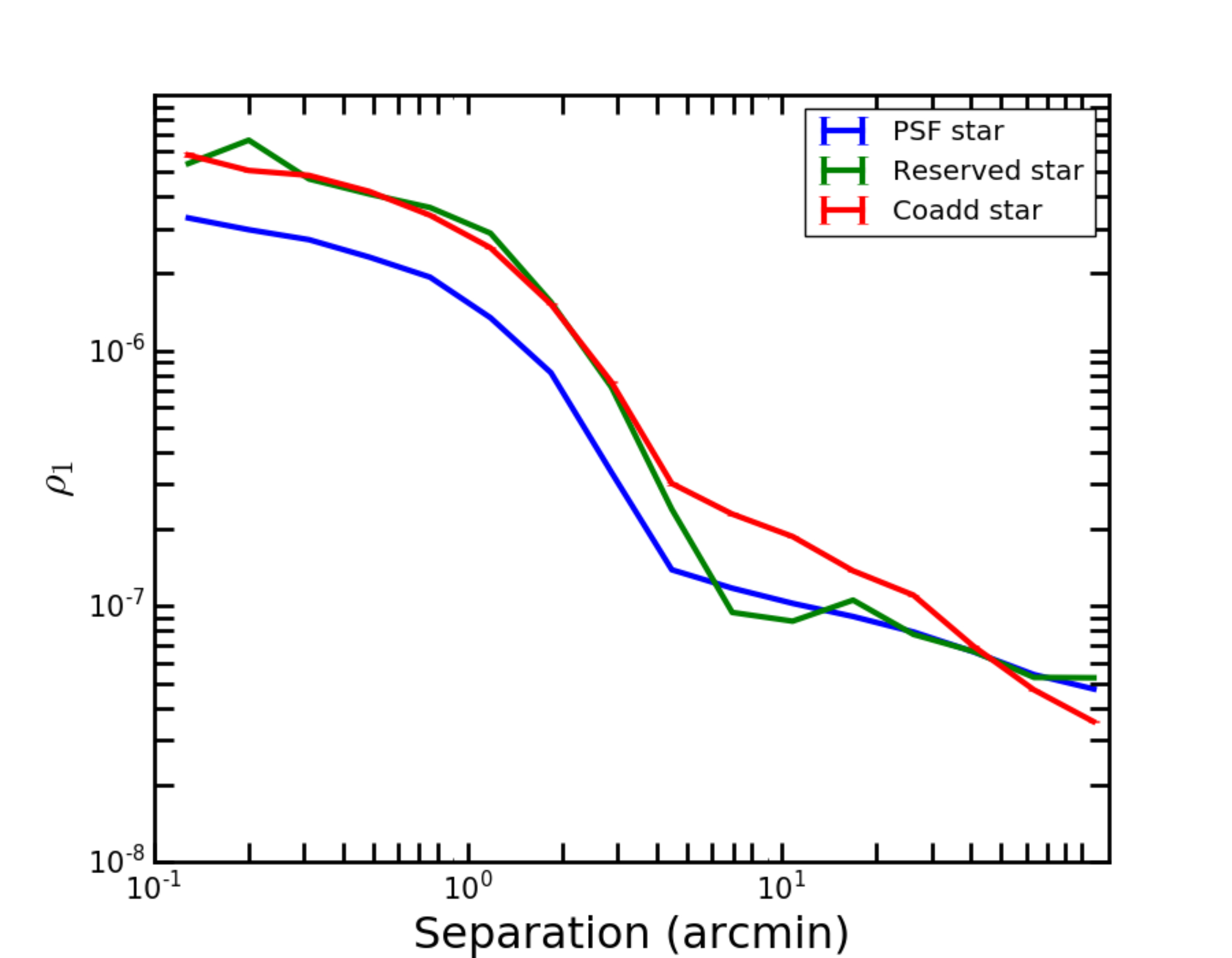}
\includegraphics[width=8cm]{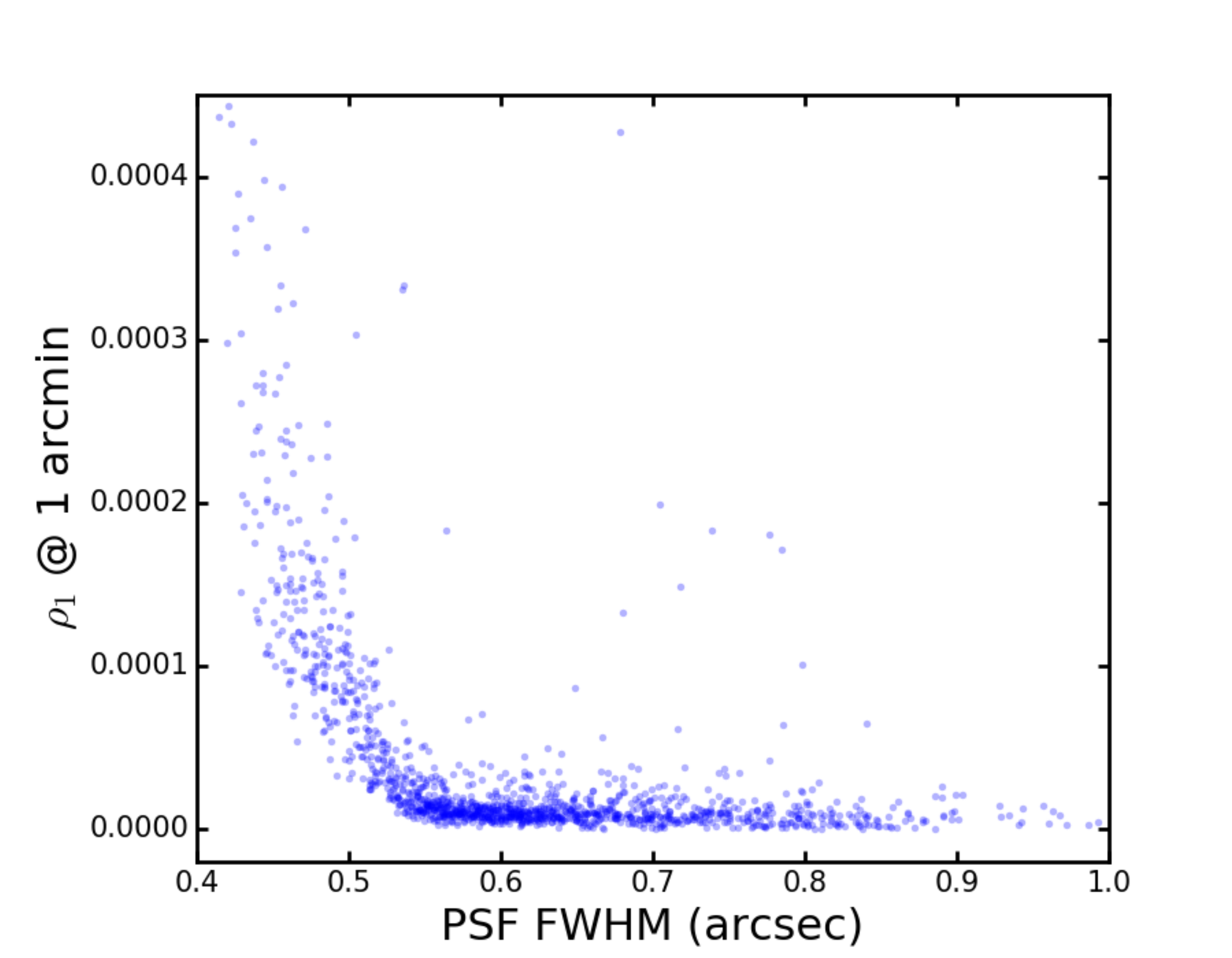}
\end{center}
  \caption{Values of $\rho_1$ as defined in \eqnref{rho1} in the text, averaged over all $i$-band stars (left) and the value of $\rho_1$ at 1\arcmin\ for all $i$-band visits.
 }
\label{fig:psf_corr}
\end{figure*}

Figure~\ref{fig:psf_hist} shows the size and ellipticity residuals for the visit PSF stars, visit reserved stars, and coadd stars.  Figure~\ref{fig:psf_focal} shows the ellipticity and the size and ellipticity residual patterns averaged over the focal plane. These plots show that there is un-modeled structure in the outer regions of the focal plane, especially for $e_1$.  Increasing the polynomial order to four removes some of this structure, but we were concerned that some visits would have too few stars to fit the additional degrees of freedom, and we have hence kept the polynomial order at 2.  In addition, the reserved stars are systematically worse than the stars used in the fit, suggesting potential problems in the interpolation or with over-fitting the data.  The coadd PSFs also perform slightly worse than the per-visit PSFs; naively, one would expect the size of the stars on the coadd to be larger relative to the stack of the PSF models due to astrometric scatter.  However, Figure~\ref{fig:psf_hist} shows that the stars on the coadd are systematically smaller, whereas the stars on the individual visits are larger.  We were unable to identify the cause of this discrepancy.  Figure~\ref{fig:psf_corr} shows the $\rho_1$ values averaged over all the objects in each star set and the $\rho_1$ values for each visit evaluated at 1\arcmin.

One serious problem we found was that the PSF modeling becomes much worse as the seeing gets better.  The lower right panel of Figure~\ref{fig:psf_hist} and the right panel of Figure~\ref{fig:psf_corr} show this explicitly: as the PSF FWHM decreases, the size residual and $\rho_1$ increase significantly.  One problem with these visits is that the initial processing (see \secref{ccd-processing}) uses an initial guess of the PSF size of 1\arcsec, which is too large for the good-seeing visits.  This causes the star selector to behave strangely and allows a large number of galaxies into the sample.  Fixing this does not resolve all problems with good-seeing data, however, and we have been unable to determine the cause of the remaining issues (though we believe them to be internal to PSFEx).  In the S15a, S16a, and PDR1 data releases \citep[see][]{2017arXiv170208449A}, we chose to reject problematic visits entirely when building the coadd.  After the processing for these releases was completed, we identified more regions and visits with good seeing data that had not originally been identified, but caused significant biases in the PSF modeling.  There is one particularly large region in the $i$-band GAMA09 field that users should be aware of.  \citet{hsc-shear} shows PSF diagnostics for individual fields in the HSC-SSP Wide layer, including which parts of these fields pass the PSF model quality cuts for weak lensing.

In addition to the overall poor performance in good seeing, the lower right panel of Figure~\ref{fig:psf_hist}
shows an abrupt change at $\sim 0.5$\arcsec\ where we change from the native pixel sampling to the oversampled
basis.  We attempted to improve the performance in worse seeing by using the oversampled basis, but found
that the additional degrees of freedom caused other problems.  In \citet{hsc-shear}, a different set of diagnostics more directly related to weak lensing analysis is performed.  In future releases, we plan to improve the PSF modeling code to handle undersampled images so that we do not need to reject our best seeing data.   We also plan to automatically reject CCDs with problems so as to eliminate the need to look by hand for problematic visits.

\subsection{Cosmic Ray Detection}
\label{sec:cr-detection}

\begin{figure}
    \includegraphics[width=0.158\textwidth]{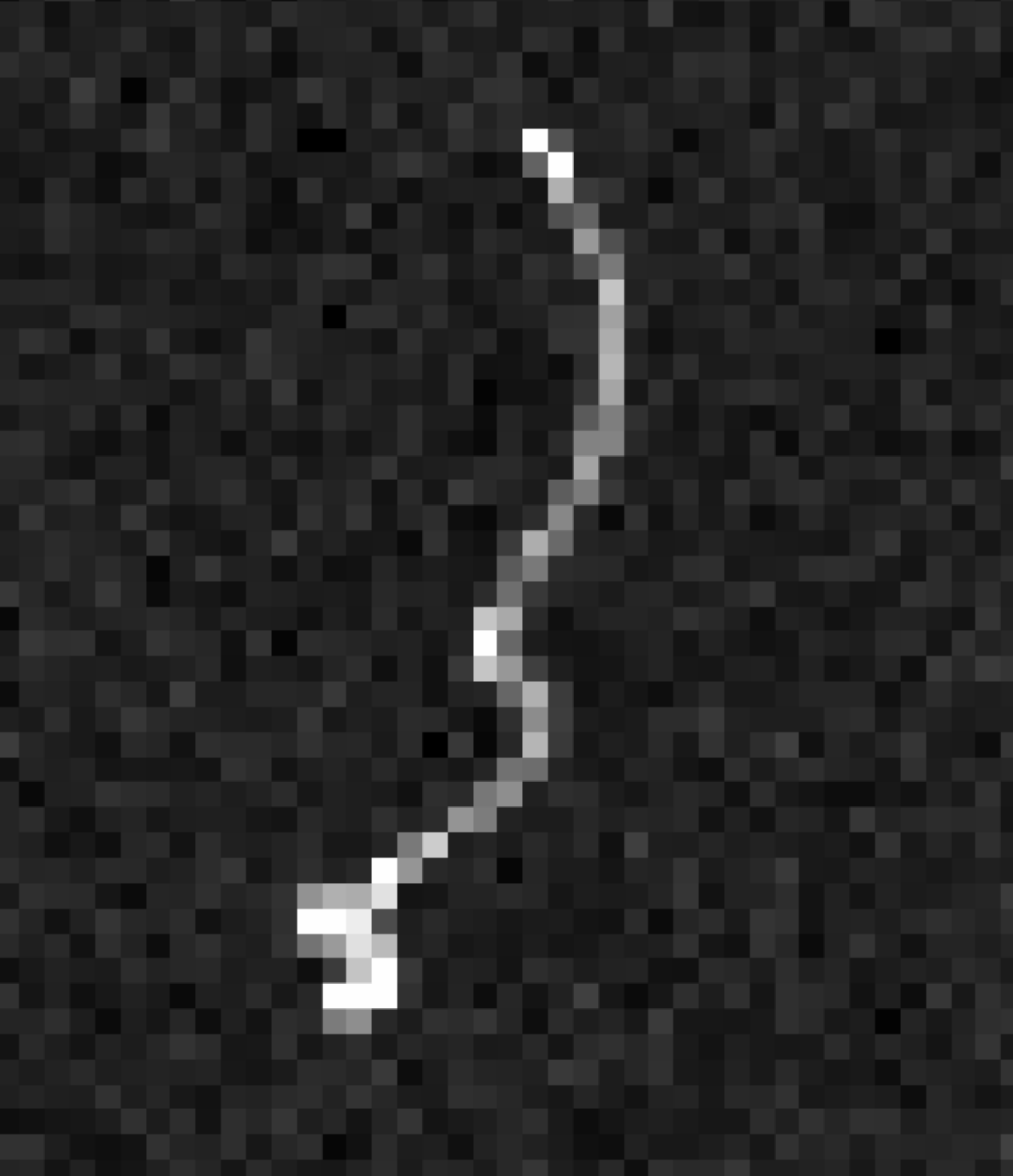}
    \includegraphics[width=0.158\textwidth]{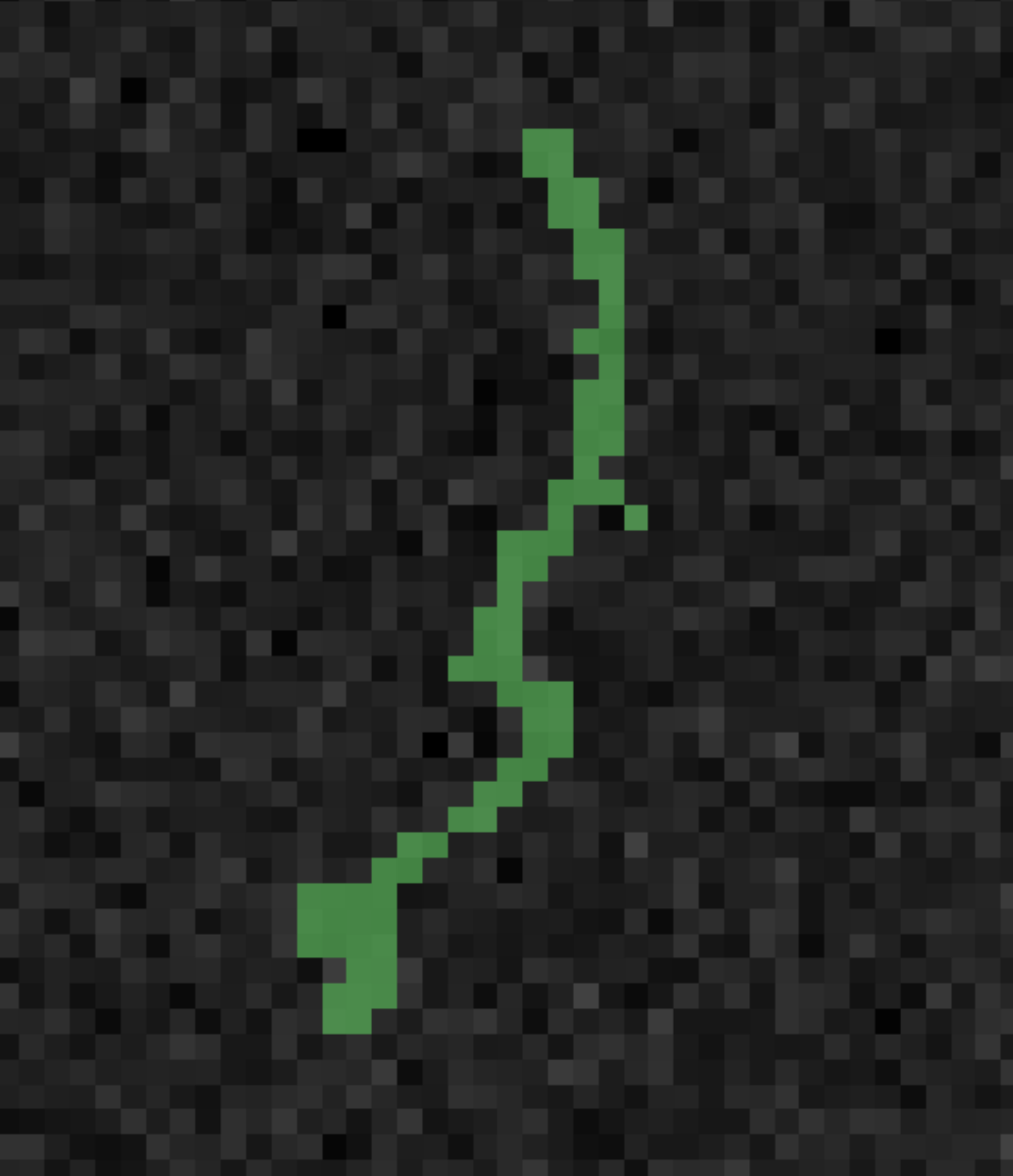}
    \includegraphics[width=0.158\textwidth]{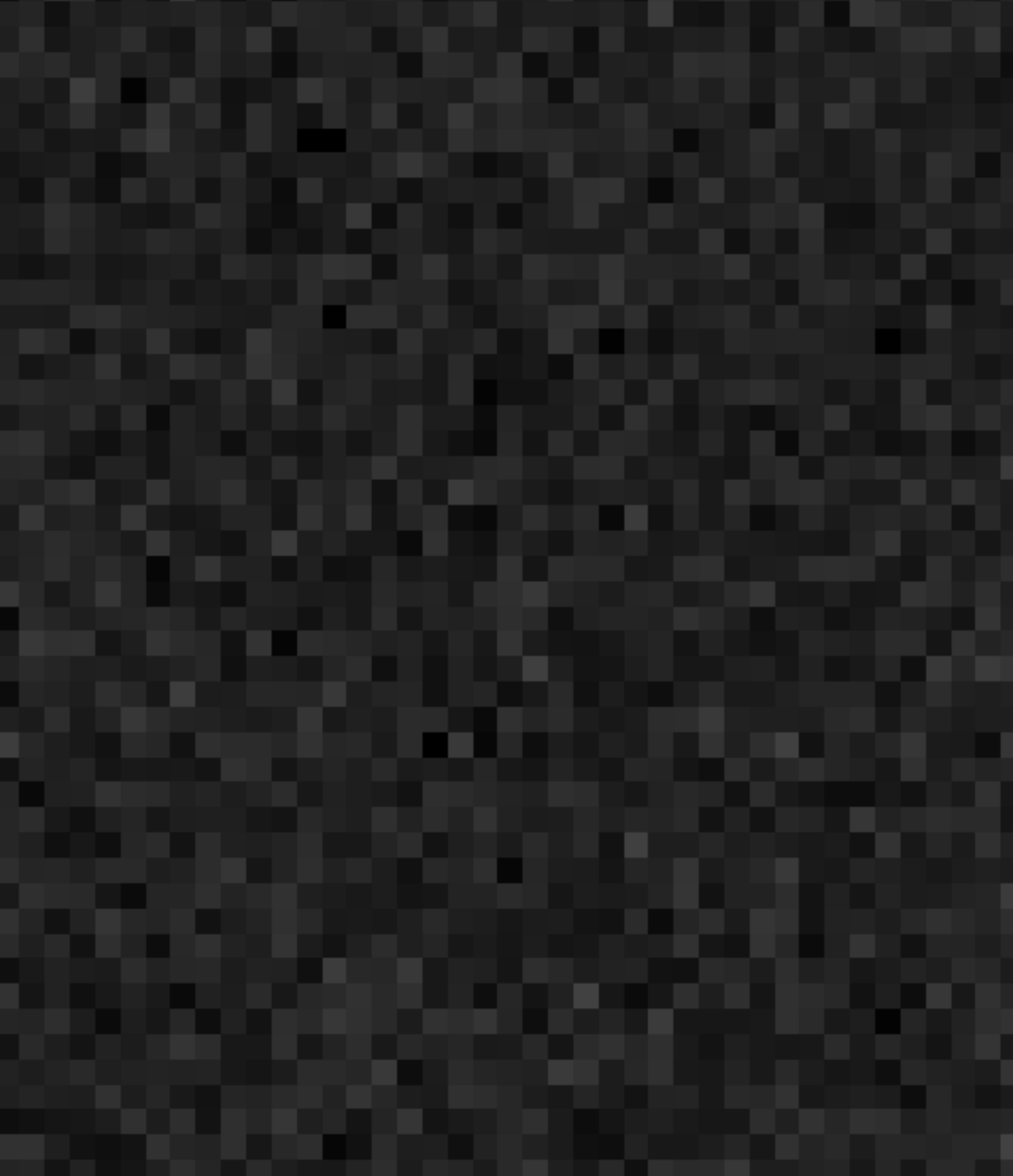} \\
    \includegraphics[width=0.158\textwidth]{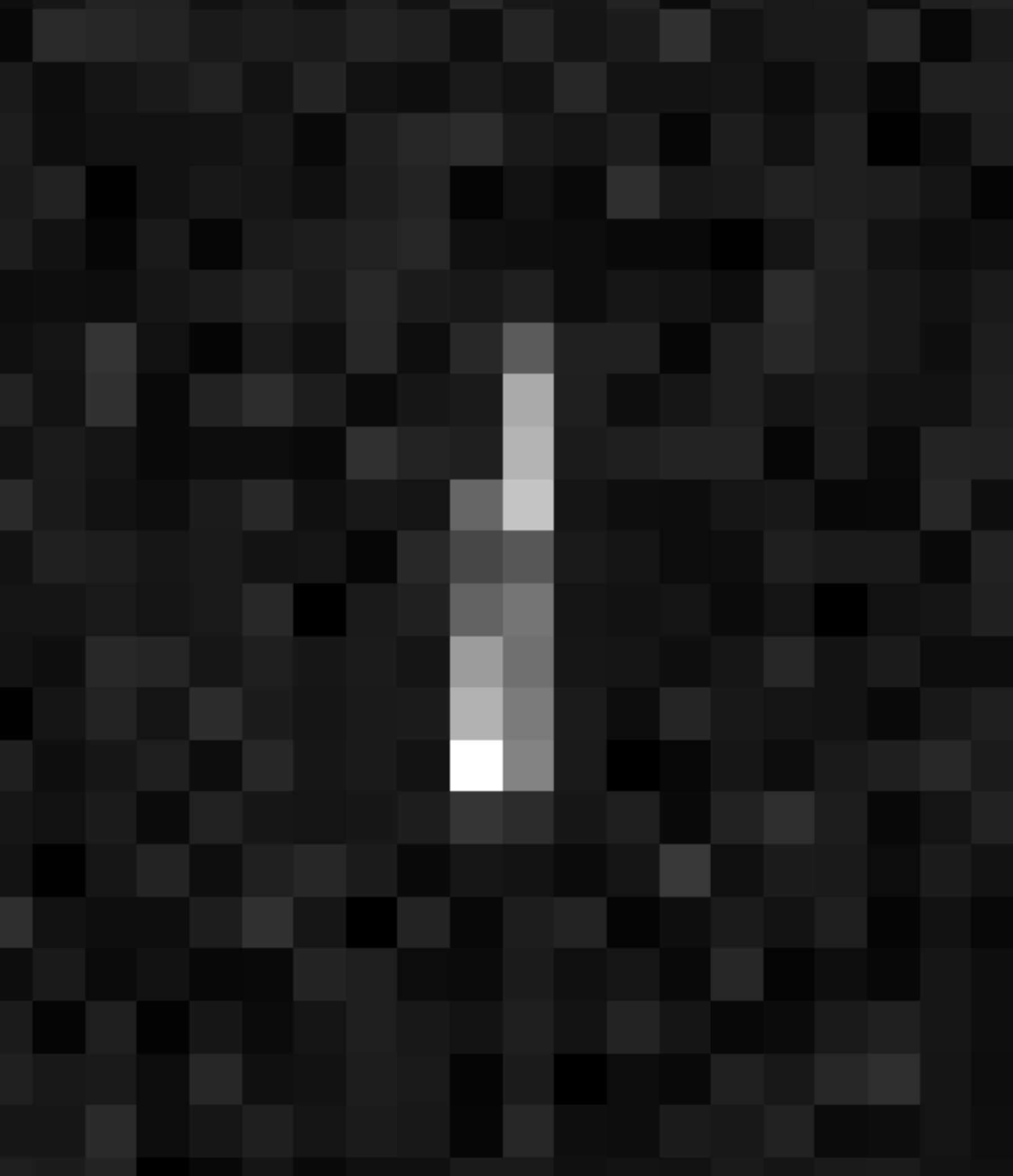}
    \includegraphics[width=0.158\textwidth]{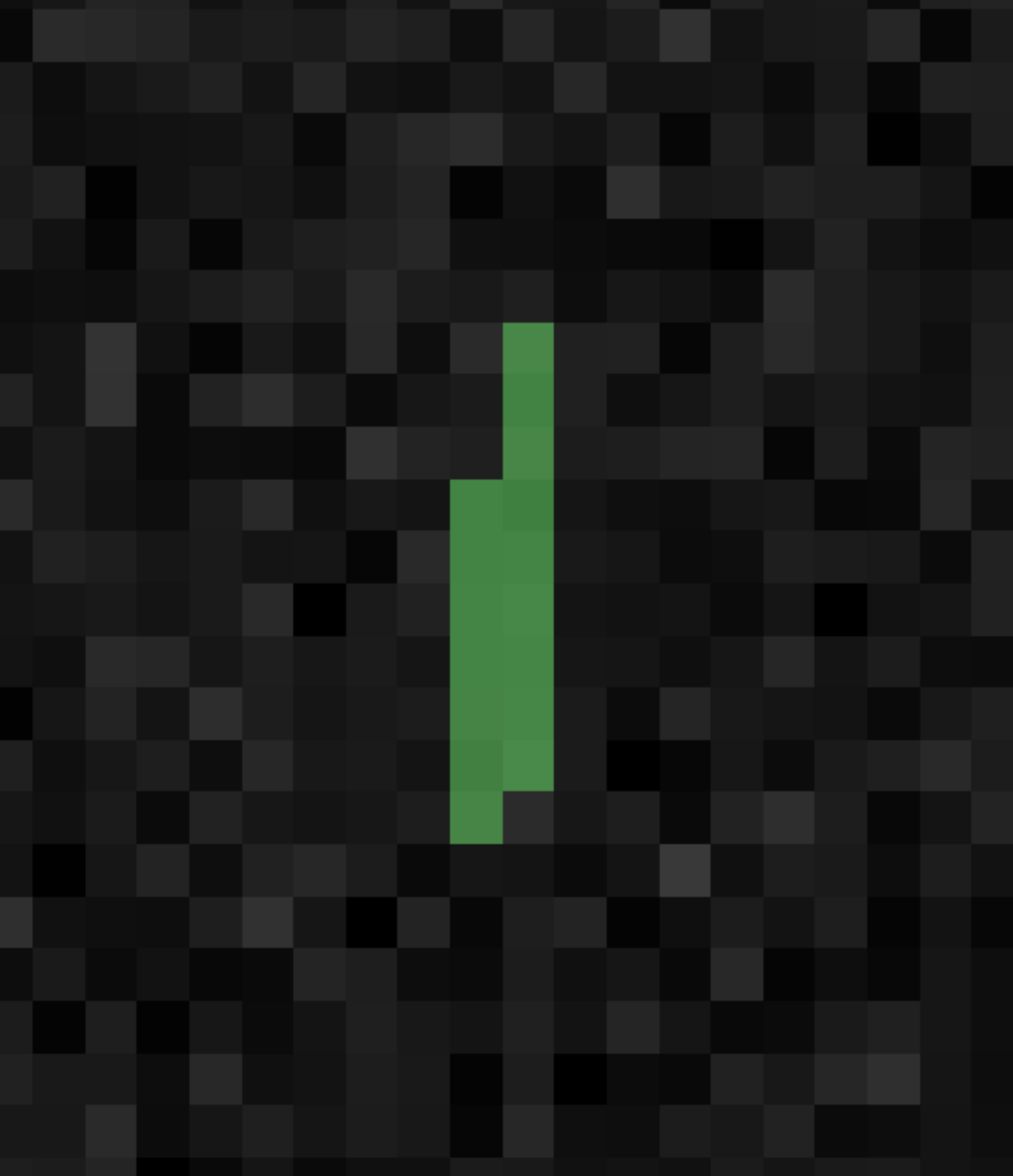}
    \includegraphics[width=0.158\textwidth]{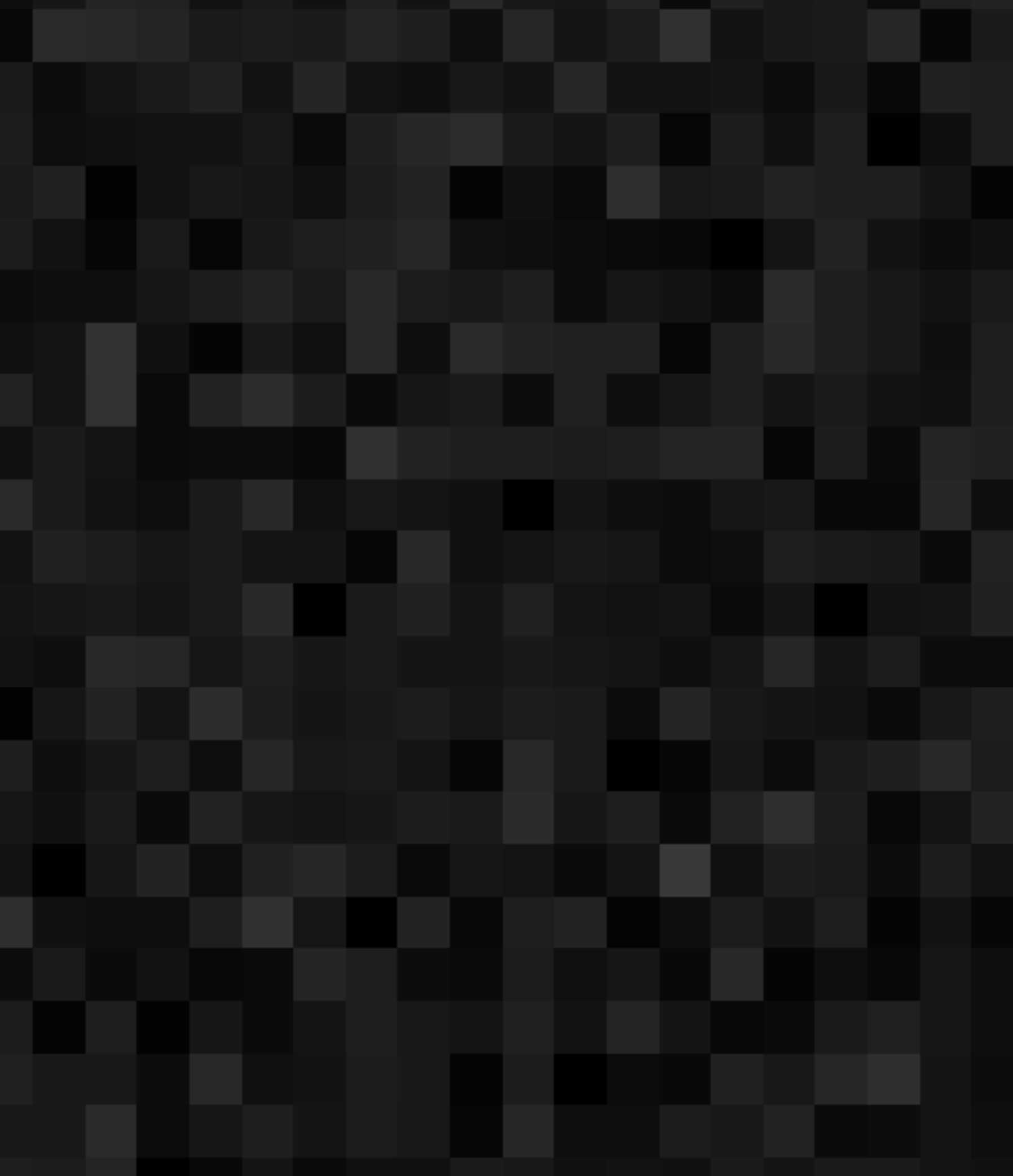}
    \caption{
        Detection and interpolation of electron (upper panels) and
        muon (lower panels) cosmic rays.  The left panels show the original images, the middle panels highlight in green the pixels identified as part of the cosmic rays, and the right panels show the results of interpolating these pixels using a Linear Predictive Code.
    }162
    \label{fig:cr-interpolation}
\end{figure}

The cosmic ray (CR) removal algorithm in HSC is based upon that used successfully in SDSS \citep{2001ASPC..238..269L}, although the CRs detected by the 200$\mu m$-thick CCDs employed by HSC \citep{hsc-instrument} have quite different morphologies.  The major sources of CRs in HSC are electrons (which have long worm-like tracks, often showing short side-branches; see the upper panel in \figref{cr-interpolation}) and muons, which are straight and wedge-like\footnote{The wider part of the track is generated at the top of the CCDs where diffusion of carriers within the silicon after photo-conversion is more important.} (lower panel in \figref{cr-interpolation}).

The first-pass algorithm is to search for all pixels which satisfy a series of conditions:

\begin{enumerate}

\item That the candidate bad pixel $p$ not be adjacent to a saturated pixel.

\item That $p$'s intensity $z$ exceed the locally-determined background (actually the mean of pairs of neighboring pixels) by $n\sigma$ where $\sigma^2$ is the sky variance.
We take $n = 6$.                    

\item That the gradients near the pixel exceed the band-limit imposed by the PSF; specifically we require that the pixel be part of a peak which is sharper than the center of a star centered in a pixel. Allowing for noise, this condition becomes
\begin{eqnarray}
z - c*N(z) > \hbox{PSF}(d) \left(\bar{z} + c N(\bar{z}))\right)
\label{CRcond3}
\end{eqnarray}
where $c$ is a constant, $N(z)$ is the standard deviation of $z$, $\hbox{PSF}(d)$ is the value of the PSF at a distance $d$ from the center of a star, and $\bar{z}$ is the average of two pixels a distance $d$ away from our pixel. We have found that in practice we must multiply $\hbox{PSF}(d)$ by some fiddle factor, $c_2 < 1$, to avoid flagging the cores of stars as `cosmic rays'.
We use $c = 2.5$ 
and $c_2 = 0.6$.      

\end{enumerate}

These conditions are applied sequentially to the pixel being studied using the four pairs of neighboring pixels (NS, EW, NW-SE, and NE-SW, $d = 1, 1, \sqrt2, \sqrt2$).  The candidate cosmic ray must exceed condition 2 for all four pairs of neighbors, and condition 3 for at least one pair. The thinking behind this choice is that while most cosmic rays contaminate more than one pixel, they pass through the CCD in a straight line so almost all pixels have at least one pair of uncontaminated neighbors.

Once a cosmic ray contaminated pixel is identified, its location is noted and its value is replaced by an interpolation based on the pair of pixels that triggered condition 3; the interpolation algorithm used is the same as that used for fixing sensor defects (\secref{bad-pixel-interpolation}). This removal of contaminated pixels as they are found makes it easier to find other pixels affected by the same cosmic ray hit.

Once the entire frame has been processed, the pixels identified individually as being contaminated by cosmic rays are assembled into cosmic ray `events' of contiguous pixels.  Each such event must contain more than a minimum number of electrons (not DN); we have adopted a threshold of 150$e^-$.                

We then go through the frame again, looking at pixels adjacent to these cosmic ray events.  Processing is identical, except that we set $c_2 = 0$ for these extra contaminated pixels.  The major difference from the algorithm used by the SDSS \textit{Photo} Pipeline is that HSC iterates, repeating this search for extra pixels 3 times. 
CR-contaminated pixels found during this process are, of course, removed before the next iteration.

\subsection{Bad Pixel Interpolation}
\label{sec:bad-pixel-interpolation}

Linear Predictive Codes \citep[LPC, \eg][]{Press:2007:NRE:1403886} are a special case of Gaussian Processes that estimate the value of a signal, in our case an image, based on other data points and a knowledge of the signal's statistical properties. Given a stationary process
\begin{equation}
y_i \equiv s_i + n_i
\end{equation}
where $s_i$ is the signal, and $n_i$ the noise, we may predict the value of the signal $s_p$ at some point $p$ as
\begin{equation}
\tilde{s}_p = \sum_j D_{p j} y_j;
\end{equation}
it is clear that the estimator is unbiased if $\sum_j D_{p j} = 1$.  It turns out that the coefficients $D_{pj}$ are determined by the autocorrelation function of the data and their signal-to-noise.

We can use LPC to interpolate over defects in images by interpreting $p$ as the index of a bad pixel. Assuming that the image consists purely of point sources, the autocorrelation is just the square of the PSF which we approximate as a Gaussian $N(0, \alpha^2)$, so the autocorrelation is $N(0,2\alpha^2)$.

In order to apply this to interpolation, we set the noise to be infinitely larger in the bad columns than the good. Considering the case of infinite signal-to-noise ratio and restricting ourselves to only $5$ terms (centered on the bad column) in estimating the $D|_{p=i}$, the resulting weights for a one-dimensional interpolation with $\alpha = 1$ are $\{-0.274, 0.774, 0.000, 0.774, -0.274\}$; the bad pixel has of course a weight of $0.000$.

Image mask bits are set to indicate both that a pixel was interpolated and the reason why (\eg~cosmic ray, sensor defect, saturation, etc.) pixels.  Generally only objects whose centers were interpolated (\texttt{flags\_pixel\_interpolated\_center}) should be considered to have unreliable measurements.

\subsection{Background Subtraction}
\label{sec:background-subtraction}

In background subtraction, we attempt to model and subtract a smooth, slowly-varying flux distribution from an image.  This flux distribution has contributions from the sky, diffuse optical ghosts from bright stars, scattered light, and sources below the detection limit.  In practice the subtracted background also includes light from astrophysical objects we would ideally like to leave in the image (or at least model and subtract explicitly), such as intra-cluster light in galaxy clusters, galactic cirrus, and the extended wings of the PSF around bright stars.  The fundamental challenge in background subtraction is to subtract smaller-scale patterns of scattered light while avoiding over-subtraction around bright astrophysical objects, which biases the photometry of both the bright object and its neighbors.

Our algorithm is a simple combination of averaging in spatial bins and model-fitting:
\begin{enumerate}
\item We average (unweighed mean, with iterative 3$\sigma$ clipping) pixels in 128$\times$128 bins, ignoring any pixels belonging to detected objects.  We also compute the variance in each bin.
\item We fit a 6th-order 2-d Chebyshev polynomial to the average pixel values over the image (typically a CCD or patch), using the average position of the non-masked pixels as the center of a bin.  We weigh each bin by its inverse variance in the fit, ensuring that heavily-masked bins cannot strongly affect the fit.
\end{enumerate}
The best-fit Chebyshev polynomial is then subtracted from the original image.  Because the polynomial is not required to exactly reproduce the value of any bin at its center, the order of the polynomial effectively sets the spatial scale (relative to the image size) of the features that are included in the background estimate.

Our code also supports using linear or spline interpolation to estimate the value of the background between bin centers.  When interpolation is used, the effective smoothing scale of the background estimate is determined entirely by the bin size; using 512$\times$512 bins yields behavior similar to the 6th-order Chebyshev fit on the 4k$\times$2k HSC CCDs.  Overall we find that Chebyshev fitting is better at avoiding over-subtraction than large-scale bins with spline interpolation, largely because it is less sensitive to the placement of bin boundaries relative to the positions of bright objects.

Because we rely on masking detected objects to prevent them from biasing the background estimate, but the detection threshold is determined relative to the background, background subtraction and source detection are not independent operations.  As described in Sections~\ref{sec:ccd-processing} and~\ref{sec:coadd-processing}, we subtract an initial background in CCD Processing before detecting any sources and then re-estimate and subtract the background every time we push our detection limit fainter (twice in CCD Processing, and once more in Coadd Processing).  The final round of background subtraction on coadds uses a single bin for the full 4k$\times$4k patch and simply subtracts a constant; we assume most spatially-varying features have been subtracted prior to coaddition, and simply scale the coadd to account for the fact that the average flux of new detections should no longer contribute to our definition of the background.

\subsection{Detection}
\label{sec:detection}

Sources are detected on both individual exposures and coadds by convolving the image with a smoothing filter and applying a simple threshold to the result.  We consider even a single pixel above threshold as a detection (unlike \eg~SExtractor; \citealt{1996A&AS..117..393B}).  When the smoothing filter is the transpose of the PSF, this approach is formally optimal (as we will show in \secref{detection-theory}) for isolated point sources in the limit where the noise is dominated by the sky and the background is known exactly.  This is at least approximately the case for the sources at the edge of our detection limit (which are mostly barely-resolved galaxies).

We use a circular Gaussian approximation matched to the RMS width of the PSF instead of the full PSF model for performance reasons.  Any decrease in sensitivity due to this approximation can easily be compensated by a small decrease in the threshold, and a cut on PSF flux S/N can then be used to select true detections at the original threshold.  In SSP processing we simply use the nominal threshold (generally $5\sigma$) directly; the loss in S/N is usually less than 1\% in the HSC-SSP Wide layer, and is always less than 3\%.

This approach is also in general non-optimal when applied naively to non-optimal coadds, as the HSC pipeline currently does -- the smoothing should instead be done separately for each image and those images combined to form a detection.  The corresponding loss in detection S/N is just the loss due to suboptimal coaddition reported in \secref{image-coaddition} ($1.7\% \pm 1.5\%$ in the HSC-SSP Wide).\footnote{A general-purpose optimal coadd can be constructed by smoothing the per-epoch images with their PSFs, averaging the images, and then deconvolving the coadded image to whiten the noise \citep[\eg][]{2015arXiv151206879Z}.  Because detection works on smoothed images, optimal coaddition for detection does not require this deconvolution but is otherwise mathematically equivalent.}

The output of our detection algorithm is a set of above-threshold regions we call \texttt{Footprints}.  In real data, object detections may of course overlap, so each \texttt{Footprint} also contains the list of peaks in the smoothed image that are found within it.  In general we assume each peak corresponds to a distinct astrophysical object that should be separated from its neighbors by the deblender (\secref{deblending}).  We also grow (dilate, in the language of mathematical morphology) the \texttt{Footprints} by the RMS width of the PSF after detection to make them better represent the region occupied by an object;  if a single smoothed pixel above threshold corresponds to a detected point source, we know that source covers approximately the area of the PSF.  If growing two or more \texttt{Footprints} causes them to overlap, they are merged (and their peak lists concatenated), so the \texttt{Footprints} in the final set are non-overlapping and individually simply-connected.

\subsubsection{Theory}
\label{sec:detection-theory}

Given an image $z(\bm{r})$, locally constant per-pixel Gaussian noise with RMS $\sigma$, and a locally constant flux-normalized PSF $\phi$, the log likelihood of a point source with flux $\alpha$ at position $\bm{\mu}$ is given by
\begin{eqnarray}
  L(\alpha,\bm{\mu}) &=& \ln P(z|\alpha, \bm{\mu}) \nonumber\\
    &\propto& -\frac{1}{2\sigma^2}
        \sum\limits_{i} \left[z(\bm{r}_i) - \alpha\,\phi(\bm{r}_i - \bm{\mu})\right]^2
\end{eqnarray}
where the index $i$ runs over pixels.

This represents a Gaussian (albeit an unnormalized one) probability in $\alpha$, which we can emphasize by expanding $L$ as
\begin{eqnarray}
  L(\alpha,\bm{\mu}) = L(\hat{\alpha}(\bm{\mu}),\bm{\mu}) +
    \frac{\partial^2 L}{\partial \alpha^2}(\alpha - \hat{\alpha}(\bm{\mu}))^2
\end{eqnarray}
with $\hat{\alpha}(\bm{\mu})$ defined to null the first-order term:
\begin{eqnarray}
  \hat{\alpha}(\bm{\mu}) = \frac{
    \sum\limits_i z(\bm{r}_i)\phi(\bm{r}_i - \bm{\mu})
  }{
    \sum\limits_j \left[\phi(\bm{\mu} - \bm{r}_j)\right]^2
  }
  = \frac{1}{A} \sum\limits_i z(\bm{r}_i)\phi(\bm{r}_i - \bm{\mu})
\end{eqnarray}
where $A$ is the effective area of the PSF:
\begin{eqnarray}
    A \equiv \sum\limits_j \left[\phi(\bm{\mu} - \bm{r}_j)\right]^2 \,.
\end{eqnarray}
Like the PSF, $A$ can be considered spatially constant here even though it may vary on scales much larger than a source.  Ignoring the term $L(\hat{\alpha}(\bm{\mu}),\bm{\mu})$ because it does not depend on $\alpha$, we can now read off the maximum likelihood flux as $\hat{\alpha}(\bm{\mu})$ and its variance as
\begin{eqnarray}
  \sigma_\alpha^2(\bm{\mu}) = \left[
    \frac{\partial^2 L}{\partial \alpha^2}
  \right]^{-1}
  = \frac{\sigma^2}{A}
\end{eqnarray}
which yields a signal-to-noise ratio of
\begin{eqnarray}
  \nu(\bm{\mu}) = \frac{\hat{\alpha}(\bm{\mu})}{\sigma_\alpha(\bm{\mu})}
  = \frac{1}{\sigma\sqrt{A}}\sum\limits_i z(\bm{r}_i)\phi(\bm{r}_i - \bm{\mu})
  .
\end{eqnarray}
We can thus form an image of $\nu$ by convolving the original image $z$ with the transpose of its PSF, divide by $\sigma\sqrt{A}$, and threshold at our desired significance.

To optimally detect extended sources, we could repeat the above derivation after replacing $\phi$ with a PSF-convolved model of the target source (or a sequence of such models designed to approximately cover the range of possible morphologies).  We do not currently include an additional detection filter for extended objects in the HSC pipeline; a very preliminary experiment with a larger filter had a small effect on the number of detections.  This possibility will be investigated more carefully in the future.

\subsubsection{Temporary Background Over-Subtraction}
\label{sec:temporary-background-oversubtraction}

For most groups of objects, the fact that our detection algorithm is derived for isolated objects is not a serious problem.  Flux from one object will artificially increase the detection significance of neighboring pixels, but this has no effect at all on the detectability of neighboring objects that would have already been above the detection threshold.  It can also lead to spurious detections due to artificially elevated noise peaks, however, and these are extremely common and problematic around the brightest stars and galaxies.

The addition of spurious objects to the catalog is not in itself a serious issue; it is relatively easy to filter out the vast majority of these spurious detections via cuts on S/N and our \textit{blendedness} parameter (\secref{blendedness-metrics}).  The problem is how they affect their real neighbors: these peaks always ``steal'' some flux from their neighbors in the deblending process, biasing their photometry low, and the large number of peaks near bright objects greatly increases the chance of more catastrophic deblender failures due to linear peak alignments (see \secref{deblending}).  It is thus imperative that we remove as many spurious peaks as possible around bright objects prior to running the deblender.

One step to remove spurious peaks has already been discussed in \secref{coadd-processing}: when merging \texttt{Footprints} across bands, we cull the faintest peaks from large blends when they are detected in only one band.  Another way to mitigate the problem is actually to do \emph{worse} (in a sense) at subtracting the background: if the spatial scale of background estimation is too small, we will subtract astrophysical flux from the wings of bright objects along with the sky background, which ensures that flux does not artificially increase the detection significance.  Doing so biases the photometry (and other measurements) of the bright objects, of course.  It also inevitably leads to \emph{over-subtraction} of the sky around bright objects, which can in turn bias the measurements of fainter objects in the over-subtracted region.

To obtain most of the benefits of aggressive background subtraction while avoiding most of the drawbacks, we perform a \emph{temporary} small-scale background subtraction just before detecting peaks (but after detecting \texttt{Footprints}); we add the temporary background back into the image after the detection step is complete.  We typically use a bin size of 64$\times$64 pixels with spline interpolation in the temporary background, as compared to a bin size of 128$\times$128 with additional polynonmial smoothing in the permanent background subtraction (see \secref{background-subtraction}).  This is not a particularly good model for the mostly circular (or at least elliptical) flux profiles of bright objects, of course, but it is sufficiently flexible to subtract off most of the flux in the outskirts of the brightest objects, and it is easy and fast to fit.

Unfortunately, due to a mistake in the configuration files, temporary local background estimation was not enabled in PDR1 or the internal releases it was derived from.  It will be enabled in all future data releases.

\subsection{Deblending}
\label{sec:deblending}

\subsubsection{Overview}
\label{sec:deblending-overview}

In a deep optical imaging survey such as the HSC-SSP, a large fraction of objects are \emph{blended} with their neighbors: they have overlapping isophotes at the surface brightness threshold used for detection.  In the Wide layer of the SSP, 58\% of all objects are members of blends; this increases to 66\% and 74\% in the Deep and UltraDeep layers, respectively (and the latter has not yet been observed to full depth).  Our detection algorithm (\secref{detection}) represents a blend as a \texttt{Footprint} containing multiple peaks.   The job of our \emph{deblender} algorithm is to apportion the flux in that \texttt{Footprint} to the peaks, creating a ``child image'' for each peak that allows it to be measured as a distinct source.  While this would ideally be done in a fully consistent way across all filters, at present our deblending algorithm processes each filter independently, with cross-band consistency enforced only by a consistent set of input peaks and \texttt{Footprints} (as described in \secref{coadd-processing}).

We store the child images in data structures we call \texttt{HeavyFootprints}.  These represent an image with irregular boundaries by combining a \texttt{Footprint} that describes the boundary with a 1-d array of child image pixel values, concatenated row-by-row.  These are used by the procedure described in \secref{neighbor-replacement} to construct a larger clean measurement image for each child object.


The hard boundaries we have defined here between detection, deblending, and measurement are themselves an algorithmic choice.  In crowded stellar fields this separation of concerns would be a poor choice, as deblending and measurement are essentially the same operation (fitting multiple point source models simultaneously) and detection must be iterated with subtracting the stars that were detected and modeled in previous steps \citep[\eg][]{1987PASP...99..191S}.  In uncrowded fields dominated by galaxies it proved very effective as part of the SDSS \textit{Photo} Pipeline \citep{2001ASPC..238..269L}, however, and we have adopted both this general approach and most of the details of the SDSS algorithm for the first generation of the LSST/HSC deblender.  As we will discuss in \secref{deblending-results}, the blending problem in HSC-SSP data is much more difficult than it was in SDSS, and it is already clear the SDSS algorithm is not adequate at typical HSC-SSP depths.  It is not yet clear whether the overall approach of separating detection, deblending, and measurement is itself a limiting factor.

\begin{figure*}
    \includegraphics[width=0.16\textwidth]{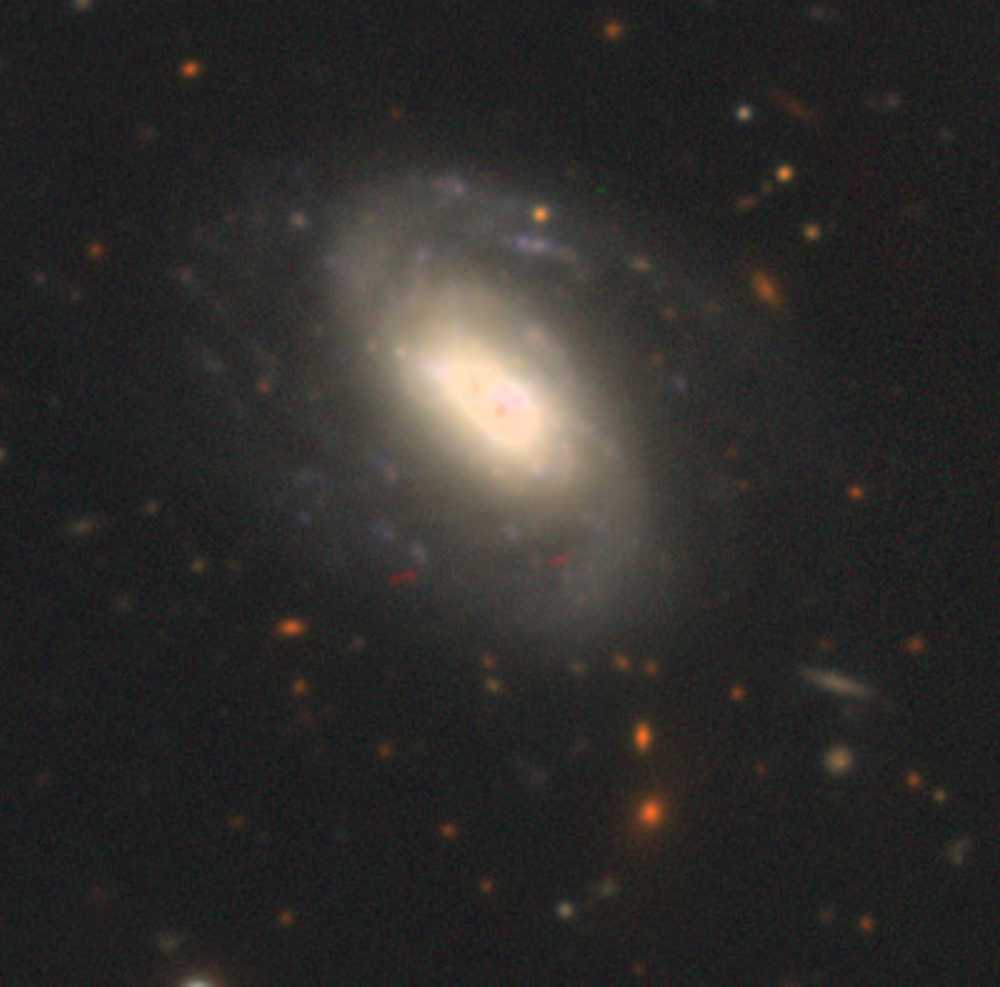}
    \includegraphics[width=0.16\textwidth]{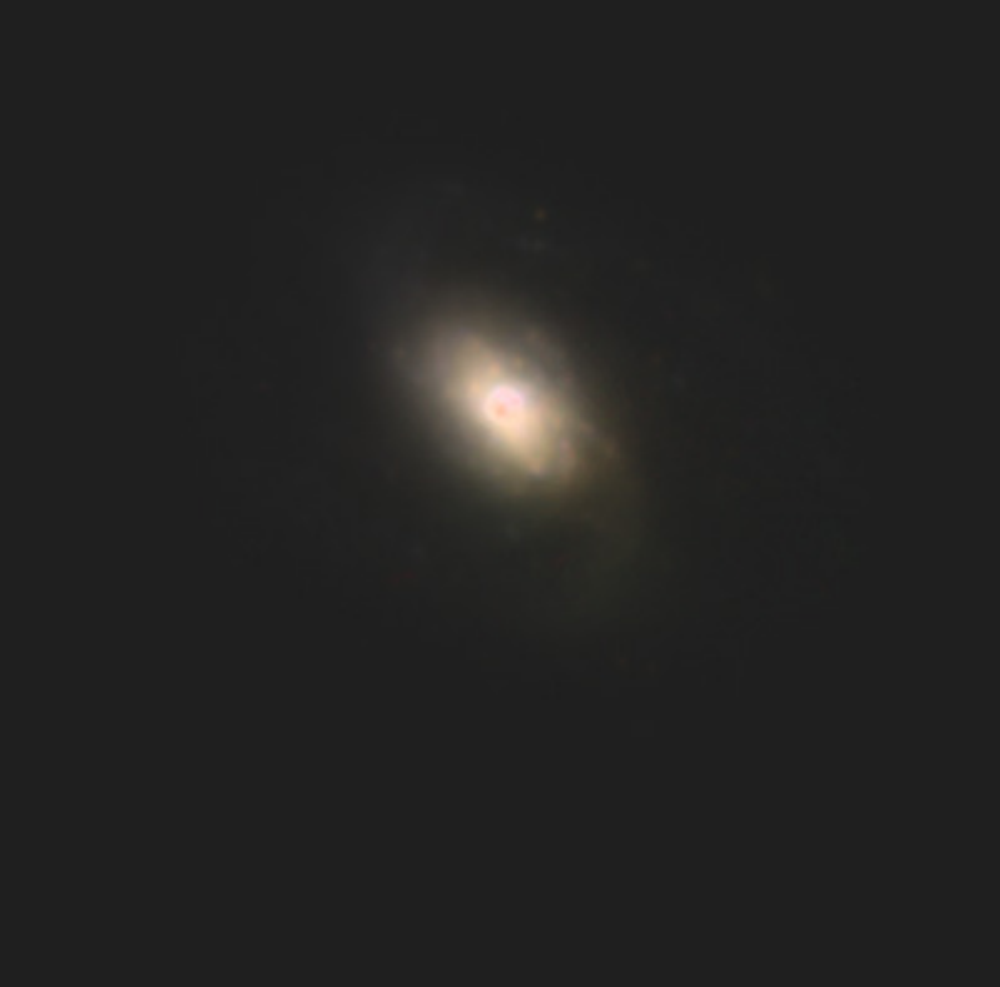}
    \includegraphics[width=0.16\textwidth]{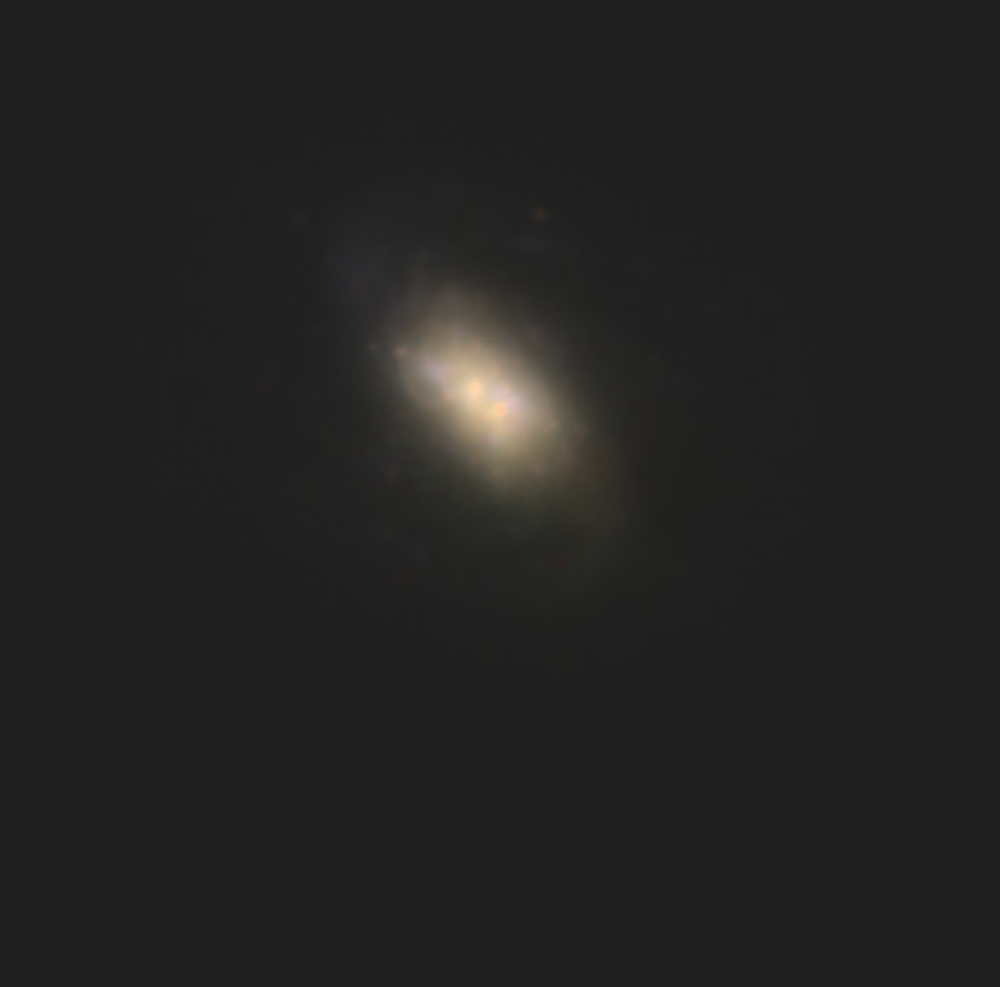}
    \includegraphics[width=0.16\textwidth]{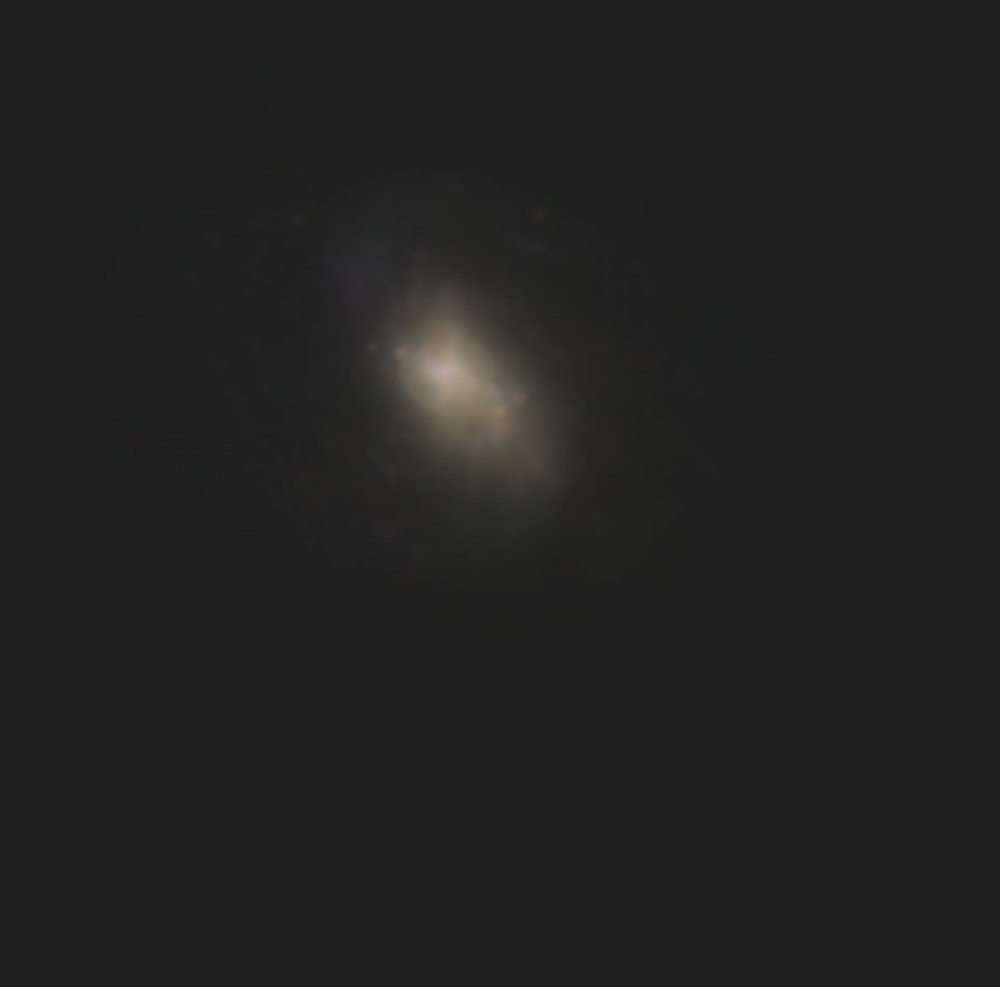}
    \includegraphics[width=0.16\textwidth]{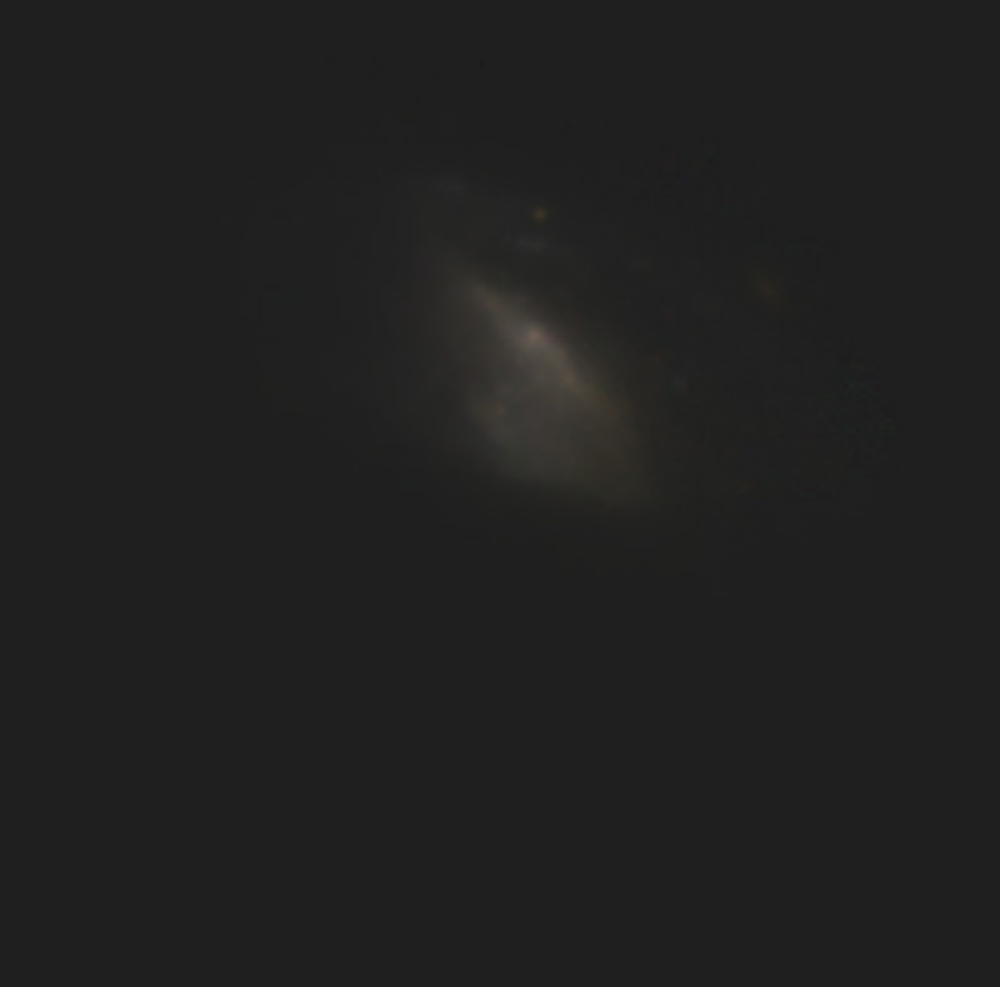}
    \includegraphics[width=0.16\textwidth]{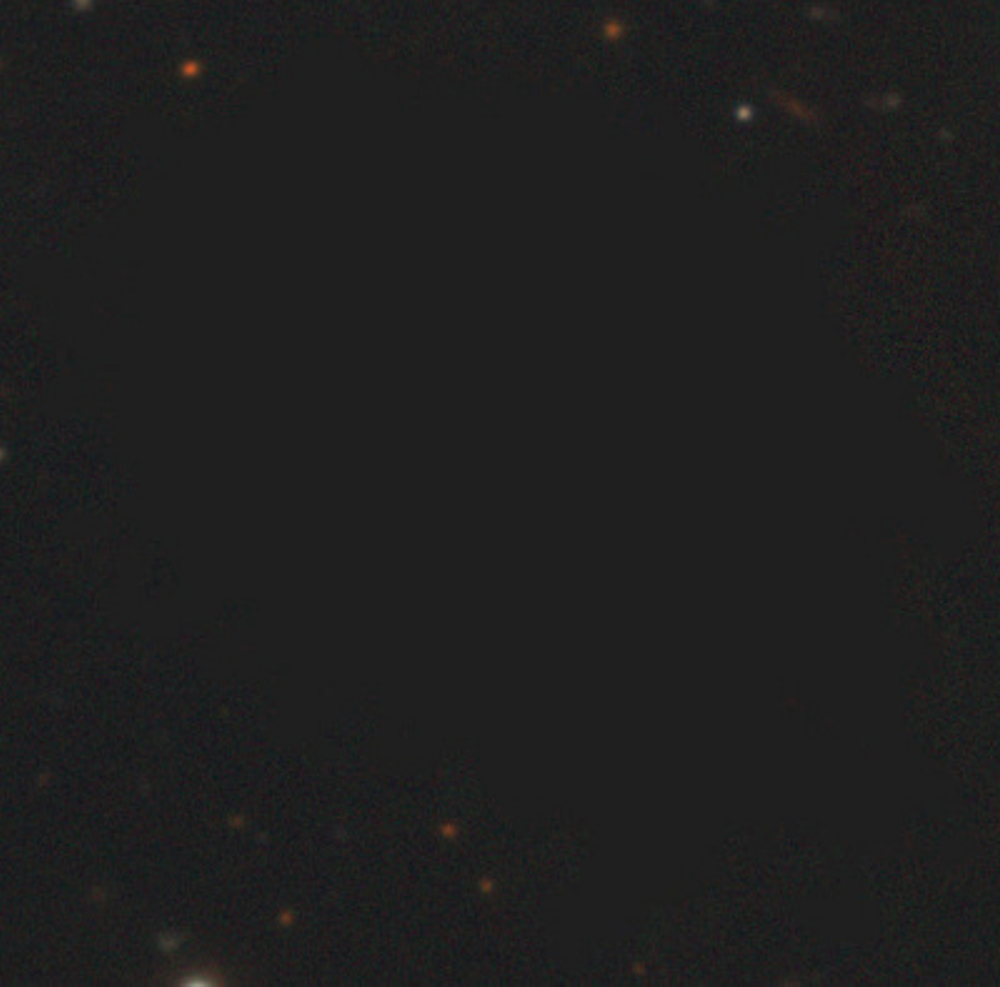}
    \caption{
        A well-resolved spiral galaxy shredded by the deblender.  From left to right: the original image, the four brightest children, and the residual after subtracting these.  Many fainter children (most of them similarly spurious) are not shown and are not subtracted in the residual image. All images are RGB=$zir$ composites.
    }
    \label{fig:deblend-shredded}
\end{figure*}

\begin{figure*}
    \includegraphics[width=0.16\textwidth]{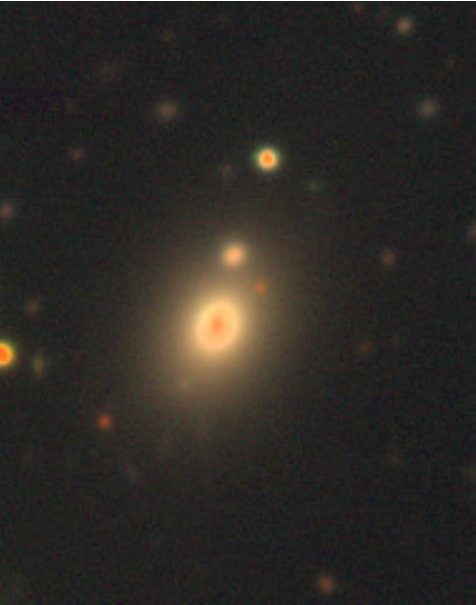}
    \includegraphics[width=0.16\textwidth]{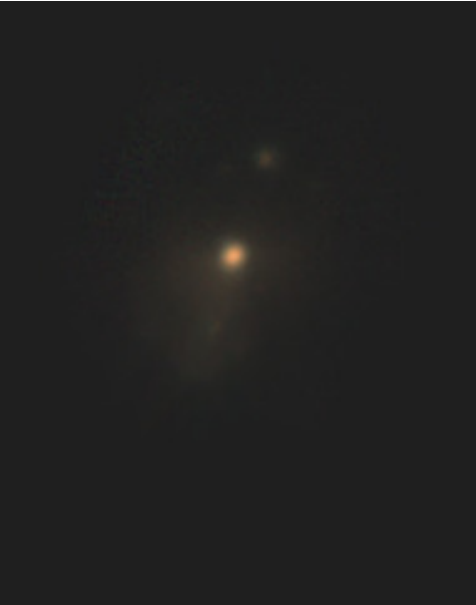}
    \includegraphics[width=0.16\textwidth]{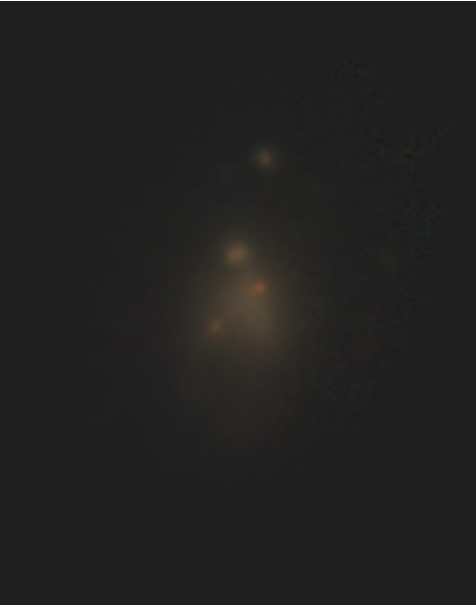}
    \includegraphics[width=0.16\textwidth]{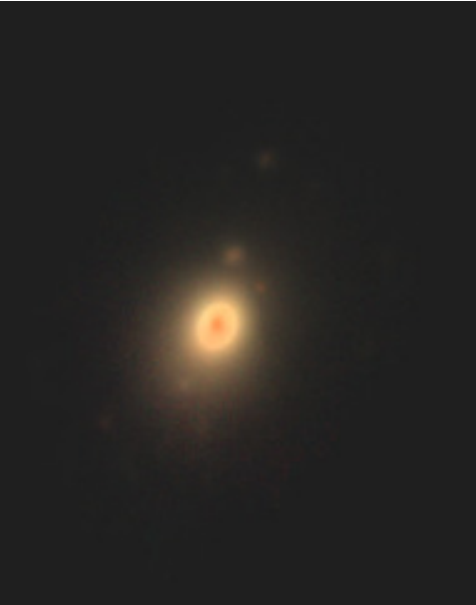}
    \includegraphics[width=0.16\textwidth]{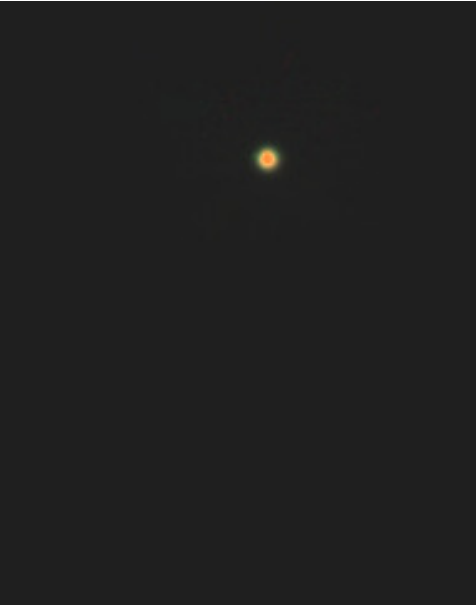}
    \includegraphics[width=0.16\textwidth]{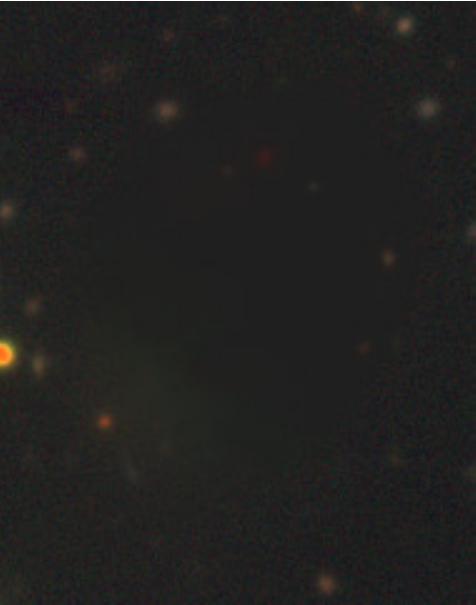}
    \caption{
        An unsuccessful deblend exhibiting the common three-in-a-row failure mode.  From left to right: the original image, four of the brighter deblended children, and the residual after subtracting these four.  The light from the largest galaxy is split up across several of the deblended children due to the linear alignment of the peaks.  All images are RGB=$zir$ composites.
    }
    \label{fig:deblend-triple}
\end{figure*}

\subsubsection{Algorithm}
\label{sec:deblending-algorithm}

The fundamental ansatz in the SDSS and HSC/LSST deblenders is that objects have approximate 180-degree rotational symmetry about their peak flux position: when some part of an object's morphology is confused with that of a neighbor, we assume we can recover it by looking on the opposite side of the peak.  More formally, we define a \emph{template} image $T_i(\bm{r})$ for peak $i$ at position $\bm{p}_i$ from the image $z(\bm{r})$ as
\begin{eqnarray}
    T_i(\bm{r}) = \min\!\left(
        z(\bm{r}),
        z(2\bm{p}_i - \bm{r})
    \right)\,.
    \label{eqn:deblender-templates}
\end{eqnarray}
Note that $2\bm{p}_i - \bm{r}$ is just the reflection of $\bm{r}$ to the opposite side of $\bm{p}_i$.  The use of the minimum reflects the fact that for objects with the assumed symmetry (and in the absence of noise) the lower of the two values is the one least affected by flux from neighbors, which is always nonnegative.

We then fit a linear combination of the templates to all pixels in the blend (assuming uniform variance), yielding a best-fit scaling parameter $\alpha_i$ for each:
\begin{eqnarray}
    \alpha_i \longrightarrow \min\limits_{\alpha_i}\!\left[
        \sum_{\bm{r}} \left(z(\bm{r}) - \sum_i \alpha_i T_i(\bm{r})\right)^2
    \right]\,.
\end{eqnarray}
It is tempting to use the scaled template $\alpha_i T_i(\bm{r})$ directly as a deblended image for peak $i$, but we can construct a better one from the relative contributions of the scaled templates to each pixel.  Our final deblended images $C_i(\bm{r})$ are defined as
\begin{eqnarray}
    C_i(\bm{r}) = \frac{\alpha_i T_i(\bm{r})}{\sum\limits_j \alpha_j T_j(\bm{r})}
        z(\bm{r})\,.
    \label{eqn:final-deblend}
\end{eqnarray}
This final step yields two important improvements relative to the scaled templates:
\begin{itemize}
    \item it exactly conserves the flux in each pixel, and by extension, the whole blend, because the sum over all $C_i$ is $z$ by construction;
    \item it allows objects that are not symmetric to retain their original morphology where it is not confused with a neighbor, because the ratio in \eqnref{final-deblend} approaches one when no neighbors contribute.
\end{itemize}

In both the SDSS and HSC/LSST deblenders, templates that closely resemble the PSF model are replaced with the PSF model before fitting for the scale factors $\alpha_i$.  This eliminates unnecessary degrees of freedom, making the final result more robust to noise and crowding.

The SDSS deblender additionally included the ability to drop templates that appeared to be too similar (\ie~their normalized dot product was too close to one).  This was important for avoiding ``shredding'', in which a large, morphologically complex galaxy is split into many spurious children (see, \eg~\figref{deblend-shredded}).  This is impossible in the current HSC processing flow, however, because the deblender is run independently in every band (see \secref{coadd-processing}) and we have no way to guarantee that the peaks removed would be consistent across bands.  Small-scale tests of this feature on HSC data suggest that it may not be as helpful for HSC data processing as it was for SDSS, overall -- while it does reduce reduce shredding in large galaxies, it also merges groups of objects that appear to be distinct.  A comparison of HSC and SDSS measurements of the same galaxies in \citet{2017arXiv170208449A} provides an estimate of the magnitude of the problem; about 15\% of galaxies with $i < 19$ appear to be shredded in HSC processing (though this analysis depends on the assumption that the SDSS deblending and measurements are correct).

Every blend processed by the deblender begins as a single parent record in our catalog with a \texttt{Footprint} with multiple peaks (deblending is a no-op for sources whose \texttt{Footprint} contain only one peak).  For each deblended peak, it adds a new child record to the catalog and attaches a \texttt{HeavyFootprint} containing the deblended pixel values $C_i(\bm{r})$.  We set two fields on all records to describe these relationships:
\begin{description}
\item[\texttt{parent}:] The ID of the record that this source was deblended from, or zero if it was not blended (either because it \emph{is} the original blend or was never blended).
\item[\texttt{deblend\_nChild}:] The number of children created by deblending this object (zero for records that represent child objects).
\end{description}
These can be used to identify two particularly useful categories of object:
\begin{itemize}
\item Isolated objects that were never blended have both \texttt{parent}=0 and \texttt{deblend\_nChild}=0.  This criterion can be used to identify a very secure (albeit small and biased) sample of objects whose measurements are definitely not affected by blending.
\item Unblended objects that were isolated or have been deblended have \texttt{deblend\_nChild}=0.  These objects can be used for most science analyses when other cuts (on \eg~blendedness; see \secref{blendedness-metrics}) are used to remove sources with especially poor deblending.  
\end{itemize}

The \texttt{detect\_is-primary} flag is set for objects that have \texttt{deblend\_nChild=0} as well as both \texttt{detect\_is-tract-inner} and \texttt{detect\_is-patch-inner} (\secref{coadd-processing}), and hence can be used to select a sample of objects with no duplicate processing.

\subsubsection{Results}
\label{sec:deblending-results}

The single biggest failure mode of the deblender occurs when three or more peaks in a blend appear in a straight line, as in \figref{deblend-triple}.  When building the template for the middle peak, the $\min$ in \eqnref{deblender-templates} will have to choose between pixels that are both affected by neighbors, yielding a poor template.  In SDSS, this alignment was sufficiently rare that the overall performance of the deblender was satisfactory.  In HSC data, this problem is dramatically more prevalent, simply because blends are both more common and more complex at HSC depths.

\begin{figure}
    \includegraphics[width=0.23\textwidth]{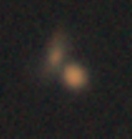}
    \includegraphics[width=0.23\textwidth]{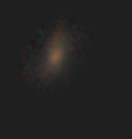}\\
    \includegraphics[width=0.23\textwidth]{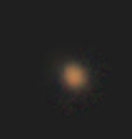}
    \includegraphics[width=0.23\textwidth]{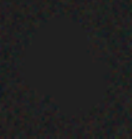}
    \caption{
        A successful deblend of a close pair of galaxies in the SSP Wide survey.  Upper left shows the original image, upper right and lower left show deblended child images, and lower right shows the original image with both child images subtracted.  All images are RGB=$zir$ composites.
    }
    \label{fig:deblend-easy}
\end{figure}

\begin{figure*}
    \includegraphics[width=0.191\textwidth]{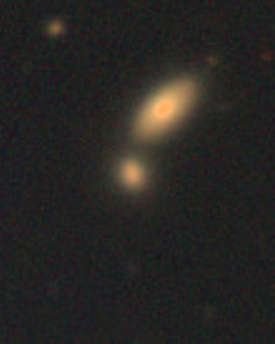}
    \includegraphics[width=0.191\textwidth]{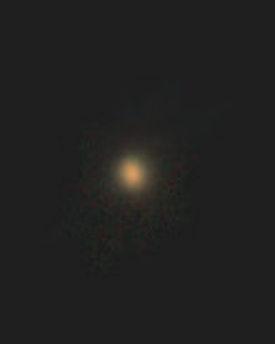}
    \includegraphics[width=0.191\textwidth]{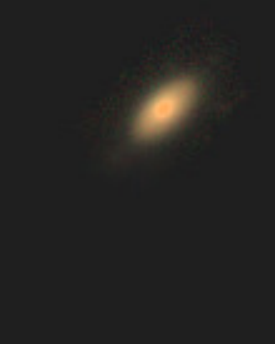}
    \includegraphics[width=0.191\textwidth]{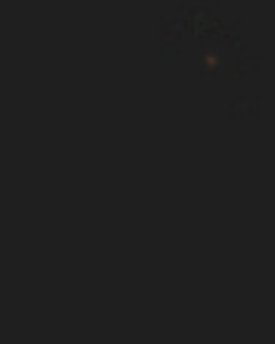}
    \includegraphics[width=0.191\textwidth]{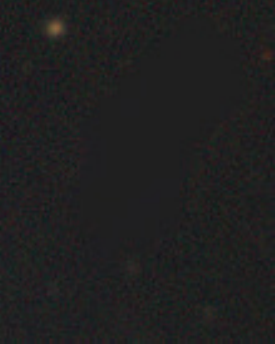}
    \caption{
        A largely successful deblend of a three-object family.  From left to right: the original image, the three deblended children, and the residual after subtracting all children.  The largest object (middle image) appears to steal a tiny small fraction of the flux from its faintest neighbor (second from the right).
    }
    \label{fig:deblend-success}
\end{figure*}

Overall, the HSC deblender performs adequately in most areas of the survey, where blending is common but not severe and most galaxies in blends are small and not obviously morphologically complex (\eg~Figures~\ref{fig:deblend-easy} and~\ref{fig:deblend-success}).  But it fails dramatically in the cores of most galaxy clusters and in the neighborhood of nearby galaxies or bright stars.  These constitute a significant fraction of our catalog, and they (especially in the case of galaxy clusters) are of particular importance to major HSC science goals.

While a new deblender is currently being developed for future releases, at present we provide several ways to work around the current deblender's problems.  Most of these are discussed elsewhere in the paper:
\begin{itemize}
\item We eliminate some less significant peaks before deblending, both via temporary local background over-subtraction in detection (\secref{temporary-background-oversubtraction}) and when merging peaks across bands (\secref{coadd-processing}).
\item We provide a \emph{blendedness} metric (\secref{blendedness-metrics}) that can be used to identify objects in the kind of severe blends the deblender does not handle reliably.
\item We run variants of the aperture photometry (for small apertures only) and Kron photometry measurements at the positions of the children using the undeblended pixels (see \secref{neighbor-replacement}).
\end{itemize}

We refer the reader to \citet{hsc-synpipe-2} for a detailed investigation of the impact of blending (and deblending) on the measurements produced by the HSC Pipeline.

\subsection{Source Measurement}
\label{sec:source-measurement}

\textit{Source measurement} here refers to the suite of algorithms that characterize the properties of individual astronomical sources.  Most of these algorithms are run in both CCD Processing (\secref{ccd-processing}) and Coadd Processing (\secref{coadd-processing}).  Our source measurement system is highly extensible; it iterates over all sources in an image and runs a configurable suite of measurement \textit{plug-ins} on each.  These plug-ins can be implemented outside the main suite of pipeline packages, allowing pipeline users to easily add new measurement algorithms to the system.  We do not attempt to classify objects prior to measurement; while some algorithms may only be appropriate for certain types of sources, we run the same measurement algorithms on all sources.

Source measurement algorithms that perform photometry can also be run in \textit{forced} mode, which fixes the position and shape parameters at values measured previously (generally on another image).  This is used in the last stage of Coadd Processing to measure consistent fluxes across all bands, with positions and shapes defined in one band (see \secref{coadd-processing}).

Most of the HSC Pipeline's algorithms measure at least one of three quantities: centroids, shapes, and fluxes.  While there is no requirement that a plug-in measure any of these, those that do can be configured to fulfill a measurement \textit{slot} -- a label that indicates a recommended version of a particular measurement, allowing it to be used by other plug-ins as an input.  We have a single centroid slot and a single shape slot, but several flux slots, reflecting the fact that different kinds of photometry are not expected to yield equivalent results.

For comparisons between different photometry algorithms and characterization of the HSC Pipeline's overall photometric performance, we refer the reader to \citet{hsc-synpipe-1} and \citet{2017arXiv170208449A}.

The details of the measurement system are described in the following subsections.  We begin by describing two aspects of the system common to all measurement algorithms: how we use deblender outputs (\secref{neighbor-replacement}) and how we measure and apply aperture corrections (\secref{aperture-corrections}).  Centroid algorithms are discussed in \secref{centroids}, shape algorithms in \secref{shapes}, and fluxes in Sections~\ref{sec:psf-photometry}-\ref{sec:cmodel-photometry}.  Two additional algorithms for classifying sources and estimating the impact of blending on an object are discussed in Sections~\ref{sec:star-galaxy-classification} and~\ref{sec:blendedness-metrics}, respectively.

\subsubsection{Neighbor Replacement}
\label{sec:neighbor-replacement}

While the deblender discussed in \secref{deblending} is responsible for apportioning flux between neighboring sources in the pixels where they overlap, we do not run source measurement plug-ins directly on these deblended pixel values.  The pixels processed by the deblender are only those included in the original detection \texttt{Footprint}, but many measurement plug-ins require pixels beyond those regions.  Some plug-ins may in fact utilize pixels belonging to \texttt{Footprint}s other than that of the source being measured.  To generate an image that permits deblended measurement beyond a source's \texttt{Footprint}, we use the following procedure:
\begin{enumerate}
\item We replace \emph{all} \texttt{Footprint}s in the image with random Gaussian noise with the same variance as the original noise in those pixels.
\item We insert the deblended pixels for a particular source to be measured back into the image (replacing the noise pixels).
\item We run all measurement plug-ins on the current source.
\item We re-replace the \texttt{Footprint} of the current source with noise.
\item We repeat steps 2-4 for all sources.
\end{enumerate}
The obvious flaw in this procedure is that the flux in pixels between \texttt{Footprint}s is counted multiple times, as it is ``assigned'' to each of the sources in its neighborhood; while this flux is by definition below the detection threshold, it cannot be considered to be purely background.  However, nearly all practical source measurement algorithms (and all the ones we use) down-weight pixels far from the center of the source, so this flaw is not nearly as damaging as it may appear at first glance.  Moreover, counting this flux multiple times is unquestionably better than not counting it at all in the limit where sources are far apart.  Finally, if multiply-counted flux beyond \texttt{Footprint} boundaries does prove a problem, a simple solution exists: we can simply grow the size of the \texttt{Footprint}s (either by a fixed radius or by extending to a lower surface brightness threshold) before running the deblender.

As a partial hedge against bad deblending (galaxy ``shredding'' in particular), we run the suite of measurement plug-ins on both child and parent sources and include both in the resulting catalog.  Child measurements use the deblender outputs to provide the per-source deblended pixel values inserted into the image in the procedure described above.  Parent measurements instead interpret all pixels within a contiguous above-threshold region as belonging to a single source.  As no deblending is needed in this case, the per-source pixel values are simply the original pixel values of the image.  We encourage most science analysis on HSC pipeline results to use only child sources, but the parent measurements may be better for extremely bright sources whose neighbors are sufficiently faint that they can simply be ignored.

We also run variants of the Kron photometry (\secref{kron-photometry}) and fixed aperture photometry (\secref{fixed-aperture-photometry}) algorithms for each child object on PSF-matched images that have not been deblended, to further guard against deblending failures.  This is described in \secref{afterburner}.

\subsubsection{Aperture Corrections}
\label{sec:aperture-corrections}

Most consumers of photometry measurements assume that they represent the total flux of each source.  Many practical photometry algorithms directly measure only a fraction of the flux, with most attempting to measure the same fraction for all sources.  When that fraction is indeed constant, the conversion to total magnitudes is implicit in the photometric calibration: the magnitude zero-point maps our (fractional) flux measurements to the (total) magnitudes in the photometric reference catalog.

When applied to stars and other point sources, the fraction of flux measured directly by most photometry algorithms depends on the PSF.  While some algorithms utilize the PSF model to limit this dependency, our PSF model only extends to $\sim\!10\sigma$, where $\sigma$ is the second-moment radius of the core of the PSF (see \secref{shapes}).  While this is sufficient for centroid, shape, and morphology measurements, it ignores a non-negligible fraction of flux at large radius.

To enforce consistent flux fractions, we measure and apply aperture corrections, using the following procedure:
\begin{enumerate}
\item We run all photometry algorithms on a sample of securely classified, moderately bright unsaturated stars (typically the same stars used for PSF modeling, as described in \secref{psf-modeling}).
\item For each star, we compute the ratios of each flux algorithm measurement to the measurement used for photometric calibration (by default, a 4\arcsec\ diameter circular aperture flux).
\item We interpolate these ratios across each CCD using 2nd-order Chebyshev polynomials.  This suite of interpolated ratios is what we actually call the aperture corrections.
\item We apply the aperture corrections to per-exposure source measurements by multiplying the fluxes measured by each algorithm with the aperture correction for that algorithm.
\item We generate aperture corrections on coadds at the position of each coadd detection by averaging the aperture corrections of the coadd's constituent CCD images at that position, using the same weights used to build the coadd.  The coadd aperture corrections are then applied to coadd flux measurements.
\end{enumerate}
The value of the aperture correction is typically about 0.97 (\ie~it represents a 3\% change in the flux) for PSF photometry (\secref{psf-photometry}) in the HSC-SSP Wide survey.  CModel (\secref{cmodel-photometry}) aperture corrections are approximately the same, while those for Kron photometry (\secref{kron-photometry}) are typically about twice as large (1.06), in the opposite direction.

Unlike PSF coaddition (\secref{psf-coaddition}), aperture correction coaddition via a simple weighted sum is not exact.  Consider two photometry algorithms $A$ and $B$, with measurements on the $i$th epoch $f_i^A$ and $f_i^B$.  The measurements we would make on the coadd generated by summing images with weights $w_i$ are then
\begin{eqnarray}
  f_{\mathrm{coadd}}^A = \sum_i w_i f_{i}^A \\
  f_{\mathrm{coadd}}^B = \sum_i w_i f_{i}^B \,.
\end{eqnarray}
If we use $B$ for photometric calibration, the per-exposure aperture correction for $A$ is
\begin{eqnarray}
  r_i = \frac{f_{i}^B}{f_{i}^A}\,.
\end{eqnarray}
Similarly, the true coadd aperture correction $r_\mathrm{coadd}$ is defined as the ratio of the coadd measurements:
\begin{eqnarray}
    r_{\mathrm{coadd}} = \frac{f_{\mathrm{coadd}}^B}{f_{\mathrm{coadd}}^A}\,.
\end{eqnarray}
We can now write the denominator in terms of the per-epoch measurements of $B$ and the per-epoch aperture corrections $r_i$:
\begin{eqnarray}
    r_{\mathrm{coadd}} = \frac{f_{\mathrm{coadd}}^B}{\sum\limits_i w_i f_{i}^A}
     = \frac{f_{\mathrm{coadd}}^B}{\sum\limits_i w_i \frac{f_{i}^B}{r_i}}
    \,.
\end{eqnarray}
But calibrating to $B$ also implies that (in the absence of noise), we measure the same value for $B$ on all images: $f_i^B = f_j^B = f_\mathrm{coadd}^B$ for all $i,j$.  That implies
\begin{eqnarray}
    r_{\mathrm{coadd}} =
        \frac{f_{\mathrm{coadd}}^B}{\sum\limits_i w_i \frac{f_{\mathrm{coadd}}^B}{r_i}}
       = \left( \sum_i \frac{w_i}{r_i} \right)^{-1}
    \,.
    \label{eqn:coadd-apcorr-exact}
\end{eqnarray}
The simple weighted sum of per-epoch actually used by the pipeline is the first-order Taylor expansion of this about $r_i = 1$:
\begin{eqnarray}
    r_{\mathrm{coadd}} \approx \sum_i w_i r_i
    \label{eqn:coadd-apcorr-approx}
    \,.
\end{eqnarray}
In practice, the difference is typically negligible ($2.5\times10^{-4}$ on average in the HSC-SSP Wide layer).  In approximately 3\% of our coadd patches (those with the largest variation in seeing among the input images) it exceeds $10^{-3}$, however, and for future data releases we will use the exact correction.

A more difficult problem in our (or any) approach to aperture corrections is galaxy photometry.  Galaxy photometry measurements can miss flux due to both the same large-radius PSF flux that affects point sources as well as per-galaxy morphology differences.  Because we have no source of ground truth for the latter, we have no way to independently measure the former.  Our approach thus far has simply been to apply the aperture corrections for stars to galaxies as well, which should adequately correct photometry for galaxies near the size of the PSF and be better than no correction for large galaxies.  In any case, we suspect (but cannot easily demonstrate) that systematic errors in galaxy fluxes due to per-source morphology and/or surface-brightness differences dominate over the errors due to incomplete correction for the wings of the PSF.

One aspect of this approach that may appear to be a serious problem (but in fact is not) is our reliance on a finite 4\arcsec aperture for calibration.  This aperture is not large enough to capture all of the flux in the wings of the PSF, and hence our aperture corrections do not force all fluxes to a constant missing flux fraction.  However, our final photometric calibration (see \secref{joint-calibration}) includes a spatially-varying per-visit multiplicative term that can ``soak up'' any differences in extra flux in the wings of the PSF beyond the aperture.  This is not merely a first-order correction; if the functional form of the spatial variation is sufficiently general, a multiplicative calibration correction and aperture correction are completely degenerate.  This does mean, however, that our photometric calibration coefficients cannot be interpreted naively as a measure of the combination of transparency and exposure time.

\subsubsection{Centroids}
\label{sec:centroids}

The first algorithm run by the HSC Pipeline's measurement system is a centroider, as the resulting position is used by later shape and photometry algorithms.  While the HSC Pipeline contains multiple centroid algorithms and implementations, we have thus far only used one in production: an approximate maximum-likelihood algorithm first developed for the SDSS \textit{Photo} Pipeline \citep{2003AJ....125.1559P}.

Basically, we note that in the limit that the noise is dominated by the background the maximum-likelihood estimate of the position of a star is the peak of the image correlated with the PSF (approximated as a Gaussian, as we do in \secref{detection}); we then use parabolic interpolation in this smoothed image to find the position of the likelihood peak.  When the object is significantly larger than the smoothing filter we bin the original image and re-smooth (effectively doubling the smoothing length); this is repeated until the smoothing filter and the object size roughly agree.

Unlike the SDSS algorithm, we do not de-bias centroid measurements to account for asymmetry in the PSF, as this is largely degenerate with the astrometric solution as long as the PSF asymmetry is approximately constant at the smallest spatial scales on which the astrometric solution varies.

\subsubsection{Shapes}
\label{sec:shapes}

HSC Pipeline's ``shape'' algorithms produce an ellipse that characterizes the extent of the object \emph{without} correcting for its convolution with the PSF.  This is obviously not the only way to characterize an object's morphology or size (or even the most useful for science), but most other characterizations are highly algorithm-specific and hence hard to interpret generically.  More importantly, this PSF-uncorrected ellipse is what is typically most useful to downstream flux measurement algorithms.  Algorithms (both those that produce this ellipse and those that do not) are free to define additional outputs, and these frequently do include PSF-corrected shapes such as those used for weak gravitational lensing.

All of the HSC Pipeline's current shape plug-ins compute the second moments of the image of the source about its centroid, using a Gaussian weight function.  Given image $z(\bm{r})$ (with position $\bm{r}$ defined about its centroid), the raw moments (a $2 \times 2$ matrix) are:
\begin{eqnarray}
  \bm{M} = \frac{
    \sum\limits_{\bm{r}} \bm{r}\, \bm{r}^T z(\bm{r}) \,
    e^{-\frac{1}{2}\bm{r}^T \bm{C}^{-1} \bm{r}}
  }{
    \sum\limits_{\bm{r}} z(\bm{r}) \,
    e^{-\frac{1}{2}\bm{r}^T \bm{C}^{-1} \bm{r}}
  },\label{eqn:raw-moments}
\end{eqnarray}
where $\bm{C}$ is the covariance matrix of the Gaussian weight function.  Using a weight function is necessary to keep the noise in the measurement from diverging, but it also biases the result.  We can see this by inserting a Gaussian image with known moments $\bm{Q}$ into \eqnref{raw-moments}, and taking the limit of small pixels (so the sums become integrals):
\begin{eqnarray}
\bm{M} =& \frac{
    \sum\limits_{\bm{r}} \bm{r}\, \bm{r}^T
    e^{-\frac{1}{2}\bm{r}^T \bm{Q}^{-1} \bm{r}}\,
    e^{-\frac{1}{2}\bm{r}^T \bm{C}^{-1} \bm{r}}
  }{
    \sum\limits_{\bm{r}} \,
    e^{-\frac{1}{2}\bm{r}^T \bm{Q}^{-1} \bm{r}}\,
    e^{-\frac{1}{2}\bm{r}^T \bm{C}^{-1} \bm{r}}
  } \\
  \approx& \frac{
    \int \bm{r}\, \bm{r}^T
    e^{-\frac{1}{2}\bm{r}^T \left(\bm{C}^{-1} + \bm{Q}^{-1}\right) \bm{r}}
    d^2\bm{r}
  }{
    \int
    e^{-\frac{1}{2}\bm{r}^T \left(\bm{C}^{-1} + \bm{Q}^{-1}\right) \bm{r}}
    d^2\bm{r}
  }\\
  =& \left(\bm{C}^{-1} + \bm{Q}^{-1}\right)^{-1} \,.
  \label{eqn:shape-weight-bias}
\end{eqnarray}
To correct for this bias (only approximately, because real sources are not Gaussian), we can solve \eqnref{shape-weight-bias} for $\bm{Q}$:
\begin{eqnarray}
  \bm{Q} = \left(\bm{M}^{-1} - \bm{C}^{-1}\right)^{-1}\,.
  \label{eqn:shape-bias-correction}
\end{eqnarray}
This provides a way to correct the measured raw moments $\bm{M}$ given the moments of the weight function $\bm{C}$, which may be a predefined constant or a quantity computed iteratively.

The three unique elements of the symmetric $2 \times 2$ matrix $\bm{Q}$ parameterize an ellipse, and this is the form used to record the outputs of our shape algorithms.  See \appendixref{ellipse-parameterizations} for information on how this parameterization relates to others.

Two of the three shape algorithms currently implemented in the HSC Pipeline use \emph{adaptive} Gaussian moments, in which the weight function's moments $\bm{C}$ are iteratively matched to the (corrected) measured moments $\bm{Q}$.  This makes the correction formula \eqnref{shape-bias-correction} particularly simple: $\bm{Q} = 2\bm{M}$.  More importantly, it produces the highest (for a Gaussian weight function) S/N measurement, assuming that the iterative procedure converges.  This is not guaranteed, especially for low surface-brightness objects, where noise can make the raw moments singular.  Nearly singular moments are in some respects even more problematic, as these can yield to spuriously large ellipses for small objects, with no straightforward indicator of failure, which can then compound by increasing the size of the weight function.  Our two adaptive moments algorithms, \texttt{SdssShape} and \texttt{HsmMoments}, differ primarily in the heuristics they apply to stabilize the iteration and test for convergence.  \texttt{SdssShape} is a reimplementation of the algorithm used in the SDSS \textit{Photo} Pipeline \citep{2001ASPC..238..269L}, while the \texttt{HsmMoments} algorithm is simply a thin wrapper around the HSM \citep{2003MNRAS.343..459H} implementation in GalSim \citep{2015A&C....10..121R}.  The two algorithms fail on approximately the same number of sources (16\% for \texttt{SdssShape} and 17\% for \texttt{HsmMoments}), though this can depend on the details of how the sample is otherwise selected.  The two algorithms fail on a different population of sources (only approximately 8\% of sources are not measured successfully by either algorithm) and hence we run both in HSC-SSP data release productions.  Most of these failures occur when the moments matrix $\bm{Q}$ becomes singular.  In addition to adaptive moments, the HSM algorithms also include several approaches to estimating PSF-corrected ellipticities for the purpose of estimating shear due to weak gravitational lensing.  These are described more fully in \citet{2003MNRAS.343..459H} and \citet{hsc-shear}.

The HSC Pipeline's third shape algorithm, \texttt{SimpleShape}, uses a circular Gaussian weight function with fixed (but configurable) radius.  This makes iteration unnecessary, removing one source of catastrophic failure.  Singular moments are still possible, of course, and the resulting measurement is more noisy than the adaptive methods.  We use \texttt{SimpleShape} primarily when processing data from HSC's 8 out-of-focus wavefront sensors as described in \citet{hsc-onsite}.

The CModel photometry algorithm (\secref{cmodel-photometry}) also measures multiple ellipses for each source (corresponding to different profiles), but these ellipses are approximately PSF-corrected (assuming analytic forms for the source morphology) and hence cannot be used as a ``slot'' shape that feeds other algorithms.

\subsubsection{PSF Photometry}
\label{sec:psf-photometry}

The algorithm for PSF photometry given a measured centroid can be derived from two different starting premises: maximum-likelihood fitting and matched-filter signal processing.  In the former, we consider an image $z_i$ with Gaussian noise $\sigma_i$ and assume a one-parameter model that simply multiplies the PSF model $\phi_i$ (shifted to the same sub-pixel centroid as the source, which we consider fixed) with an amplitude parameter $\alpha$.  The likelihood is then
\begin{eqnarray}
  P(\bm{z}|\alpha) = \left(\prod_i \sqrt{2\pi}\sigma_i\right)^{-1} \!
    e^{-\sum\limits_i\frac{\left(z_i - \alpha \phi_i\right)^2}{2\sigma_i^2}}
    \,.
\end{eqnarray}
Differentiating the logarithm of $P$, setting the result to zero, and some algebra yields the familiar solution for linear least squares:
\begin{eqnarray}
  \alpha_{\mathrm{ML}} = \frac{
    \sum\limits_i \phi_i z_i / \sigma_i^2
  }{
    \sum\limits_i \phi_i^2 / \sigma_i^2
  }
  \,.
  \label{eqn:psf-photometry-ml}
\end{eqnarray}
In the matched-filter version, we simply compute the inner product of the data (the image) and the filter (the PSF model, shifted to the centroid of the source, as before), and divide by the effective area of the filter:
\begin{eqnarray}
  \alpha_\mathrm{MF} = \frac{
    \sum\limits_i \phi_i z_i
  }{
    \sum\limits_i \phi_i^2
  }
  \,.
  \label{eqn:psf-photometry-mf}
\end{eqnarray}

When the per-pixel noise is constant, as is the common case for faint objects where noise from the sky dominates, these two derivations agree.  When the per-pixel noise is not constant, as is the case for bright objects whose own photon noise dominates, both versions yield the correct result, as can be seen if we insert $z = \alpha\phi_i$ into Eqns.~(\ref{eqn:psf-photometry-ml}) and~(\ref{eqn:psf-photometry-mf}).  The noise properties of the measurements are not the same, however; the variance in the maximum-likelihood measurement is
\begin{eqnarray}
  \sigma_{\mathrm{ML}}^2 = \left(\sum\limits_i \phi_i^2 / \sigma_i^2\right)^{-1}
\end{eqnarray}
while for the matched filter estimate it is
\begin{eqnarray}
  \sigma_{\mathrm{MF}}^2 = \frac{
    \sum\limits_i \phi_i^2 \sigma_i^2
  }{
    \left(\sum\limits_i \phi_i^2\right)^2
  }\,.
\end{eqnarray}
These once again reduce to the same estimate when sky noise dominates, but the maximum-likelihood measurement has lower noise when it does not.  In the bright-object regime, however, systematic errors from incorrect PSF modeling are a much bigger concern than noise.  To illustrate the effect such an error has on these estimates, we consider a true image
\begin{eqnarray}
  z_i = \alpha(\phi_i + \epsilon_i)
\end{eqnarray}
where $\epsilon$ represents the deviation of our model from the true PSF, along with noise of the form
\begin{eqnarray}
  \sigma_i^2 = b + \alpha(\phi_i + \epsilon_i)
\end{eqnarray}
where $b$ is the level of the background (before it is subtracted).  If we work in units of electrons, this is the natural combination of Poisson noise from both the source and the background in the limit where both are large enough that we can approximate the Poisson distributions as Gaussian.  Inserting these into \eqnref{psf-photometry-ml} and \eqnref{psf-photometry-mf} yields, respectively,
\begin{eqnarray}
  \alpha_{\mathrm{ML}} = \alpha \frac{
    \sum\limits_i \phi_i \left(\phi_i + \epsilon_i\right)
        / \left(b + \alpha\left[\phi_i + \epsilon_i\right]\right)
  }{
    \sum\limits_i \phi_i^2
        / \left(b + \alpha\left[\phi_i + \epsilon_i\right]\right)
  }
\end{eqnarray}
and
\begin{eqnarray}
  \alpha_\mathrm{MF} = \alpha\frac{
    \sum\limits_i \phi_i^2 + \phi_i\epsilon_i
  }{
    \sum\limits_i \phi_i^2
  }
  \,.
\end{eqnarray}
Both formulae yield the incorrect flux, of course, but the damage done to the maximum-likelihood estimate is much more severe: it no longer responds linearly to the true flux.  The matched-filter estimate, in contrast, remains linear, and can be corrected \emph{completely} by aperture corrections if the error in the PSF model is constant or smoothly-varying over the image.

As the above discussion suggests, PSF photometry in the HSC Pipeline always uses the matched-filter formula \eqnref{psf-photometry-mf}, and considers the per-pixel noise only when estimating the uncertainty in the flux.  We also treat the centroid as fixed at the position measured by the centroider (see \secref{centroids}) and do not attempt to fold centroid uncertainties into the PSF flux uncertainty; even when the S/N is low, additional uncertainty introduced by the centroid is subdominant.  We use Lanczos interpolation to shift the PSF model image to the centroid of the source, and when the PSF model is defined on an oversampled pixel grid relative to the image, we perform the shift before down-sampling to the image pixel grid.

\subsubsection{Fixed Aperture Photometry}
\label{sec:fixed-aperture-photometry}

The HSC Pipeline measures a set of aperture fluxes with fixed radii.  The radii are configurable, but by default are set to am approximately logarithmically-spaced sequence from 1-23\arcsec\ in diameter.  These fluxes are by default not aperture-corrected, because their intent is to measure a crude, PSF-uncorrected radial profile, not a total magnitude.

However, not applying aperture corrections to these fluxes also puts them in a different photometric system from our other photometry.  As discussed in \secref{aperture-corrections}, we use the 4\arcsec\ diameter aperture for photometric calibration, and hence define it to be the total flux.  Apertures larger than 4\arcsec\ thus typically appear to have more flux than the ``total'' flux measured in the aperture-corrected photometric system.  In the future, we plan to address this by estimating the curve-of-growth from the brightest stars and interpolating this in the same way we interpolate our current aperture corrections; we could then tie the radial profile to the aperture corrections at infinity.  This would require using saturated stars to get a sufficiently large signal-to-noise ratio in the largest aperture fluxes, which would in turn require more sophisticated algorithms for aperture photometry.  An approach that worked in SDSS but has not yet been implemented for HSC/LSST is to divide each aperture both radially and azimuthally into segments of annuli; this allows discrepant segments (due to \eg~bleed trails) to be removed with outlier rejection and more robust uncertainties to be estimated from the statistics of the segments.

Instead, aperture photometry measurements in the HSC Pipeline are simply direct integrals over the circular area.  For small radii, they are indeed integrals, not simple sums -- we use the approach of \citet{2013MNRAS.431.1275B} to exactly integrate over sub-pixel regions, avoiding the discretization artifacts that can otherwise affect small-aperture fluxes.  We just use naive sums for larger apertures, where the discretization noise is negligible compared to photon noise and the performance difference between the two approaches is significant.

\subsubsection{Kron Photometry}
\label{sec:kron-photometry}

The Kron radius \citep{1980ApJS...43..305K} of a galaxy is defined as
\begin{eqnarray}
\label{eqn:KronRadius}
R_{\mbox{\tiny Kron}} \equiv \frac{\int_{\cal A} r z(r) \, 2\pi r\, dr}{\int_{\cal A} z(r) \, 2\pi r\, dr}.
\end{eqnarray}
where $z$ is the surface brightness and $r$ is the distance to the center of the galaxy.

Ideally, the integration area $\cal A$ would extend to infinity, but in practice the integrals do not converge very well (the numerator diverges logarithmically if the profile falls as $r^{-3}$).

Once $R_{\mbox{\tiny Kron}}$ has been determined, the Kron flux is the flux enclosed in a region of radius $f_{\mbox{\tiny Kron}} R_{\mbox{\tiny Kron}}$; a common choice is $f_{\mbox{\tiny Kron}} = 2.5$.

If the object is elliptical things are a bit more complicated.  We can still measure $R_{\mbox{\tiny Kron}}$, but its meaning is unclear; HSC uses 2nd-moment shapes (\secref{shapes}) to set the axis ratio and position angle, and then defines the \textit{Kron ellipse} to have area $\pi R_{\mbox{\tiny Kron}}^2$.

The HSC algorithm begins by defining the area, ${\cal A}$, over which we evaluate $R_{\mbox{\tiny Kron}}$. We start with the adaptive moments defined in \secref{shapes} and take the initial $\cal A$ to be the ellipse with the same shape and orientation but with area $\texttt{nSigmaForRadius}\times \sqrt{\det\bm{Q}}$ (by default \texttt{nSigmaForRadius = 6}).  The adaptive moments code fails for some objects; in these cases we take the initial ellipse to be that of the PSF model.

With this definition of $\cal A$ we evaluate \eqnref{KronRadius} and adjust $\cal A$ to have area $\pi \times \texttt{nSigmaForRadius}^2 \times R_{\mbox{\tiny Kron}}^2$. In theory we could iterate, but in practice we have found that a single update is sufficient.

There are some failure modes to keep track of; in particular if $\cal A$ reaches the edge of the image we give up and set \texttt{kron\_flags\_edge}.  If either the numerator or denominator in \eqnref{KronRadius} is non-positive (as can happen as $z$ is not guaranteed to be positive in the presence of noise) or the estimated $R_{\mbox{\tiny Kron}}$ is less than the PSF's $R_{\mbox{\tiny Kron}}$ we set \texttt{kron\_flags\_usedPsfRadius} and set $R_{\mbox{\tiny Kron}}$ to the Kron radius of the PSF.

With a value of $R_{\mbox{\tiny Kron}}$ in hand we can measure the Kron flux as the flux included in an elliptical aperture with determinant radius (see \appendixref{ellipse-parameterizations}) $\texttt{nRadiusForFlux}\times R_{\mbox{\tiny Kron}}$; by default $\texttt{nRadiusForFlux} = 2.5$ as suggested above. The only error that can be encountered at this stage is when this aperture extends outside the image, in which case \texttt{kron\_flags\_edge} is set and no flux is reported.  The error on the Kron flux is purely that due to the pixel noise (\ie~we do not attempt to propagate the error in estimating the radius, which would be correlated with the errors in the pixel values).

\subsubsection{Afterburner Photometry: PSF-Matched Apertures Without Deblending}
\label{sec:afterburner}

As we discussed in \secref{deblending}, our deblender performs adequately when the number of objects in a blend and the amount of overlap between them are both small.  In dense regions such as galaxy clusters, however, it can produce catastrophically bad results that make it essentially impossible to accurately photometer the deblended children.  As an alternative, our \emph{afterburner}\footnote{The ``afterburner'' name is a historical artifact referring to when these algorithms were run relative to the rest of the pipeline, but it has caught on in the HSC collaboration and is used in other HSC papers \citep[\eg][]{2017arXiv170208449A}, and hence it is helpful to use it here as well for clarity.} photometry algorithms perform measurements at the position of each child object on the original (not deblended) image.  We run both Kron photometry and fixed circular aperture photometry in this mode.

The measurement image used by the afterburner algorithms is not the original image, however; using aperture photometry to measure consistent colors requires images with the same PSF in all bands.  We thus approximately match the coadd images in each band to predefined circular Gaussian PSFs with full-width at half maximums (FWHMs) of 3.5, 5.0, and 6.5 pixels (0.58, 0.84, 1.09\arcsec), by convolving with a Gaussian matching kernel derived by assuming the original PSF is Gaussian.  More precisely, if the target PSF is defined as a Gaussian with covariance matrix
\begin{eqnarray}
\bm{Q}_\mathrm{target} =
\left[
    \begin{array}{ c c }
        1 & 0 \\
        0 & 1
    \end{array}
\right]
\frac{\mathrm{FWHM}_\mathrm{target}}{8 \ln 2}
\end{eqnarray}
and the 2nd moments matrix (see \secref{shapes}) of the PSF model at the position of the source to be measured is $\bm{Q}_\mathrm{psf}$, then the matching kernel we use is simply a Gaussian with covariance $\bm{Q}_\mathrm{match} = \bm{Q}_\mathrm{target} - \bm{Q}_\mathrm{psf}$.  When $\det{\bm{Q}_\mathrm{match}} \le 0$ (\ie~the original PSF is larger than the target PSF) we do not match the image at all; the fact that we match to a sequence of increasingly larger target PSFs ensures that at least one set of afterburner measurements should exist for nearly all objects.  The matching is performed locally for each object, so we assume the PSF (and hence the matching kernel) is constant over the pixels to be used in the measurement.  We never produce a PSF-matched image that covers the full area of a patch.

Because the matching process makes the PSF broader, it actually increases the degree to which objects are blended, and hence the only measurements even approximately unaffected by blending are those measured with a very small aperture.  These will of course only measure the colors of the cores of objects, and even they rely on blending not being too severe, but they nevertheless appear to be our most reliable measurements of galaxy colors \citep{2017arXiv170405988T,2017arXiv170208449A}.  Larger fixed aperture and Kron photometry measurements on these PSF-matched images are likely to be strongly affected by blending, but can still be quite useful for objects that were not blended or whose flux dominates that of their neighbors.  They are also useful when the blending involves galaxies with similar colors at the same redshift, as is the case in the galaxy cluster cores where our deblender's performance is at its worst.

\subsubsection{CModel Photometry}
\label{sec:cmodel-photometry}

Our primary algorithm for general galaxy photometry is a modified version of the approach used in the SDSS \textit{Photo} Pipeline \citep{2004AJ....128..502A}.  It fits multiple PSF-convolved galaxy models in a sequence designed to approximate a bulge-disk decomposition or S\'{e}rsic model \citep{1963BAAA....6...41S} fit while minimizing the number of degrees of freedom in any individual fit.  Roughly speaking, the sequence is:
\begin{enumerate}
  \item Fit an elliptical model with an exponential profile (a S\'{e}rsic profile with index $n=1$) to the source, with both the ellipse parameters and the amplitude free.
  \item Fit a de Vaucouleurs model (\citealt{1948AnAp...11..247D}; a S\'{e}rsic profile with index $n=4$) with the same parameters free (allowing the fit to yield different values for these parameters).
  \item Fit both models simultaneously, keeping their ellipses fixed at the results from the previous two fits, and allowing only the two amplitudes to vary.
\end{enumerate}
The amplitudes from all three of these fits are useful flux measurements.  The last is not well-motivated statistically, but worked well empirically in SDSS as a crude linear interpolation between exponential and de Vaucouleurs models.  This model is typically a worse fit to real galaxy morphologies than a S\'{e}rsic profile, but it can be fit much more robustly and efficiently.  This is especially true for low S/N and/or poorly-resolved galaxies.

Because all of these models are convolved with the PSF of the image being measured, CModel fluxes are PSF-corrected to the extent the models are sufficient to describe the underlying galaxy morphology.  Given the nearly limitless range of morphologies real galaxies have, the models are obviously inadequate at some level, but in practice we have seen no indication that biases due to this sort of mismatch are significantly affecting the CModel colors.   This may be masked by other failures in CModel, however; as our photometry algorithm with the most degrees of freedom, it has a tendency to respond very poorly to even minor deblending failures.

We also expect any model biases in CModel photometry to be much less significant for colors (ratios of fluxes) than the fluxes themselves.  It is difficult to decide how far out to extrapolate an analytic galaxy profile even for nearby objects that are being modeled with significant human intervention, and these poorly-constrained tails can have a large influence on the integrated flux.  The best we can hope for in fully-automated fitting of faint, poorly-resolved sources is some degree of consistency, and even this is difficult to verify because we lack any kind of ground truth.  Colors need not correspond to total fluxes, however; a color measurement that systematically gives too much or too little weight to some stellar populations in a galaxy will still correspond to some physically meaningful star formation history as long as the measurements are consistent across bands (note that this is true even in the presence of large population gradients).  We can \emph{almost} guarantee this in forced photometry: we use the same model in all bands, allowing only the amplitude parameters to vary.  If we had the same PSF in every band or complete models, consistent colors could be guaranteed; because the PSF is not the same in all bands and we correct for these differing PSFs only by convolving our (imperfect) models, colors may still be slightly inconsistent.

While the CModel algorithm produces PSF-corrected ellipse shapes, the fitting algorithm has been tuned for photometry, and we do not recommend using these ellipses for weak lensing shear measurements.  Similarly, half-light radii derived from these ellipses are probably not reliable for objects close to the PSF size, and even the radii of well-resolved objects are more sensitive than fluxes to our simplistic choice of models.

The technical details of the CModel algorithm are described more fully in \appendixref{cmodel-details}.  This includes discussion of a serious bug in PDR1 that strongly distorts the structural parameters measured by CModel (without significantly damaging -- and in some cases probably improving -- measured colors for most objects).

An analysis of CModel performance on simulated galaxies in \citet{hsc-synpipe-1} indicates that CModel performs well for photometry down to $i\sim 25$ despite the algorithm's aforementioned limitations.  As noted by \citet{2017arXiv170208449A}, however, CModel colors are outperformed by the PSF-matched small-aperture ``afterburner'' photometry (\secref{afterburner}) in yielding a lower scatter in the red sequence of galaxies in clusters \citep[Figure~19]{2017arXiv170208449A}, which is one of the only tests of galaxy photometry we have that does not rely on simulations.  The afterburner photometry also seems to yield better photometric redshifts \citep{2017arXiv170405988T}.  It is unclear whether this is due to CModel's sensitivity to deblender failures (which the afterburner fluxes avoid), the number of degrees of freedom in the algorithms (CModel has many more free parameters that must be constrained), or astrophysical differences between the stellar populations in the cores of galaxies (which the afterburner measure) and those in the outskirts (which are also included in the CModel photometry).

\subsubsection{Star/Galaxy Classification}
\label{sec:star-galaxy-classification}

Separating stars from galaxies in images relies chiefly on a measure of how different an object's morphology is from that of a point source.  Our primary measure of this is called \emph{extendedness}, and it is defined as the difference between the PSF magnitude and the CModel magnitude. For a star, the expected value of this difference is zero because the scale size of the best-fit galaxy model will approach zero before PSF convolution and hence the PSF-convolved model will be very close to the PSF itself.  For resolved galaxies the expected value of the difference in magnitudes is significantly different from zero, and in most cases it is positive, because PSF photometry typically underestimates the flux of extended objects.

\begin{figure}[htb]
    \includegraphics[width=0.48\textwidth]{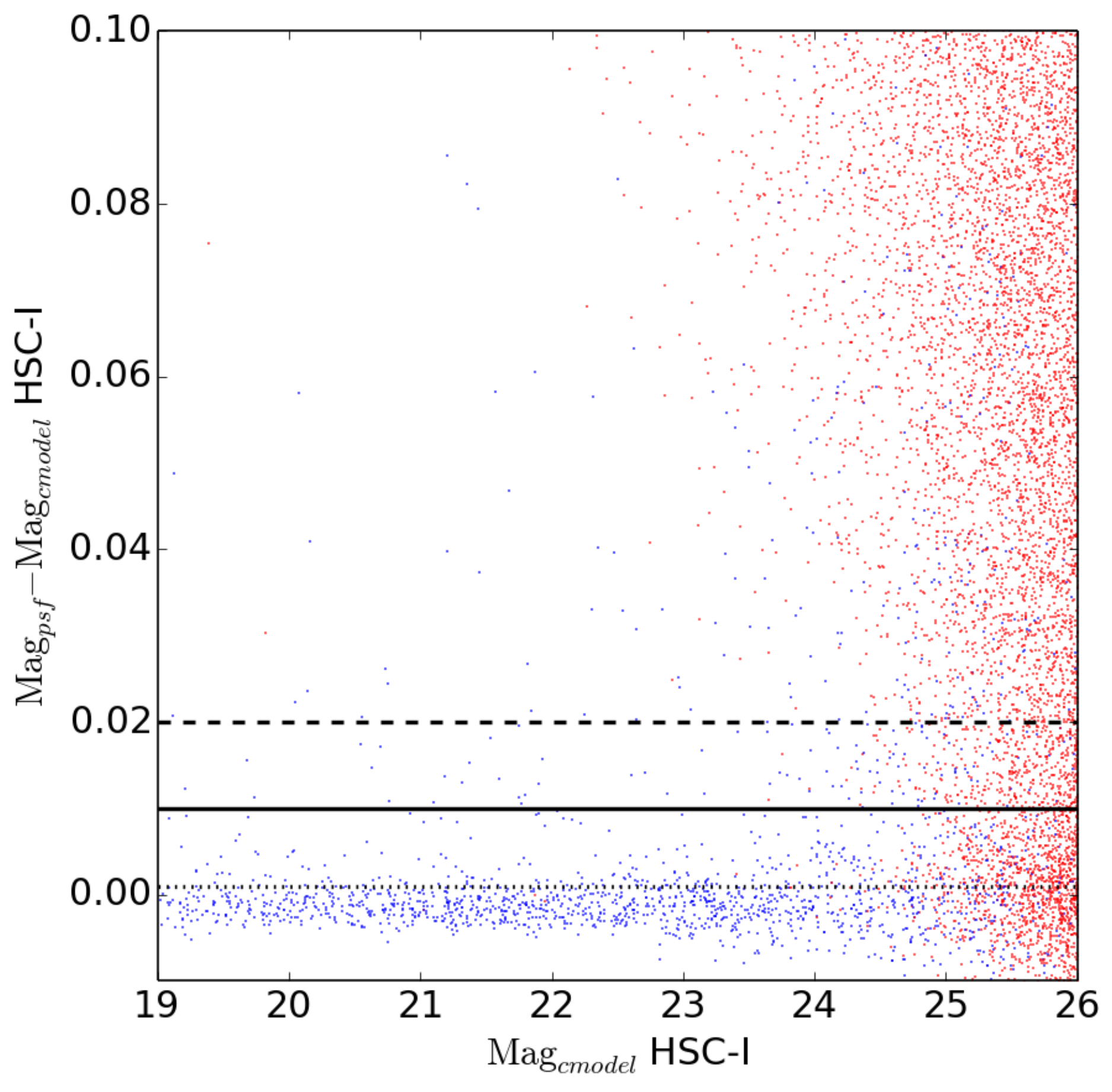}
    \caption{$\textrm{Mag}_{\mathrm{psf}}-\textrm{Mag}_{\mathrm{cmodel}}$ star/galaxy cuts in $i$ with objects according their classifications in the HST/ACS catalog of \citet{2007ApJS..172..219L}. Red dots correspond to galaxies and blue dots to stars. The horizontal lines are three extendedness cuts chosen by eye: the dashed line strives for high completeness in stars, the solid line for a balanced separation, and the dotted line for high purity in stars. \figref{extCutScoresI}shows the corresponding scores in magnitude bins.
    }
  \label{fig:extCutsI}
\end{figure}

\begin{figure*}[htb]
    \includegraphics[width=1.0\textwidth]{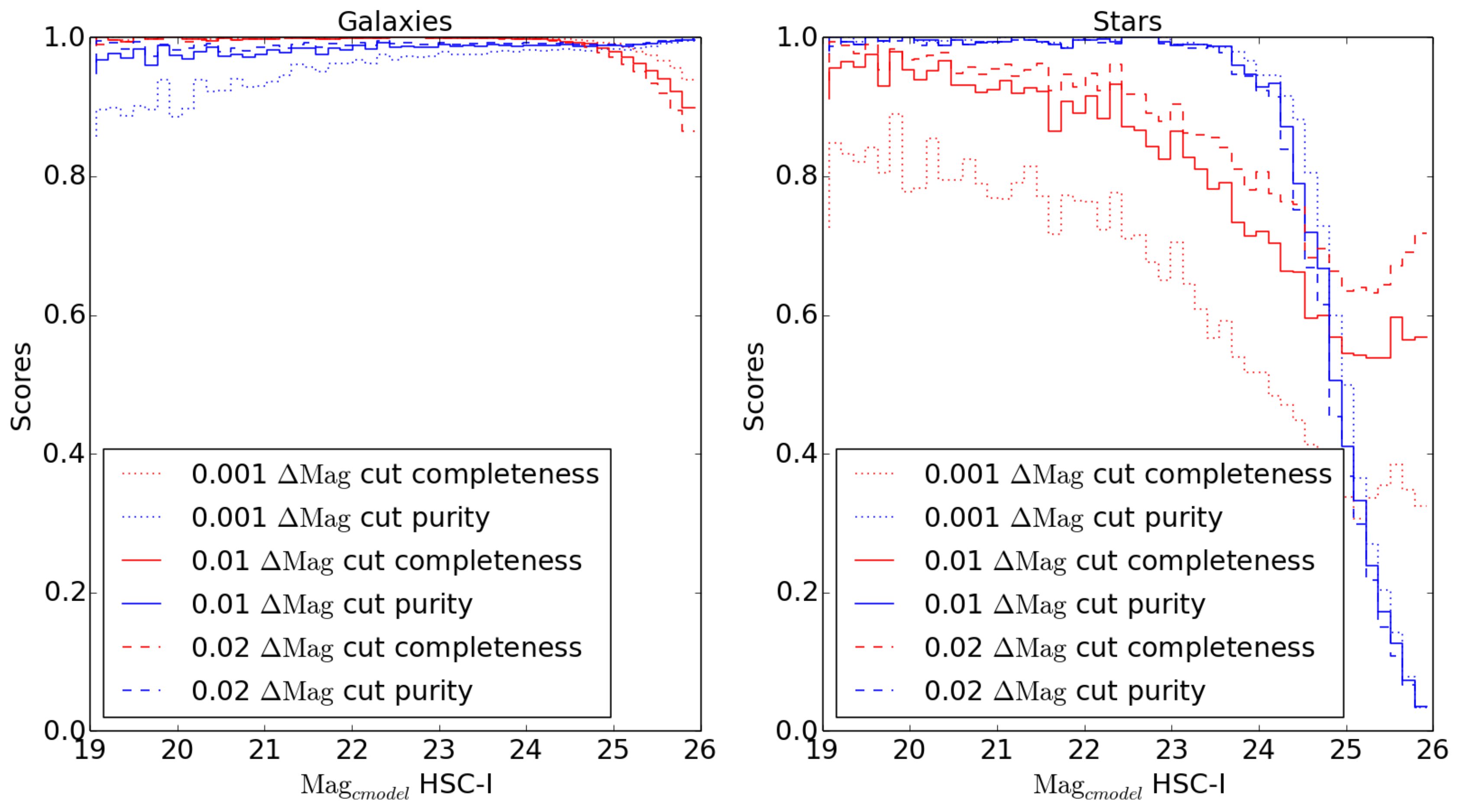}
    \caption{Purity and completeness for galaxies (left) and stars (right) obtained from hard cuts on extendedness ($\Delta\mathrm{Mag}$; the difference between the PSF and CModel magnitudes) in $i$.  Blue lines are for purity, and red lines for completeness. Dotted lines correspond to a very conservative cut (avoiding galaxy contamination on stars), solid lines correspond to a typical cut chosen by eye, and dashed lines to a very permissive cut (avoiding missing stars).
    }
  \label{fig:extCutScoresI}
\end{figure*}

To illustrate extendedness-based star/galaxy classification, \figref{extCutsI} shows a diagram of the PSF-CModel magnitude difference vs. CModel magnitude in $i$ for objects detected in the COSMOS UltraDeep field. These objects were matched to the HST/ACS catalog of \citet{2007ApJS..172..219L}.  With significantly higher resolution (0.1\arcsec\ FWHM PSF) and comparable depth, we can effectively use this HST/ACS catalog to provide ``ground truth'' star/galaxy labels: the stars in the figure are colored blue, and the galaxies red. \figref{extCutScoresI} shows the purity and completeness of the samples of stars (right panel) and galaxies (left panel) obtained from the cuts in extendedness denoted by the horizontal lines in \figref{extCutsI}.  Extendedness seems to work well down to $i\sim 24$, where the ratio of stars to galaxies shrinks dramatically and most galaxies become smaller than the PSF.  The \texttt{classification\_extendedness} column in our catalogs is obtained by setting a hard cut (at $m_{\mathrm{psf}}-m_{\mathrm{cmodel}}=0.0164$) on this magnitude difference, but users of the catalog may be able to obtain a more suitable classifier for their science by making a different cut.

To produce more accurate star/galaxy classifications, we combine colors measured from CModel fluxes with extendedness via supervised machine learning, using the HST/ACS catalog of \citet{2007ApJS..172..219L} to provide labels for the training set.  We estimate probability densities for both stars and galaxies using Extreme Deconvolution \citep[XD;][]{2011ApJ...729..141B}, an algorithm based on mixtures of Gaussians that estimates the probability density from multivariate data, taking into account measurement errors. The probability densities are then combined with a flat prior to compute a posterior probability of being a star.

\begin{figure*}[htb]
    \includegraphics[width=1.0\textwidth]{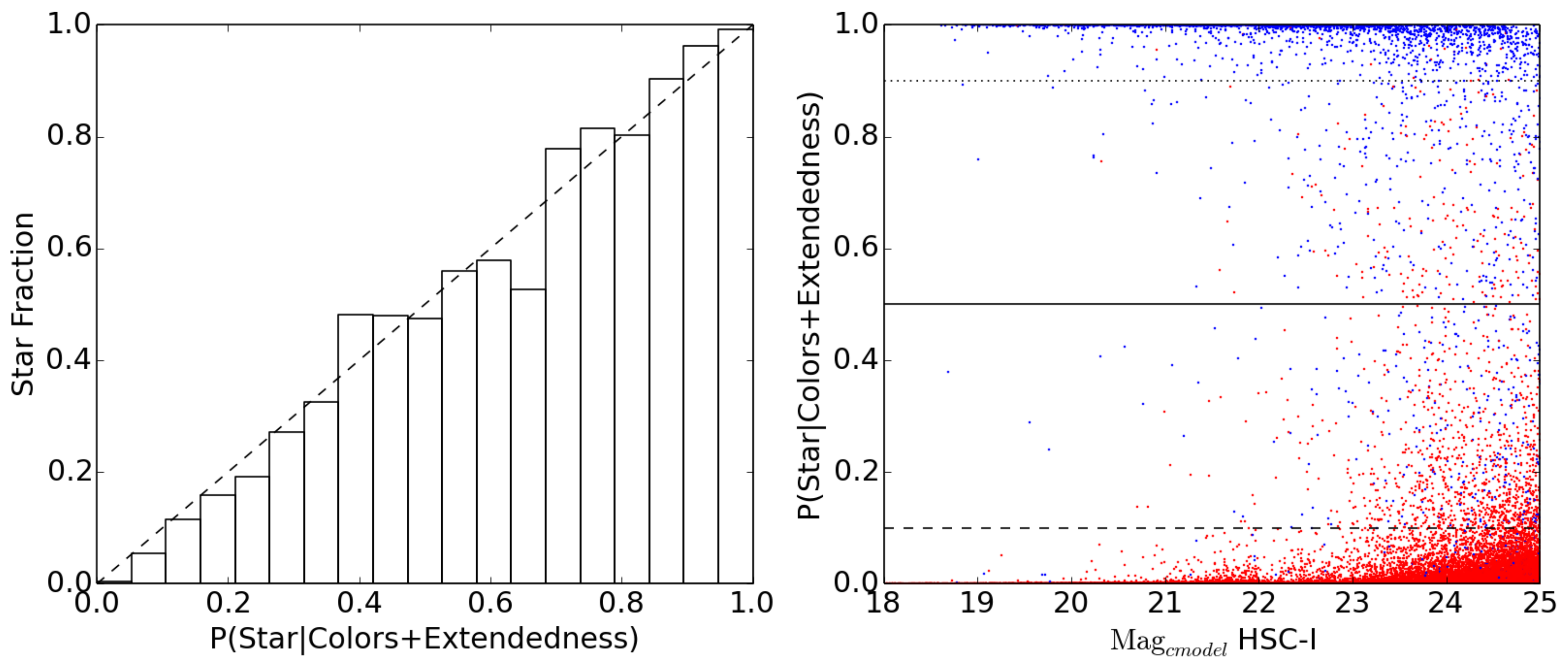}
    \caption{The left panel plots the fraction of true stars in $P(\textrm{Star}|\textrm{Colors}+\textrm{Extendedness})$ bins, verifying that this quantity behaves like a probability. The right panel plots $P(\textrm{Star}|\textrm{Colors}+\textrm{Extendedness})$ against $\textrm{Mag}_{cmodel}$ $i$ and colors the points according to the labels from the HST classifications of \citet{2007ApJS..172..219L}: blue for stars, and red for galaxies. The black lines denote the cuts we use to compute the scores in \figref{xdColExtScores}.}
  \label{fig:xdColExtPosts}
\end{figure*}

\begin{figure*}[htb]
    \includegraphics[width=1.0\textwidth]{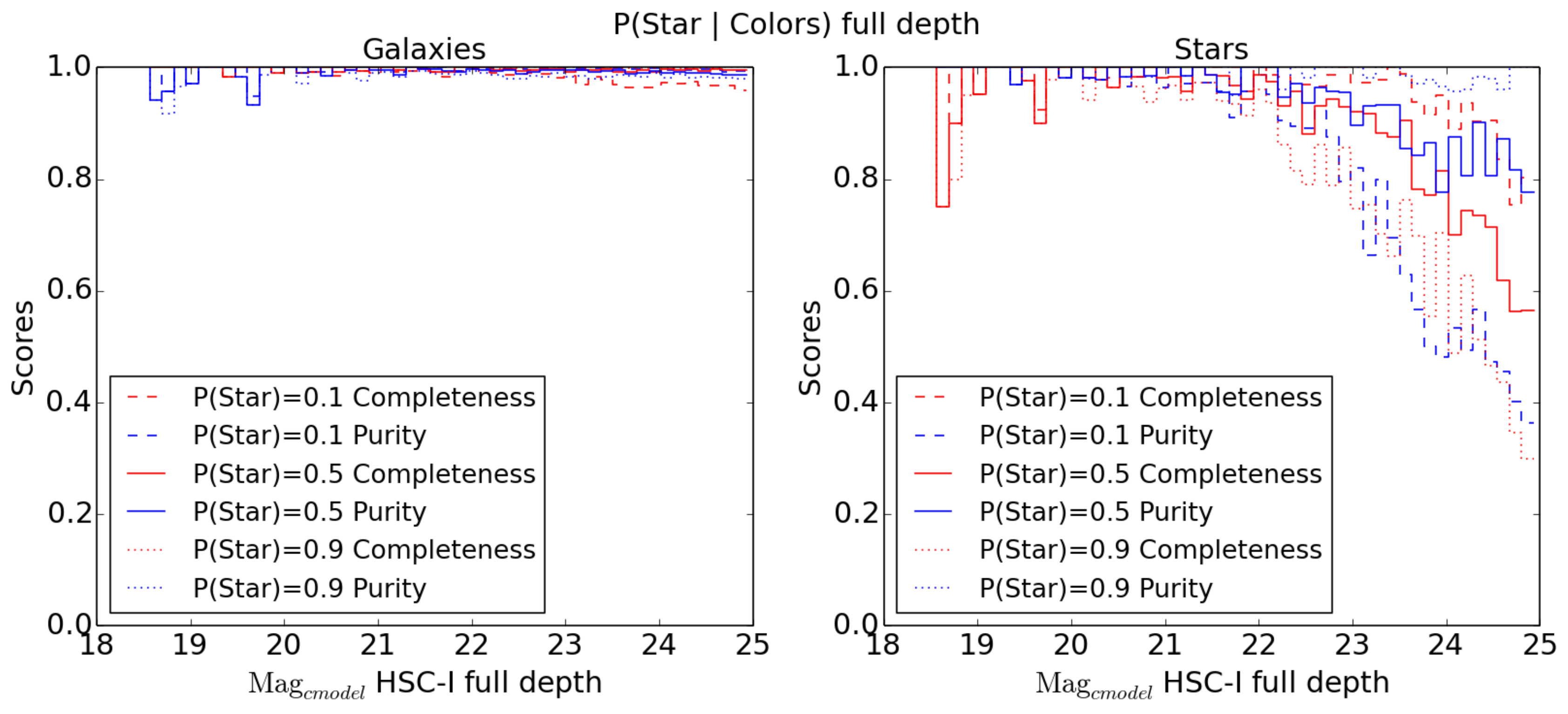}
    \caption{Scores in magnitude bins of
    $P(\textrm{Star}|\textrm{Colors}+\textrm{Extendedness})$ using the cuts shown in the right panel of \figref{xdColExtPosts}.}
    \label{fig:xdColExtScores}
\end{figure*}

Because the true probability of begin a star is a strong function of magnitude (largely because galaxy counts rise faster with magnitude than star counts), we divide the catalog 4 magnitude bins in $i$: [18, 22], [22, 24], [24, 25], and [25, 26].  An XD fit in colors and extendedness is then produced in each bin independently. T he left panel in \figref{xdColExtPosts} shows that the posterior probabilities inferred from the XD density distributions do behave like a probability; the right panel plots the posterior probability vs CModel magnitude in $i$ for objects detected in the COSMOS UltraDeep layer. \figref{xdColExtScores} shows the purity and completeness of samples of stars (right panel) and galaxies (left panel) obtained from cuts in the posterior. While most internal data releases have included a \texttt{pstar} column with the posterior probability of belonging to the star class estimated with this procedure, the machine learning classifier has not yet been run on PDR1 and hence the \texttt{pstar} column is not yet available; it will be added in a future incremental release.

\subsubsection{Blendedness Metrics}
\label{sec:blendedness-metrics}

Given the central importance (and fallibility) of the deblender, it is useful to provide some measure of the extent to which each object is blended.  The metric we have developed for this purpose essentially measures the fraction of the total flux in the neighborhood of a source that belongs to  its neighbors.  This neighborhood does not have a hard boundary; it is defined by a Gaussian weight function with its ellipse taken from the source's shape (see \secref{shapes}).

More formally, given an image $z(\bm{r})$ and the source moments matrix $\bm{Q}$, the Gaussian-weighted flux is
\begin{eqnarray}
  g(z,\bm{Q}) = \sum_{\bm{r}} \frac{z(\bm{r})}{k(\bm{Q})} \, e^{-\frac{1}{2}\bm{r}^T \bm{Q}^{-1} \bm{r}}
\end{eqnarray}
where $k(\bm{Q})$ is an ellipse-dependent normalization we can safely ignore because it will cancel out in the next step.  With $z_c$ the child image (with neighbors replaced with noise as per \secref{neighbor-replacement}) and $z_p$ the parent (original) image, the raw blendedness $b_\mathrm{raw}$ is defined as:
\begin{eqnarray}
  b_\mathrm{raw} = 1 - \frac{g(z_c, \bm{Q})}{g(z_p, \bm{Q})}\,.
\end{eqnarray}
Note that we use the same shape $\bm{Q}$ in both measurements, and that this shape is itself derived from the child image.

To make sure the quantity is well-behaved in the presence of noise, we also compute what we call the absolute blendedness $b_{
\mathrm{abs}}$, in which we use the absolute value of the image along with an attempt to correct the bias this introduces:
\begin{eqnarray}
  b_\mathrm{abs} = 1 - \frac{
    g(\mathrm{max}(\mathrm{abs}(z_c) - d(z_c), 0), \bm{Q})
  }{
    g(\mathrm{max}(\mathrm{abs}(z_p) - d(z_p), 0), \bm{Q})
  }
\label{eqn:absblendedness}
\end{eqnarray}
where $d(z)$ is a de-biasing term derived in \appendixref{blendedness-debiasing}.

Blendedness behaves as one would expect, given its name: for isolated objects, $z_c = z_p$, and $b=0$; for a faint object close to a much brighter object, $b$ approaches unity.  It is also obviously dependent on the actual performance of the deblender, which is responsible for determining $z_c$ from $z_p$ for each child.  This is most problematic when the deblender assigns too much flux to a child, which both artificially increases that child's signal-to-noise ratio and decreases its blendedness, making it less likely that cuts on either will successfully remove poor measurements.  While this is a common deblender failure mode, we have nevertheless found that a conservative blendedness cut can effectively mitigate the problem, because it occurs most frequently for faint objects on the outskirts of extremely bright objects, where $b$ is high even if it is underestimated.  \citet{hsc-shear} include only objects with $\log_{10} b < -0.375$, for instance, in the HSC-SSP weak lensing shear catalog.  We refer the reader to \citet{hsc-synpipe-2} for extensive tests of the blendedness parameter.

\section{Future Work and Conclusions}
\label{sec:future-work-and-conclusions}

Like the SSP survey, the HSC Pipeline is a work in progress.  It is in nearly all respects a state-of-the-art optical imaging data reduction pipeline, and it has already enabled an impressively broad array of scientific discoveries (as demonstrated by the rest of this PASJ special volume).  But it has also in some contexts been the limiting factor in taking full advantage of the exquisite data produced by the HSC camera and the Subaru telescope, highlighting the algorithmic challenges posed by extremely deep, ground-based optical imaging data (and in some cases the challenges posed by excellent seeing).

The biggest algorithmic challenge for HSC going forward is clearly deblending.  Our algorithm, which was quite adequate at the brighter magnitude limit of the SDSS and is still probably the most sophisticated and ambitious approach currently used in production by any many survey, is simply not good enough to handle the more severe blending prevalent in deeper ground-based data.  A completely new algorithm is currently in development that should give us the ability to experiment with much more varied (and frequently more restrictive) constraints than rotational symmetry, such as monotonicity (in the radial profile) and consistency across bands.

PSF modeling is also a serious concern, as we do not currently obtain PSF models accurate enough for weak lensing in the best-seeing images from PSFEx, and we do not believe this can be fixed with configuration changes.  Instead of a purely empirical model that uses a pixel basis for PSF images and polynomial interpolation, our next-generation PSF modeling involves attempting to separate the optical and atmospheric components of the PSF in the spirit of \citet{2016SPIE.9906E..68D}.  This should provide better interpolation and more accurate PSF images (especially in good seeing, where the contribution from the optics is more important).

These issues also highlight the role of HSC as a pathfinder for LSST; LSST's 18,000 $\mathrm{deg}^2$ wide survey will exceed the SSP Wide depth before its fourth data release, and its deepest fields may exceed SSP UltraDeep depths even sooner.  As a result, its blending problems will ultimately be even more severe.  Physical PSF models will also be important for LSST; its requirements on PSF accuracy are significantly tighter, and LSST's optical design will produce larger chromatic PSF effects \citep{2015ApJ...807..182M}, which can probably only be modeled well with a more physically-motivated approach.  Overall, by essentially commissioning an early prototype of LSST pipelines on HSC data and using the results for cutting-edge research, the HSC-SSP has revealed and clarified (and will continue to reveal and clarify) the most challenging algorithmic problems for LSST in a way that ``data challenges'' cannot; there is no substitute for having a collaboration of engaged scientists actually use the outputs of a pipeline to do research.

Efforts to improve many other aspects of the HSC pipeline are also underway, including new algorithms for background subtraction, artifact masking, and galaxy photometry.  The many straightforward bugs noted in this paper that affected the processing of PDR1 are our highest priority; most of these are already fixed in the current version of the pipeline and will be included in the next internal release.  With significant HSC data now available to the full LSST DM team from PDR1 and the algorithmic improvements made on the HSC Pipeline now integrated back into the LSST codebase, we expect this collaboration to be even more productive.  The HSC Pipeline will continue to improve as the SSP survey progresses, advancing the state-of-the-art in image processing algorithms to enable cutting-edge science from one of the highest-quality datasets in astronomy.

\begin{ack}

The Hyper Suprime-Cam (HSC) collaboration includes the astronomical communities of Japan and Taiwan, and Princeton University.  The HSC instrumentation and software were developed by the National Astronomical Observatory of Japan (NAOJ), the Kavli Institute for the Physics and Mathematics of the Universe (Kavli IPMU), the University of Tokyo, the High Energy Accelerator Research Organization (KEK), the Academia Sinica Institute for Astronomy and Astrophysics in Taiwan (ASIAA), and Princeton University.  Funding was contributed by the FIRST program from Japanese Cabinet Office, the Ministry of Education, Culture, Sports, Science and Technology (MEXT), the Japan Society for the Promotion of Science (JSPS), Japan Science and Technology Agency  (JST),  the Toray Science Foundation, NAOJ, Kavli IPMU, KEK, ASIAA, and Princeton University.

The HSC Pipeline and SSP processing would not have been possible without both the software developed by LSST Data Management and the continuing support of DM scientists and engineers who are not members of the HSC collaboration.  We thank the LSST Project for making their code available as free software at http://dm.lsstcorp.org.

The Pan-STARRS1 Surveys (PS1) have been made possible through contributions of the Institute for Astronomy, the University of Hawaii, the Pan-STARRS Project Office, the Max-Planck Society and its participating institutes, the Max Planck Institute for Astronomy, Heidelberg and the Max Planck Institute for Extraterrestrial Physics, Garching, The Johns Hopkins University, Durham University, the University of Edinburgh, Queen's University Belfast, the Harvard-Smithsonian Center for Astrophysics, the Las Cumbres Observatory Global Telescope Network Incorporated, the National Central University of Taiwan, the Space Telescope Science Institute, the National Aeronautics and Space Administration under Grant No. NNX08AR22G issued through the Planetary Science Division of the NASA Science Mission Directorate, the National Science Foundation under Grant No. AST-1238877, the University of Maryland, and Eotvos Lorand University (ELTE) and the Los Alamos National Laboratory.

This paper is based on data collected at the Subaru Telescope and retrieved from the HSC data archive system, which is operated by Subaru Telescope and Astronomy Data Center at National Astronomical Observatory of Japan.

HM is supported by the Jet Propulsion Laboratory, California Institute of Technology, under a
contract with the National Aeronautics and Space Administration.  RM is supported by the Department
of Energy Early Career Award Program. SM is supported by the Japan Society for Promotion of Science grants
JP15K17600 and JP16H01089.
\end{ack}

\bigskip

\bibliographystyle{astroads}
\bibliography{references}

\appendix

\section{Joint Calibration Mathematical Details}

\label{sec:jointcal-details}

In this appendix we describe the detailed mathematical form of the model fit by our joint astromeric and photometric calibration pipeline (\secref{joint-calibration}).  In the following, super- or subscript $s$, $e$, and $c$ denote variables that vary for each star, exposure, or CCD chip, respectively.

The astrometric model consists of several composed transforms.  The first of these relates the focal plane coordinates $(u,v)$ (in units of nominal pixels, \ie~15$\mu m$) for a star to the CCD coordinates $(x,y)$ via the CCD offsets $X_c$ and $Y_c$ and rotations $\theta_c$ by
\begin{eqnarray}
\left[
\begin{array}{c}
u^{s,e} \\
v^{s,e}
\end{array}
\right]
&=&
\left[
\begin{array}{cc}
\cos \theta_c & -\sin \theta_c \\
\sin \theta_c & \cos \theta_c
\end{array}
\right]
\left[
\begin{array}{c}
x^{s,e,c} \\
y^{s,e,c}
\end{array}
\right]
+
\left[
\begin{array}{c}
X_c \\
Y_c
\end{array}
\right]\nonumber\\
&=&
\left[
\begin{array}{c}
x^{s,e,c}\cos\theta_c-y^{s,e,c}\sin\theta_c+X_c \\
x^{s,e,c}\sin\theta_c+y^{s,e,c}\cos\theta_c+Y_c
\end{array}
\right]\,.
\end{eqnarray}
We map also celestial coordinates $(\alpha,\delta)$ for each star to gnomonic projected coordinates $(\xi, \eta)$, with the projection centered at sky position $(p_\alpha, p_\delta)$:
\begin{equation}
\left[
\begin{array}{c}
\xi^{s,e} \\
\eta^{s,e}
\end{array}
\right]
=
\left[
\begin{array}{c}
\xi(\alpha^{s}, \delta^{s}, p_\alpha^{e}, p_\alpha^{e}) \\
\eta(\alpha^{s}, \delta^{s}, p_\alpha^{e}, p_\delta^{e})
\end{array}
\right]\,.
\end{equation}
For sources not matched to a reference catalog object we solve for $\alpha^s$ and $\delta^s$ as part of the fit.  For matched sources $(\alpha, \delta)$ are held fixed to the reference catalog positions.

We relate $(u,v)$ to $(\xi,\eta)$ via 2-d polynomials with coefficients $a$ for $\xi$ and $b$ for $\eta$; the differences between the model and data in $(\xi_,\eta)$ are thus
\begin{eqnarray}
    A_\xi^{s,e} &\equiv&
        \xi^{s,e}-\sum_{j,k} a_{j,k}^e [u^{s,e}]^{j} [v^{s,e}]^{k} \\
    A_\eta^{s,e} &\equiv&
        \eta^{s,e}-\sum_{j,k} b_{j,k}^e [u^{s,e}]^{j} [v^{s,e}]^{k}
\end{eqnarray}
and the sum of differences yields the full $\chi^2$ for the astrometric fit:
\begin{equation}
    \chi^2 = \sum_{e,s} \left[ \left(A_\xi^{s,e}\right)^2 + \left(A_\eta^{s,e}\right)^2 \right]\,.
\end{equation}
We assume all centroid uncertainties are the same.

To minimize, we linearize each residual in all of the free parameters ($a^e$, $b^e$, $X_c$, $Y_c$, and $\theta_c$, as well as $\alpha^s$ and $\delta^s$ for sources not matched to the reference catalog), yielding the matrix equation
\begin{eqnarray}
    \left[
    \begin{array}{c}
        A_\xi \\
        A_\eta
    \end{array}
    \right]
    =
    \bm{J}^T
    \bm{\Delta r}
    \label{eqn:mosaic-block-matrix}
\end{eqnarray}
with $\bm{r}$ the vector of parameters defined block-wise as
\begin{eqnarray}
    \bm{r} \equiv \left[
        \begin{array}{c}
            a^e \\
            b^e \\
            X_c \\
            Y_c \\
            \theta_c \\
            \alpha^s \\
            \delta^s
        \end{array}
    \right]
\end{eqnarray}
and $\bm{J}$ the matrix of first derivatives with blocks
\begin{eqnarray}
    \bm{J} \equiv \left[
    \begin{array}{c c}
        \frac{\partial A_\xi^{s,e}}{\partial r} &
        \frac{\partial A_\eta^{s,e}}{\partial r} \\
    \end{array}
    \right]
    =
    \left[
    \begin{array}{ c c }
    -P^{s,e} & 0 \\
     0 & -P^{s,e} \\
    -B_\xi^{s,e} & -B_\eta^{s,e} \\
    -C_\xi^{s,e} & -C_\eta^{s,e} \\
    -D_\xi^{s,e} & -D_\eta^{s,e} \\
    \frac{\partial\xi^{s,e}}{\partial\alpha^s} &
     \frac{\partial\eta^{s,e}}{\partial\alpha^s}\\
    \frac{\partial\xi^{s,e}}{\partial\delta^s} &
     \frac{\partial\eta^{s,e}}{\partial\alpha^s}
    \end{array}
    \right]
\end{eqnarray}
with (dropping most $s,e$ superscripts for brevity)
\begin{eqnarray}
P_{j,k} &=& u^j v^k \\
B_\xi &=&  \sum_{j,k} a_{j,k} \, j \, u^{j-1} v^{k} \\
B_\eta &=& \sum_{j,k} b_{j,k} \, j \, u^{j-1} v^{k} \\
C_\xi &=& \sum_{j,k} a_{j,k} \, k \, u^{j} v^{k-1} \\
C_\eta &=& \sum_{j,k} b_{j,k} \, k \, u^{j} v^{k-1} \\
D_\xi &=& \sum_{j,k} a_{j,k} \, u^{j-1} v^{k-1}
    \Biggl[-j\, v \left(x^s \sin\theta_c + y^s\cos\theta_c\right)\Biggr.
        \nonumber\\
         & & \quad\quad\quad \Biggl. + \, k\, u \left(x^s\cos\theta_c - y^s\sin\theta_c\right) \Biggr] \\
D_\eta &=& \sum_{j,k} b_{j,k} \, u^{j-1} v^{k-1}
    \Biggl[-j\, v\left(x^s\sin\theta_c+y^s\cos\theta_c\right)\Biggr.
        \nonumber\\
         & & \quad\quad\quad \Biggl. + \, k\, u\left(x^s\cos\theta_c-y^s\sin\theta_c\right) \Biggr]
    \,.
\end{eqnarray}
Note that $P$, $a$, and $b$ are flattened into vectors in the block matrix form; the ordering of the two polynomial dimensions and the index over exposures is unimportant as long it is consistent for all three quantities.

We initialize all free parameters with the values from the single-visit astrometric fit performed in CCD Processing (\secref{ccd-processing}), and fit by iteratively solving \eqnref{mosaic-block-matrix}.

In the photometric fit, we solve for polynomial coefficients $f_{j,k}$, per-exposure magnitude offsets $dm^e$, and per-source true magnitudes $m_0^s$, with the latter held fixed at the reference catalog magnitude for sources matched to the reference catalog.  We minimize
\begin{eqnarray}
\chi^2 &=& \sum_{e,s} \left[ m_0^s - \Biggl( m^{s,e} + dm^e + \Biggl. \right.
    \nonumber\\
    &\,& \quad\quad\quad \left. \Biggl.
    \sum_{j+k\le n} f_{j,k} T_j(u^{s,e}) T_k(v^{s,e}) \Biggr) \right]^2
\end{eqnarray}
where $T_n$ is the nth-order Chebyshev polynomial of the first kind.  As with the astrometric fit, we initialize with the values from CCD Processing and proceed by iteratively solving a linearized version of the difference between the true magnitude $m^s_0$ and the per-exposure measured magnitudes $m^{s,e}$.

Both the astrometric and photometric fits are performed independently for each tract, using inputs from only those CCDs that overlap the tract; CCDs that overlap multiple tracts are fit multiple times.  In the HSC-SSP Wide layer, each iteration of the astrometric fit involves an LU factorization of an approximately 6000$\times$6000 square matrix.  The matrix for each photometric fit iteration is approximately 2500$\times$2500 (again in the Wide layer).

\section{CModel Algorithm Details}
\label{sec:cmodel-details}

The elliptical exponential and de~Vaucouleurs models fit in the CModel algorithm have the precise form
\begin{eqnarray}
  f(\bm{r}, \bm{S}, \alpha) =
    \alpha \, k\!\left(\left\Vert\bm{S}^{-1}\bm{r}\right\Vert\right)
\end{eqnarray}
where $\bm{r}$ is the pixel position (relative to the source centroid), $\alpha$ is the amplitude parameter, $k(r)$ is the radial profile, and $\bm{S}$ is the ``generating transform'' (see \appendixref{ellipse-parameterizations}) of the isophotal ellipse that contains half the flux of the model.  The matrix $\bm{S}$ depends on the ellipse parameters.  We use the $\{\eta_1, \eta_2, \ln r_{\mathrm{tr}}\}$ parameterization of the ellipse (again, see \appendixref{ellipse-parameterizations}) because any triple in $\mathbb{R}^3$ is a valid ellipse in this parameterization and hence we do not need a numerical optimizer that can handle parameter constraints.

The profiles $k(r)$ themselves are approximations to the exponential profile
\begin{equation}
k(r) \propto e^{-r/s}
\end{equation}
and de~Vaucouleurs profile
\begin{equation}
k(r) \propto e^{-7.67[(r/s)^{1/4} - 1]}
\end{equation}
formed from a linear combination of Gaussians with different widths, as described in \citet{2013PASP..125..719H}.  More specifically, we use their six-Gaussian \texttt{lux} and eight-Gaussian \texttt{luv} profiles.  These are matched to the modified exponential and de~Vaucouleurs profiles used in SDSS model fitting, which have softened cores and truncated wings (at four or eight half-light radii, respectively).  Both of these modifications allow the multi-Gaussian approximations to better approximate the ideal profiles in the range of radii over which they can be best constrained by our data, and the truncation in particular helps to prevent spuriously large fluxes due to extrapolation.

The primary advantage of these multi-Gaussian approximations is that they allow fast convolution, \emph{if} the PSF is approximated by a compatible analytic function.  Rather than follow \citet{2013PASP..125..719H} and \citet{2014MNRAS.444L..25S} in using sums of Gaussians for the PSF, however, we use a sum of two Gauss-Hermite (``shapelet'') expansions with different scales. As demonstrated by \citet{2010AJ....140..870B}, using multiple Gauss-Hermite expansions provides significantly more flexibility than using a single one in representing functions with broader-than-Gaussian tails while maintaining fast convolution properties.  We fit the Gauss-Hermite PSF approximation to our image-based PSF model (either the single-epoch PSFEx model or the effective coadd PSF described in \secref{psf-coaddition}) only after evaluating the image-based model at the position of the source.

When fitting for ellipse parameters in the separate exponential and de Vaucouleurs fits (steps 1 and 2 in \secref{cmodel-photometry}), we combine the likelihood with a simple Bayesian prior on the ellipse parameters to form the objective function that we optimize.  The prior is flat in amplitude and position angle, and is separable in $\ln r_{\mathrm{tr}}$ and $\eta = \sqrt{\eta_1^2 + \eta_2^2}$.  The prior on $\ln r_{\mathrm{tr}}$ is piecewise, transitioning from flat to a Student's $t$ distribution at $\sim 0.3$\arcsec\ ($\ln \frac{r_\mathrm{tr}}{\mathrm{arcsec}} = -1$), which is also the peak of the Student's $t$.  The prior on the flux is intentionally flat, and we do not condition the prior for the ellipse parameters on the flux.  Doing so may be worthwhile, but is potentially dangerous; flux measurements are expected to respond strictly linearly to changes in the true underlying brightness, and a prior on flux could disrupt that scaling.

\begin{figure}
\includegraphics[width=0.48\textwidth]{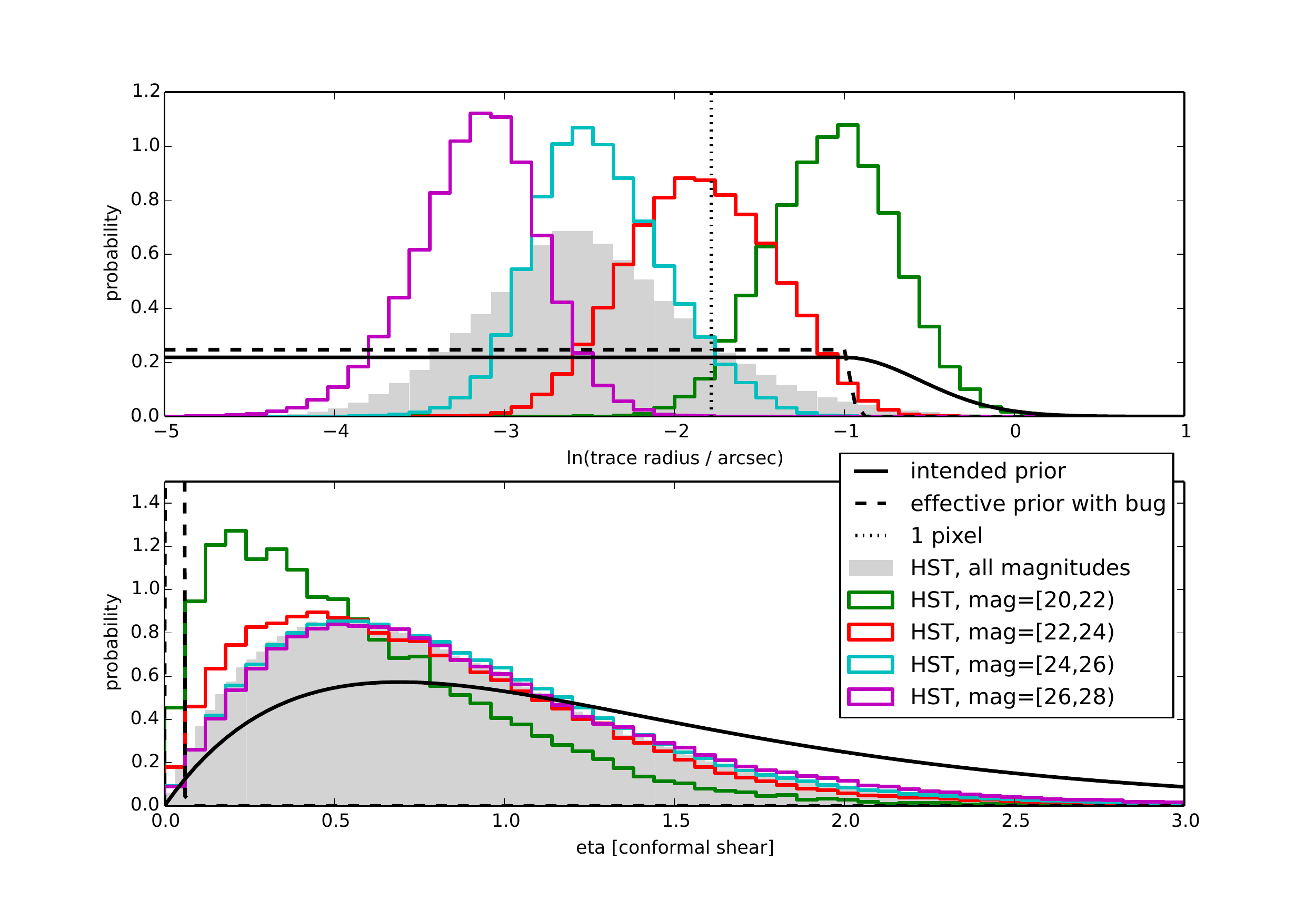}
\caption{
    Bayesian priors used in CModel fitting and the histograms of HST COSMOS sizes and ellipticities they were designed to support.  The effective prior in the presence of the bug discussed in the text has a significantly sharper cutoff in radius (yielding the ``pile-up'' in \figref{cmodel-prior-bug} at PSF-CModel $\sim 0.6$), and the effective ellipticity prior forces low S/N galaxies to have nearly circular models.  Because most galaxies (in the fainter magnitude bins) are already below the cutoff imposed by the radius prior, the bug should in fact move them closer to their true radii.
    \label{fig:cmodel-priors-cosmos}
}
\end{figure}

\begin{figure*}
\includegraphics[width=\textwidth]{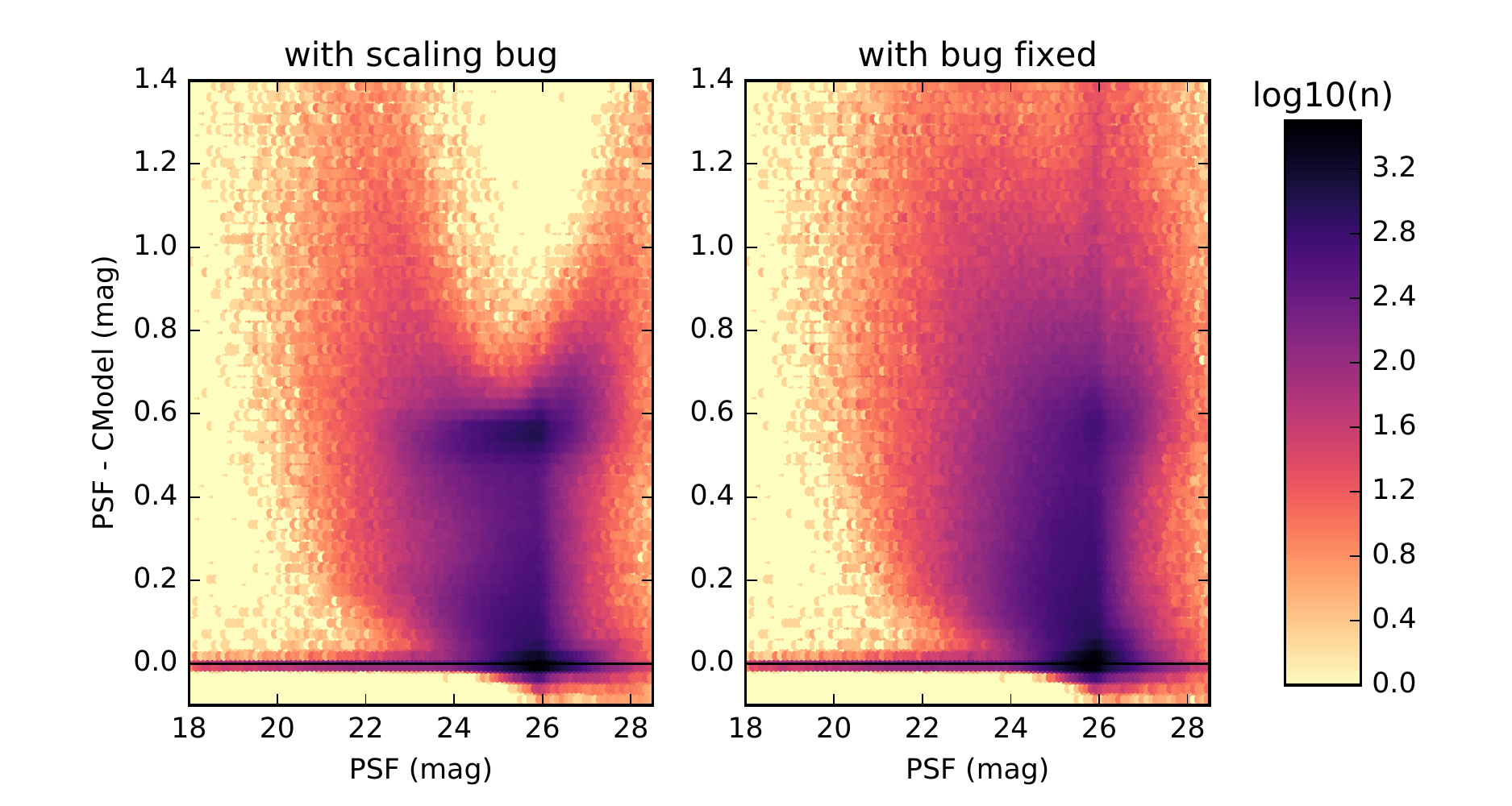}
\caption{
    Comparison of PSF and CModel magnitudes both with (left) and without (right) the likelihood-prior weighting bug discussed in the text.  Faint galaxies have seem to have their fluxes artificially decreased to $m_\mathrm{psf}-m_\mathrm{cmodel} \sim 0.6$ by the bug.  Many of these have true magnitudes that are probably even closer to their PSF magnitudes, so the bug moves them in the right direction (albeit inconsistently).
  \label{fig:cmodel-prior-bug}
}
\end{figure*}

The distribution in ellipticity and the outer piece of the distribution in radius are designed to be broader than the empirical distributions of galaxy sizes and ellipticities in galaxies in the HST COSMOS catalog of \citet{2007ApJS..172..219L}, as shown in \figref{cmodel-priors-cosmos}.  Unfortunately, a bug in the relative weighting of the likelihood and prior in the version of the code used in the first HSC-SSP public data release resulted in the effective prior being much stronger than intended.  The most significant result of this bug is a very sharp cutoff in the radius prior at the $\sim 0.37$\arcsec\ transition point.  Ellipticities are also biased low (more circular), but this has a small effect on the flux.  The radius bias yields an artificial ``pile-up'' of galaxies at $m_{\mathrm{psf}}-m_{\mathrm{cmodel}} \sim 0.6$ as shown in \figref{cmodel-prior-bug}.  In at least one context, however, this biased flux is actually be an improvement; even minor deblend failures can cause CModel radii for faint galaxies to be spuriously large, and the sharp effective prior suppresses this effect.  Because all bands are approximately equally affected by this bug, galaxy colors from CModel do not appear to be biased nearly as strongly as fluxes.  As a result, we have been reluctant to fix this bug until we have another approach in place to address CModel sensitivity to deblender failures.  A radius prior that is more informative but matched to the true distribution of faint galaxy radii is the most likely candidate.  This would probably have to be at least conditioned on flux to allow bright (and typically larger) galaxies to see a different radius prior, and as we have noted above, it is not clear whether such a prior can yield a linear response of the measurement to changes in intrinsic brightness.  Given the dependency of the measured flux on the best-fit radius and the difficulty of determining unbiased galaxy sizes, it is clear our current CModel fluxes are already at some level nonlinear.  We believe this is true of all existing approaches to measuring galaxy photometry that infer a model or aperture size from the data.

\section{Ellipse Parameterizations}
\label{sec:ellipse-parameterizations}

Many of the algorithms for photometry and shapes in the HSC Pipeline rely on different parameterizations of ellipse, including the second moments discussed in \secref{shapes}, the Kron photometry algorithm discussed in \secref{kron-photometry}, and the CModel algorithm (\secref{cmodel-photometry} and \appendixref{cmodel-details}).  We describe here the relationships between these parameterizations and our conventions for them for clarity and reader convenience.

For our purposes, all ellipse parameterizations are three numbers, as we generally consider the position of the ellipse (another two numbers) as a distinct quantity.  The most intuitive parameterization is a combination of semi-major radius ($A$), semi-minor axis ($B$), and position angle ($\theta$), which we define here as the angle from the x-axis to the semi-major axis, measured counterclockwise.

In addition to its intuitive simplicity, the $\{A, B, \theta\}$ parameterization makes it easy to compute what we will call the ellipse's ``generating transform'' $\bm{S}$: a linear transform that maps the unit circle to the ellipse:
\begin{eqnarray}
  \bm{S} = \left[
    \begin{array}{ c c }
      \cos \theta  & \sin \theta \\
      -\sin \theta & \cos \theta
    \end{array}
  \right]
  \left[
    \begin{array}{ c c }
      A & 0 \\
      0 & B
    \end{array}
  \right]
\end{eqnarray}
The generating transform is not unique; it may be multiplied on the right by any orthogonal matrix.  In addition to the form we have defined above, a symmetric form where this matrix is the inverse of the rotation on the left is also particularly common.  As described in \appendixref{cmodel-details}, the ellipse-generating transform can be used to map a 1-d radial profile to an elliptical model.

Most algorithms based on the moments of images (see \secref{shapes}) instead use the three unique elements of a symmetric positive definite $2 \times 2$ matrix we will call $\bm{Q}$.  For the 1$\sigma$ contour of an elliptical 2-d Gaussian, $\bm{Q}$ is just the covariance matrix.  The three unique elements of $\bm{Q}$ are most easily related to the $\{A, B, \theta\}$ parameterization by the fact that $\bm{S}$ is the matrix square root of $\bm{Q}$:

{
\footnotesize
\begin{eqnarray}
  \bm{Q} &=& \bm{S}\bm{S}^T \nonumber\\
  &=&
  \left[
    \begin{array}{ c c }
      \cos \theta  & \sin \theta \\
      -\sin \theta & \cos \theta
    \end{array}
  \right]
  \left[
    \begin{array}{ c c }
      A^2 & 0 \\
      0 & B^2
    \end{array}
  \right]
  \left[
    \begin{array}{ c c }
      \cos \theta  & -\sin \theta \\
      \sin \theta & \cos \theta
    \end{array}
  \right]
  \,.
\end{eqnarray}
}

This also means that $A^2$ and $B^2$ are the eigenvalues of $\bm{Q}$, and the columns of the rotation matrix for $\theta$ are its eigenvectors.

Weak gravitational lensing frequently uses definitions of ellipticity comprised of two real numbers that can be interpreted as a single complex number.  The magnitude of the complex ellipticity is some function of $A$ and $B$, and its phase is $2\theta$.  Common functions to define the magnitude include the conformal shear
\begin{eqnarray}
  \eta = \tanh\!\left(\frac{a}{b}\right),
\end{eqnarray}
the distortion
\begin{eqnarray}
  \delta = \frac{A^2 - B^2}{A^2 + B^2},
\end{eqnarray}
and what we will call the shear\footnote{Both $\delta$ and $g$ are confusingly referred to as both ellipticity and eccentricity in the literature; we use ``distortion'' and ``shear'' because, to our knowledge, $\delta$ is never called shear and $g$ is never called distortion.}
\begin{eqnarray}
  g = \frac{A - B}{A + B}.
\end{eqnarray}
All ellipticities in this paper use the distortion definition.  The conformal shear $\eta$ is defined on $(0, \infty)$, and hence its real and imaginary parts
\begin{eqnarray}
  \eta_1 &=& \eta \cos 2\theta \\
  \eta_2 &=& \eta \sin 2\theta
\end{eqnarray}
are defined on $(-\infty, \infty)$, making them very useful parameters for working with unconstrained optimizers or probability distributions with infinite extents.  $\delta$ and $g$ are defined on $(0, 1)$ and their real and imaginary parts on the unit circle, making them less useful for these purposes, but they are easier to connect to gravitational lensing observables.

To represent a full ellipse using any of these definitions of complex ellipticity, we need to add a radius parameter.  Either $A$ or $B$ may be used in this role, but we tend to instead use two definitions related to the determinant and trace of $\bm{Q}$
\begin{eqnarray}
  r_{\mathrm{det}} &=& \det(\bm{Q})^{\frac{1}{4}} = \sqrt{AB} \\
  r_{\mathrm{tr}} &=&  \sqrt{\frac{\mathrm{tr}(\bm{Q})}{2}}
      = \sqrt{\frac{A^2 + B^2}{2}}
\,.
\end{eqnarray}
The \textit{determinant radius} $r_{\mathrm{det}}$ is proportional to the square root area of the ellipse, but we prefer the \textit{trace radius} $r_{\mathrm{tr}}$ (or its logarithm) for fitting, as it remains nonzero (finite) when $B$ approaches zero but $A$ does not, which is common when fitting noisy, unresolved objects as in the CModel algorithm.

\section{Correcting for Bias in Absolute Blendedness}
\label{sec:blendedness-debiasing}

In order to compute a blendedness parameter that is well behaved in the presence of noise, we use the absolute value of the image, rather than the image itself while computing the blendedness parameter in \eqnref{absblendedness}. For objects which are well above the noise level, this procedure should have no effect on the blendedness parameter, but using the absolute value will bias the blendedness for objects which are close to the noise level. We correct for such a bias using the following formalism.

Consider an object which results in a photon count $z(\bm{r})$ in a pixel located at ${\bm r}$ in the image plane. Under the assumption that the photon counts are Gaussian-distributed with mean $\mu$ and dispersion $\sigma$, the total integrated flux is given by
\begin{equation}
	F = \int {\rm d}{\bm r} \, z({\bm r}) = A \mu\,,
\end{equation}
where $A$ is the area over which the integral is performed, and we have assumed a uniform profile for the object. Instead, if we integrate $|z|$ over the same area, we obtain a biased answer
\begin{eqnarray}
	F' &=& \int {\rm d}{\bm r} \, |z({\bm r})| \nonumber \\
	   &=& A \left[ \sqrt{\frac{2}{\pi}} \sigma \left( \exp\left[-\frac{\mu^2}{2\sigma^2}\right] \right) + \mu\,{\rm erf}\left(\frac{\mu}{\sqrt{2}\sigma}\right) \right] \,.
\end{eqnarray}
This bias can be corrected if instead of $|z|$, we integrate $|z|-d$, where $d$ is given by
\begin{equation}
	d = \sqrt{\frac{2}{\pi}} \sigma \left( \exp\left[-\frac{\mu^2}{2\sigma^2}\right] \right) - \mu\,{\rm erfc}\left(\frac{\mu}{\sqrt{2}\sigma}\right)\,.
\end{equation}

Because that the objects in our images will have a non-uniform profile, we will use $d(z)$ instead of a constant $d$ in \eqnref{absblendedness}.

Given the photon count $z(\bm{r})$ and the dispersion $\sigma$, we thus need to estimate $\mu$ in order to compute $d(z)$. We compute $\mu(z, \sigma)$ by imposing the condition that the expected photon count be greater than 0. Therefore,
\begin{eqnarray}
	\mu(z, \sigma) &=& \int_{0}^{\infty} {\rm d}z' z' P(z'|z, \sigma) \nonumber \\
	   &=& \frac{\sigma}{\sqrt{2\pi}} \exp\left[ -\frac{z^2}{2\sigma^2} \right] + \frac{z}{2}{\rm erfc}\left[-\frac{z}{\sqrt{2}\sigma}\right]\,.
\end{eqnarray}

The original implementation (used in PDR1) of the bias correction for blendedness used an incorrect formula, resulting primarily in a remapping of the blendedness values for low S/N objects (though it was not strictly a remapping).  Because all users of the blendedness parameters have relied on calibrating its values directly via e.g. via simulations \citep{hsc-synpipe-2} instead of relying on its theoretical interpretation, we do not expect this bug to significantly affect science analyses.

\end{document}